\documentclass[%
 reprint,
 superscriptaddress,
 nobibnotes,
 amsmath,amssymb,
 aps,
 prx,
 longbibliography,
]{revtex4-2}
\usepackage{graphicx}
\usepackage{dcolumn}
\usepackage{bm}
\usepackage[colorlinks=true,allcolors=blue]{hyperref}

\usepackage{amsmath,amssymb,amsfonts}%
\usepackage{xcolor}%
\usepackage[version=3]{mhchem}
\usepackage{natbib}

\begin{document}

\preprint{APS/123-QED}

\title{Attosecond X-ray Core-level Chronoscopy of Aromatic Molecules}

\newcommand{\AffilETH}{Laboratory of Physical Chemistry, ETH Z\"urich, Zurich, Switzerland}
\newcommand{\AffilSLAC}{SLAC National Accelerator Laboratory, Menlo Park, CA, USA}
\newcommand{\AffilPULSE}{Stanford PULSE Institute, SLAC National Accelerator Laboratory, Menlo Park, CA, USA}
\newcommand{\AffilLUXS}{Laboratory for Ultrafast X-ray Sciences, EPFL, Lausanne, Switzerland}
\newcommand{\AffilCSU}{California State University, Maritime Academy, Vallejo, CA, USA}
\newcommand{\AffilZJU}{State Key Laboratory of Extreme Photonics and Instrumentation, College of Optical Science and Engineering, Zhejiang University, Hangzhou, China}
\newcommand{\AffilKSU}{J. R. Macdonald Laboratory, Department of Physics, Kansas State University, Manhattan, KS, USA}
\newcommand{\AffilLBL}{Lawrence Berkeley National Laboratory, Berkeley, CA, USA}
\newcommand{\AffilUCD}{Department of Chemistry, University of California, Davis, CA, USA}
\newcommand{\AffilTHK}{Department of Chemistry, Tohoku University, Sendai, Japan}
\newcommand{\AffilSHT}{School of Physical Science and Technology, ShanghaiTech University, Shanghai, China}


\author{Jia-Bao Ji}\thanks{These authors contributed equally.}
\affiliation{\AffilETH}

\author{Zhaoheng Guo}\thanks{These authors contributed equally.}
\affiliation{\AffilSLAC} 
\affiliation{\AffilPULSE} 
\affiliation{\AffilLUXS}

\author{Taran Driver} 
\affiliation{\AffilSLAC} 
\affiliation{\AffilPULSE} 

\author{Cynthia S. Trevisan} 
\affiliation{\AffilCSU}

\author{David Cesar} \affiliation{\AffilSLAC}
\author{Xinxin Cheng} \affiliation{\AffilSLAC}
\author{Joseph Duris} \affiliation{\AffilSLAC}
\author{Paris L. Franz} \affiliation{\AffilSLAC} \affiliation{\AffilPULSE}
\author{James Glownia} \affiliation{\AffilSLAC}
\author{Xiaochun Gong} \affiliation{\AffilETH} \affiliation{\AffilZJU}
\author{Daniel Hammerland} \affiliation{\AffilETH}
\author{Meng Han} \affiliation{\AffilETH} \affiliation{\AffilKSU}
\author{Saijoscha Heck} \affiliation{\AffilETH}
\author{Matthias Hoffmann} \affiliation{\AffilSLAC}
\author{Andrei Kamalov} \affiliation{\AffilSLAC}
\author{Kirk A. Larsen} \affiliation{\AffilSLAC} \affiliation{\AffilPULSE}
\author{Xiang Li} \affiliation{\AffilSLAC}
\author{Ming-Fu Lin} \affiliation{\AffilSLAC}
\author{Yuchen Liu} \affiliation{\AffilLBL} \affiliation{\AffilUCD}
\author{C. William McCurdy} \affiliation{\AffilLBL} \affiliation{\AffilUCD}
\author{Razib Obaid} \affiliation{\AffilSLAC}
\author{Jordan T. O'Neal} \affiliation{\AffilPULSE}
\author{Thomas N. Rescigno} \affiliation{\AffilLBL}
\author{River R. Robles} \affiliation{\AffilSLAC} \affiliation{\AffilPULSE}
\author{Nicholas Sudar} \affiliation{\AffilSLAC} \affiliation{\AffilPULSE}
\author{Peter Walter} \affiliation{\AffilSLAC}
\author{Anna L. Wang} \affiliation{\AffilSLAC} \affiliation{\AffilPULSE}
\author{Jun Wang} \affiliation{\AffilSLAC} \affiliation{\AffilPULSE}
\author{Thomas J. A. Wolf} \affiliation{\AffilSLAC} \affiliation{\AffilPULSE}
\author{Zhen Zhang} \affiliation{\AffilSLAC}
\author{Kiyoshi Ueda} \affiliation{\AffilETH} \affiliation{\AffilTHK} \affiliation{\AffilSHT}

\author{Robert R. Lucchese} 
\email{rlucchese@lbl.gov} 
\affiliation{\AffilLBL}

\author{Agostino Marinelli}\email{marinelli@slac.stanford.edu}
\affiliation{\AffilSLAC}
\affiliation{\AffilPULSE} 

\author{James P. Cryan}\email{jcryan@stanford.edu}
\affiliation{\AffilSLAC}
\affiliation{\AffilPULSE} 

\author{Hans Jakob W\"orner}\email{hwoerner@ethz.ch}
\affiliation{\AffilETH}

\begin{abstract}
Attosecond photoemission or photoionization delays are a unique probe of the structure and the electronic dynamics of matter. However, spectral congestion and spatial delocalization of valence electron wave functions set fundamental limits to the complexity of systems that can be studied and the information that can be retrieved, respectively. Using attosecond X-ray pulses from LCLS, we demonstrate the key advantages of measuring core-level delays: the photoelectron spectra remain atom-like, the measurements become element specific and the observed scattering dynamics originate from a point-like source. We exploit these unique features to reveal the effects of electronegativity and symmetry on attosecond scattering dynamics by measuring and calculating the photoionization delays between N-1s and C-1s core shells of a series of aromatic azabenzene molecules. Remarkably, the delays increase with the number of nitrogen atoms in the molecule and reveal multiple resonances. We identify two previously unknown mechanisms regulating the associated attosecond dynamics, namely the enhanced confinement of the trapped wavefunction with increasing electronegativity of the atoms and the decrease of the coupling strength among the photoemitted partial waves with increasing symmetry. This study demonstrates the unique opportunities opened by measurements of core-level photoionization delays for unraveling attosecond electron dynamics in complex matter.  
\end{abstract}

\maketitle


\section{Introduction}
Tracking attosecond multi-electron dynamics with atomically resolved spatial information is the current focus of numerous research efforts worldwide. X-ray spectroscopy has established itself as a powerful solution to breaking the complexity barriers faced by valence-shell spectroscopies. These barriers are particularly pronounced in time-resolved studies of the photoelectric effect, known as attosecond chronoscopy \cite{schultze10a,kluender11a,pazourek15a}. This technique has integrated time-domain access to multi-electron dynamics into the powerful framework of electron-scattering physics which has created a flourishing research field with applications to molecules \cite{huppert16a,vos18a,biswas20a,kamalov20a,nandi20a,heck21a}, clusters \cite{gong22a,heck22a}, liquids \cite{jordan20a} and solids \cite{cavalieri07a,neppl15a,tao16a,siek17a}. Such studies have traditionally made use of attosecond pulses in the extreme ultraviolet (up to 120 eV), which has so far prevented access to atomic core levels. 

Here, we demonstrate the unique opportunities that arise from advancing the research field of attosecond chronoscopy from the extreme-ultraviolet to the X-ray domain. This adds element specificity and site selectivity to the method, while simultaneously confining the emission site of the photoelectron wave to a point-like source. Moreover, core-shell photoelectron spectra have a much simpler structure than valence-shell spectra, which facilitates the extension of the method to complex forms of matter. This advance was made possible by groundbreaking progress in free-electron-laser (FEL) science that has led to the generation of attosecond pulses in the X-ray domain through the X-ray laser-enhanced attosecond pulse generation (XLEAP) technique \cite{duris20a}. The long-standing challenge of temporal jitter between the FEL pulses and laser pulses was solved by using attosecond angular streaking with a circularly-polarized infrared laser pulse and detecting the photoelectron momentum distribution in the polarization plane \cite{li18a,li22a,driver24a}. 

\begin{figure*}
\centering\includegraphics[width=\textwidth]
{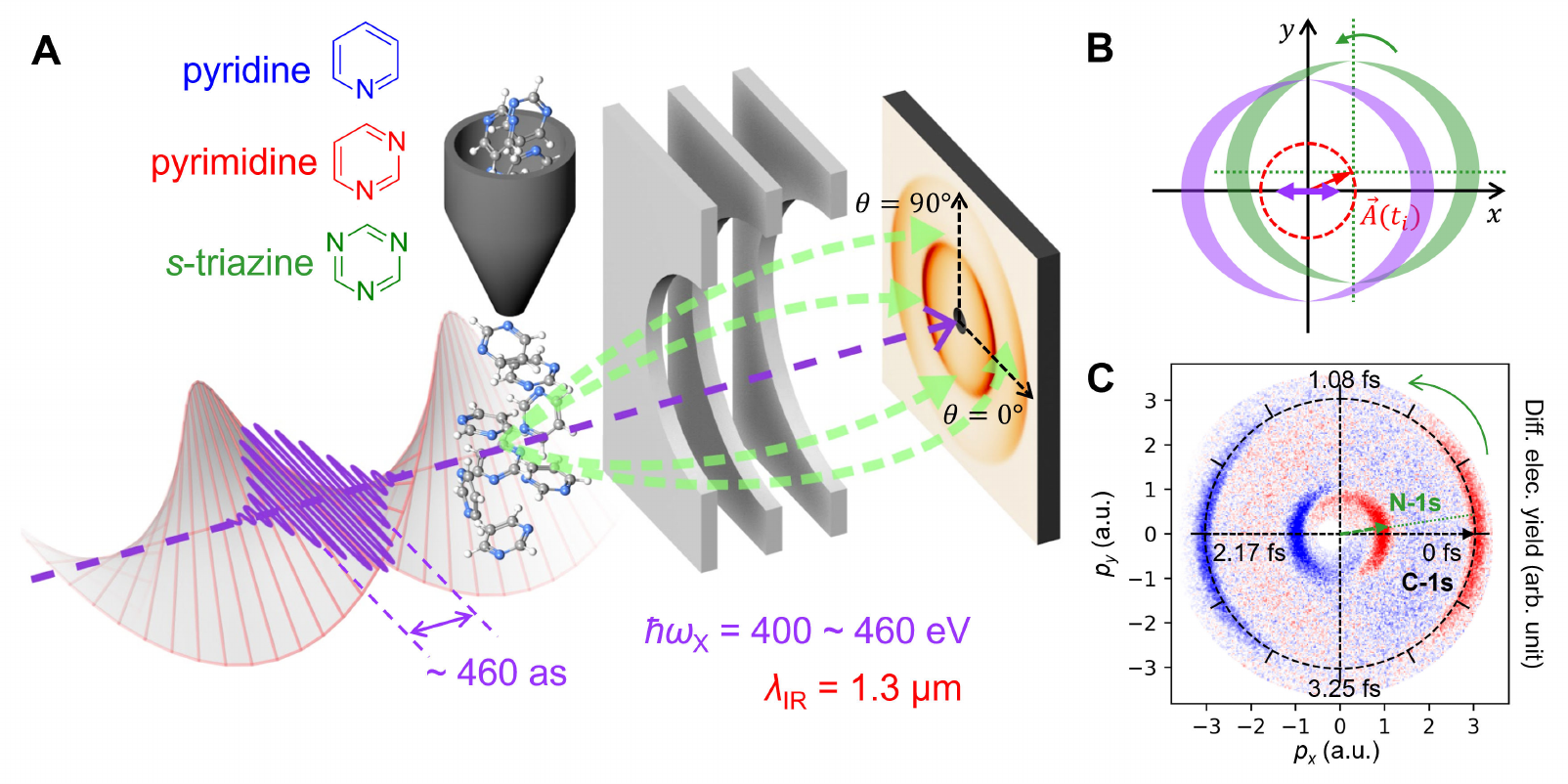}
\caption{\textbf{The principle of measuring core-level photoionization delays with an XFEL.} 
(A) Gas-phase azabenzene molecules (pyridine, pyrimidine or $s$-triazine) are ionized by the attosecond X-ray free-electron laser (XFEL) pulses in the presence of a $1.3~\mu$m circularly polarized streaking laser field. The photoelectrons are collected with a c-VMI spectrometer, with $\theta$ as the polar angle on the detector. (B) The initial photoelectron momentum distribution (purple) is shifted by the vector potential of the IR field $\Vec{A}(t_i)$ at the instant of ionization $t_i$, where the purple double arrow indicates the X-ray polarization direction. (C) Difference image between streaking within a narrow range of angles and the average over all streaking angles, revealing the angular offset of the N-1s and C-1s electrons, which directly encodes their time delay. }
\label{fig:1}
\end{figure*}

The present experiments establish the unique capabilities of X-ray chronoscopy, recently demonstrated in a diatomic molecule \cite{driver24a}, at unraveling the mechanisms of attosecond electron dynamics in complex systems, notably in aromatic molecules (i.e., benzene derivatives). Specifically, our study reveals the multiple-scattering phenomena that lead to transient trapping of photoelectrons following their release from specific atoms. 
Our results reveal two previously unknown mechanisms regulating such attosecond scattering dynamics. First, we find that substituting the ionized functional group with a more electronegative one (C-H $\rightarrow$ N) increases the trapping times. Second, we show that increasing the symmetry of the molecular scaffold also increases the trapping times by suppressing the coupling to the less-trapped partial waves with lower angular momentum. To demonstrate these effects, we chose the azabenzene molecules pyridine, pyrimidine and $s$-triazine, depicted in Fig. \ref{fig:1}A. 
We produced attosecond X-ray pulses with energies just above the nitrogen K-edge located at $\sim$405~eV and measured the photoionization delays between the electrons emitted from the nitrogen K-shell~(N-1s) and those emitted from the carbon K-shell~(C-1s). Since the latter have kinetic energies of more than 110~eV, the C-1s electrons are emitted with negligible photoionization delays ($\lesssim 5$~as), defining a self-referenced technique for accessing absolute photoionization delays of the N-1s electrons. We measured N-1s photoionization delays of up to $\sim$300~as close to the nitrogen K-shell threshold, which generally decrease with increasing kinetic energy but display local maxima at 3-15~eV above the threshold. The measured delays indeed markedly increase from pyridine to pyrimidine. These experimental results are compared with state-of-the-art core-level photoionization calculations, which reveal the presence of several shape resonances in this low kinetic-energy region. The comparison of experiment and theory reveals that the photoelectrons originating from the nitrogen K-shell are trapped with increasing efficiency in the molecular plane in the sequence pyridine, pyrimidine, $s$-triazine, i.e., with increasing number of nitrogen atoms. Additionally, the increasing symmetry of the molecular scaffold (from C$_{\rm 2v}$ in pyridine to D$_{\rm 3h}$ in $s$-triazine) reduces the coupling among the photoelectron partial waves, which frustrates the decay of the trapped partial waves with high angular momentum. 
Pyrimidine is a building block of the nucleobases thymine, cytosine and uracil. All three azabenzene molecules are widespread structural motifs in biomolecules, drugs and molecular optoelectronics \cite{ling21a,selvam15a,sharma21a,moliton10a}. The discovered mechanisms therefore suggest guidelines for extending attosecond chronoscopy to biomolecules and molecules used in optoelectronics.

\section{Experimental setup and data analysis}
Figure \ref{fig:1} illustrates the principle of the present experiments, carried out at the time-resolved atomic, molecular, and optical science~(TMO) experimental hutch~\cite{Walter22a} of the Linac Coherent Light Source (LCLS). Linearly polarized attosecond X-ray pulses with a tunable central energy were obtained using the XLEAP technique \cite{duris20a} and were superimposed with circularly polarized short-wave infrared (IR, 1.3~$\mu$m, $\sim$40~fs) pulses in the interaction region of a co-axial velocity-map-imaging (c-VMI) spectrometer. A continuous molecular beam containing pyridine, pyrimidine or $s$-triazine intersected with the overlapping X-ray and IR pulses in the interaction region. The photoelectrons created from nitrogen K-shell ionization of the sample were streaked by the IR pulses and projected onto a position-sensitive imaging detector with a central hole. A photoelectron image recorded by ionizing pyrimidine with X-ray pulses centered at 454~eV is shown in Fig.~\ref{fig:1}A, whereby the inner (outer) ring corresponds to ionization of the nitrogen~(carbon) K-shell electrons. 
Figure \ref{fig:1}C shows how the relative photoionization delays between the N-1s and C-1s photoelectrons were determined. Because each photoelectron was angularly streaked into the direction of the IR-laser vector potential at the instant of photoionization (Fig.\ref{fig:1}B), the relative streaking angle between N-1s and C-1s electrons directly encodes their relative photoionization delay. The conversion from relative streaking angle $\Delta \vartheta$ to the relative photoionization delay $\Delta\tau$ is simply given by $\Delta\tau = T_L\times \Delta \vartheta/2\pi$, where $T_L$ is the period of the streaking laser \cite{zhao05a,eckle08b,kazansky16a,kazansky19a,hartmann18a,kheifets_ionization_2022,serov23a}.

\begin{figure*}
\centering\includegraphics[width=0.8\textwidth]{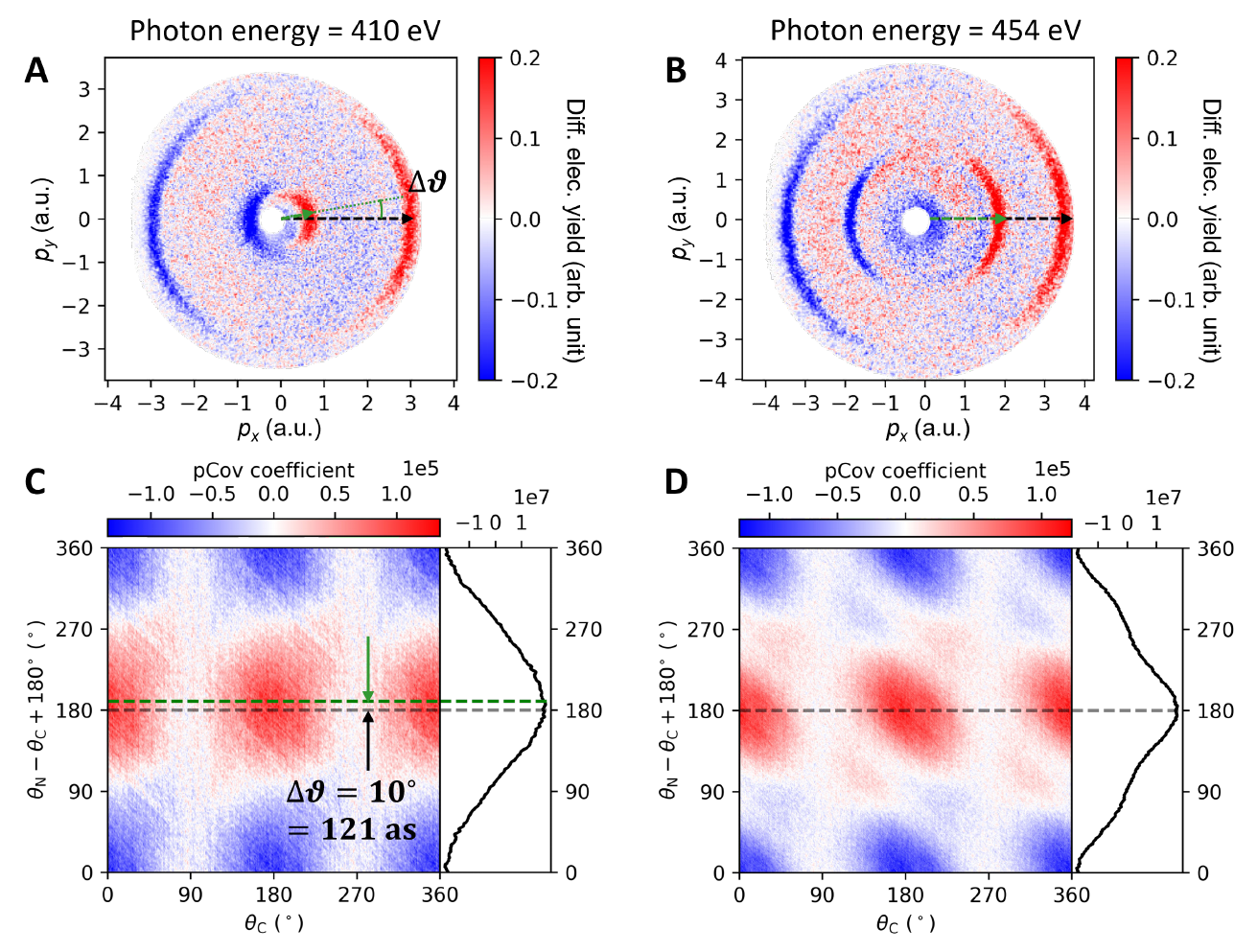}
\caption{\textbf{Using the partial covariance analysis to access the differential streaking angles.} (A,B) Difference images obtained as outlined in Fig.~\ref{fig:1}, illustrating a rather large delay (at 410~eV) and a negligible delay (at 454~eV). The inner (outer) line corresponds to N-1s (C-1s) photoelectrons from pyrimidine. 
(C,D) \textcolor{black}{Rotated partial covariance coefficient maps between streaking-induced signal fluctuations in N-1s and C-1s photoelectrons, where the right panels correspond to the sum of the partial covariance coefficients along the $\theta_{\rm C}$-axis. 
Given a photoionization delay $\Delta\tau = T_L \times \Delta \vartheta/2\pi$ where $T_L$ is the streaking laser period, the streaking-induced signal fluctuation in C-1s photoelectrons at $\theta$ has the most positive covariance with streaking-induced signal fluctuation in N-1s photoelectrons around $\theta+\Delta \vartheta$. 
The global shift in most-correlated region in panel C represents a global differential angle $\Delta \vartheta$ corresponding to $\sim120~$as photoionization time delay between the N-1s and C-1s electrons at $410~$eV X-ray photon energy. 
As a comparison, no global shift is observed in panel D, indicating a negligible time delay at $454~$eV X-ray photon energy.}
}
\label{fig:2}
\end{figure*}

\begin{figure*}
\centering\includegraphics[width=0.9\textwidth]{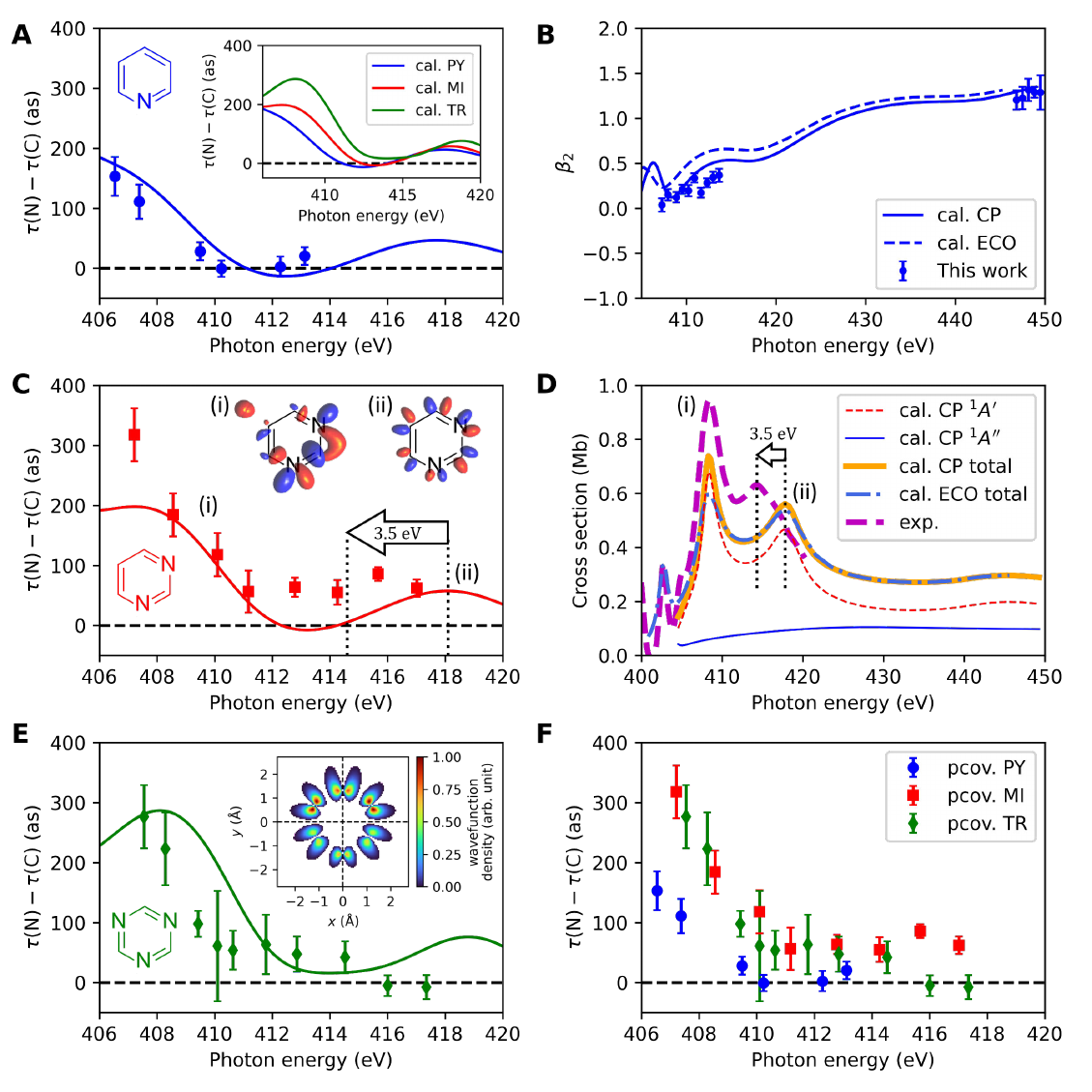}
\caption{\textbf{Energy-dependent photoionization delays and shape resonances} 
(A, C, E) Measured relative photoionization delays ($\tau =\tau_{\rm N1s}-\tau_{\rm C1s}$, symbols) and calculated absolute photoionization delays including CLC delays (full lines) for pyridine (A), pyrimidine (C), and $s$-triazine (E), as a function of the central photon energy of the attosecond FEL pulses. 
Panel (F) and the inset panel in (A) compare the relative photoionization delays extracted by the partial covariance method (pcov.) and the CP calculations (cal.), respectively. 
The resonant one-electron wavefunctions of the two local maxima (i) and (ii) are shown as insets in (C). 
(B) Photoelectron asymmetry parameter of pyridine measured in the present work, compared to the CP and ECO calculations.
(D) The photoionization cross sections of pyrimidine from calculation (cal.) using the ECO and CP methods and experiment (exp.) \cite{vall08a}. 
The decomposition into partial cross sections corresponding to final states of $^1$A$^{\prime}$ and $^1$A$^{\prime\prime}$ symmetry of the C$_{\rm s}$ point group are also shown for the CP calculation. 
The arrows in (C) and (D) indicate the shift between the calculated and observed position of the higher-lying shape resonance.
The inset panel of (E) illustrates the calculated wavefunction density of the $s$-triazine at photon energy of 417.7 eV, where the N atom with the core hole is located at $x=0.00$ {\AA}, $y=-1.35$ {\AA}.
}
\label{fig:3}
\end{figure*}

Figure \ref{fig:2} illustrates how the relative streaking angles $\Delta \vartheta$ were obtained from the experimental data. Due to the large temporal jitter ($\sim$500~fs) between the arrival time of XFEL and IR laser pulses~\cite{glownia2010time}, the streaking directions of the core-shell photoelectrons were completely random from shot to shot. 
The information of relative photoionization time delay is therefore only encoded in the relative streaking angle of the N-1s and C-1s electrons, which is recorded on a shot-by-shot basis. 
Figure~\ref{fig:2}A shows the difference image between the average over all shots corresponding to a narrow range of streaking directions sorted by post-selection of the images and the average over all streaked images without any sorting and selection. 
An angular shift $\Delta \vartheta$ between the streaking directions of C-1s (outer) and N-1s (inner) electrons can be observed, which directly reflects the relative photoionization delays, as described above. 
To accurately determine the relative streaking angle, we employ a partial covariance analysis that utilizes the large shot-to-shot variation of streaking directions to directly extract $\Delta \vartheta$ without estimating and sorting the streaking directions of C-1s and N-1s electrons in every single image. 
For a given relative streaking angle $\Delta\vartheta$, the streaking-induced signal fluctuations in C-1s electrons at $\theta$ have the most positive covariance with those in N-1s electrons around $\theta+\Delta \vartheta$. 
By correctly calculating the partial covariance coefficients between streaking-induced signal changes in N-1s and C-1s electrons, the differential streaking angle $\Delta \vartheta$ manifests itself as a global shift in the partial covariance maps, as shown in Fig. \ref{fig:2}C. Additional details on the partial covariance analysis are given in Section 2.2.1 of the supplemental material (SM) and systematically explained in Ref.~\cite{wang2024covariance}.
By comparing Figs.~\ref{fig:3}A,C to B,D, a relative photoionization delay of up to $120~$as is observed at $410~$eV, as opposed to a negligible delay at $454~$eV. We have also developed a complementary data-analysis method (the so-called ``center-of-mass'' method), which is described in Section 2.2.2 of the SM. These two independent analysis methods yielded consistent photoionization delays (Fig.~S14), which confirms the reliability and robustness of both methods.

Figure \ref{fig:3} summarizes all photoionization delays measured in this work. Panels A, C, and E show the measured photoionization delays between the N-1s and C-1s photoelectrons for the three molecules, which are excellent approximations to the absolute photoionization delays of the N-1s electrons. These delays are compared to the absolute N-1s photoionization delays from core-level photoionization calculations using the Schwinger variational principle with a correlation-polarization potential, summed with the Coulomb-laser coupling (CLC) delays. Details on the calculations are given in Section 3 of the SM. The comparison of the experimental and theoretical delays among the three molecules can be found in panel F and the inset of panel A, respectively. The delays globally decrease with increasing kinetic energy, which reflects the decreasing sensitivity of faster electrons to the potential of the molecular cations. The delays of all three molecules feature local maxima between 408 and 420~eV, which are in reasonable agreement with the local maxima in the calculations. These local maxima in both experiment and theory originate from shape resonances, which correspond to a transient trapping of the photoelectron for a few tens to hundreds of attoseconds before it escapes from the molecule.

\section{Theoretical calculations}
Since such shape resonances are notoriously sensitive to electron correlations and molecular structure \cite{piancastelli99a,huppert16a,heck21a,nandi20a}, we have carefully benchmarked the core-level photoionization calculations against measured photoionization cross sections (Figs.~\ref{fig:3}D and S22), as well as photoelectron angular distributions measured in the present work (Figs.~\ref{fig:3}B and S23). Comparing the calculations using the equivalent-core (ECO) method with the experimental photoionization cross section (Fig.~\ref{fig:3}D) revealed very good agreement for the resonance lying below the nitrogen K-edge and the lowest-lying shape resonance after shifting the calculations to lower energies by 4.3~eV. In comparison, the correlation-polarization (CP) calculations only needed to be shifted by 0.6~eV to agree with the experimental cross sections and the calculated width of the lowest-lying shape resonance was in better agreement with the experiment than the ECO calculations. The CP calculations also agree better with the angular asymmetry parameters measured in this work (Figs.~\ref{fig:3}B and S23). Since both calculations overestimate the energetic position of the upper shape resonance by $\sim$3.5~eV, we therefore identified the CP calculations as the more accurate ones and from hereon exclusively discuss these results.

\begin{figure*}
\centering\includegraphics[width=0.95 \textwidth]{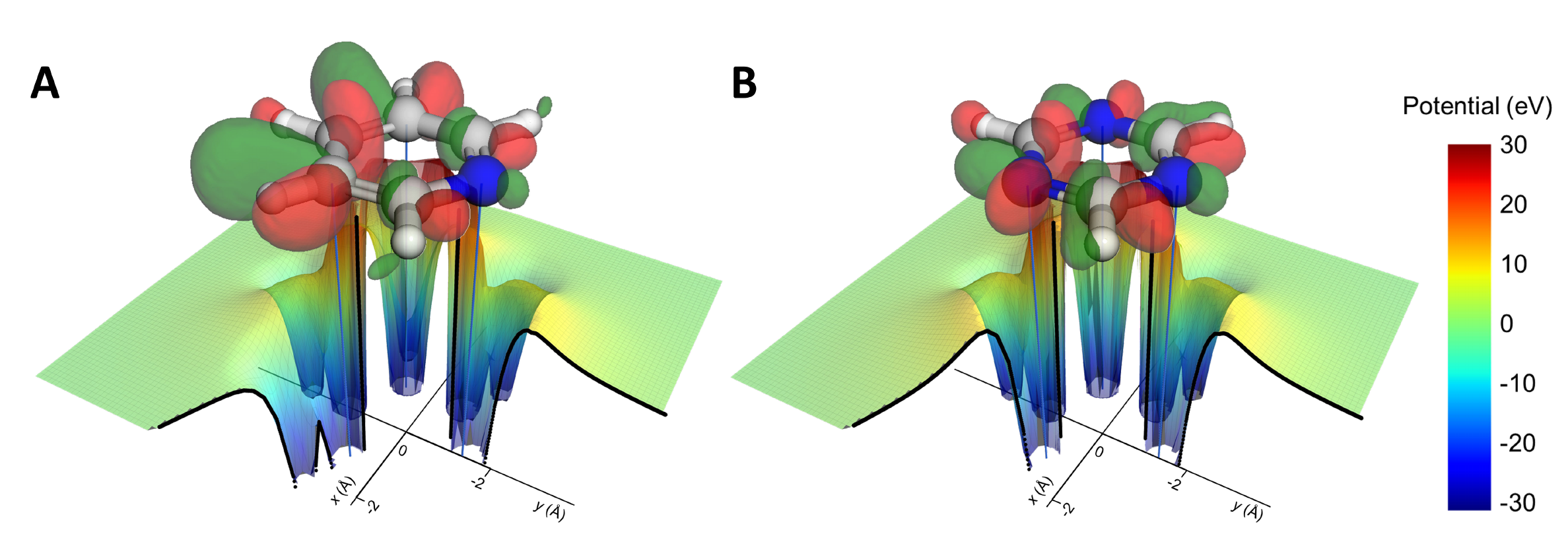}
\caption{\textbf{Illustration of the mechanism of electron trapping.}
\textcolor{black}{
(A) The electrostatic potential of the pyridine cation summed with the centrifugal potential corresponding to $L=5$ is plotted as the false-colored surface.
The shape of the resonant orbital at about 410 eV is illustrated with the ball-stick model (Kapur-Peierls state with isosurfaces of +0.14 (red) and -0.14 (green)). The vertical blue lines map the N atom and the two meta-C atoms to their coordinates on the $xy$-plane. The black curves correspond to the tunneling barriers that trap the resonant wave function along the N atom and the meta-C atom. (B) The same plot for the $s$-triazine cation, where the meta positions are N atoms.}}
\label{fig:new4}
\end{figure*}

Figures~\ref{fig:3}C,D demonstrate the two shape resonances of pyrimidine within the energy range, which lead to the local maxima of the cross section. The $^1$A$^{\prime}$ symmetry component is dominating in the cross section, which displays an anti-bonding $\sigma^*$ character and corresponds to the electron trapping that occurs in the molecular plane in both resonances. The insets in Fig.~\ref{fig:3}C show the one-electron representations of the corresponding resonant wavefunctions (Kapur-Peierls states, see SM for definition). 
The lower-lying shape resonance features 5 nodal planes and the upper one 6, showing that partial waves of high angular momentum are dominating the photoelectron trapping in both cases. The narrower width of the lower resonance nicely correlates with a longer photoemission delay and the broader linewidth of the upper one correlates with the shorter delay. The partial cross section corresponding to a total final-state wavefunction of $^1$A$^{\prime\prime}$ symmetry, i.e., anti-symmetric with respect to the molecular plane and thus a $\pi/\pi^*$-type continuum wavefunction, does not feature any sharp shape resonances 
above the ionization threshold ($\sim$404 eV) in the CP calculation.  Note however, the shifted ECO calculation does have a $\pi/\pi^*$  narrow feature slightly below the threshold, as shown in Fig. \ref{fig:3}D, which is in agreement with the autoionization feature seen in a previous experimental study \cite{vall08a}.  Thus we conclude that in the continuum above the nitrogrn K-edge the photoelectron trapping is dominated by the $\sigma$-type resonances in the molecular plane of pyrimidine. 
The probability density of the resonant wavefunction of $s$-triazine at $\sim$418 eV is illustrated in the inset of Fig. \ref{fig:3}E and in Fig. \ref{fig:new4}. Similar resonances occur in all three molecules, as shown in Fig.~S20.

\begin{figure*}
\centering\includegraphics[width=0.85\textwidth]{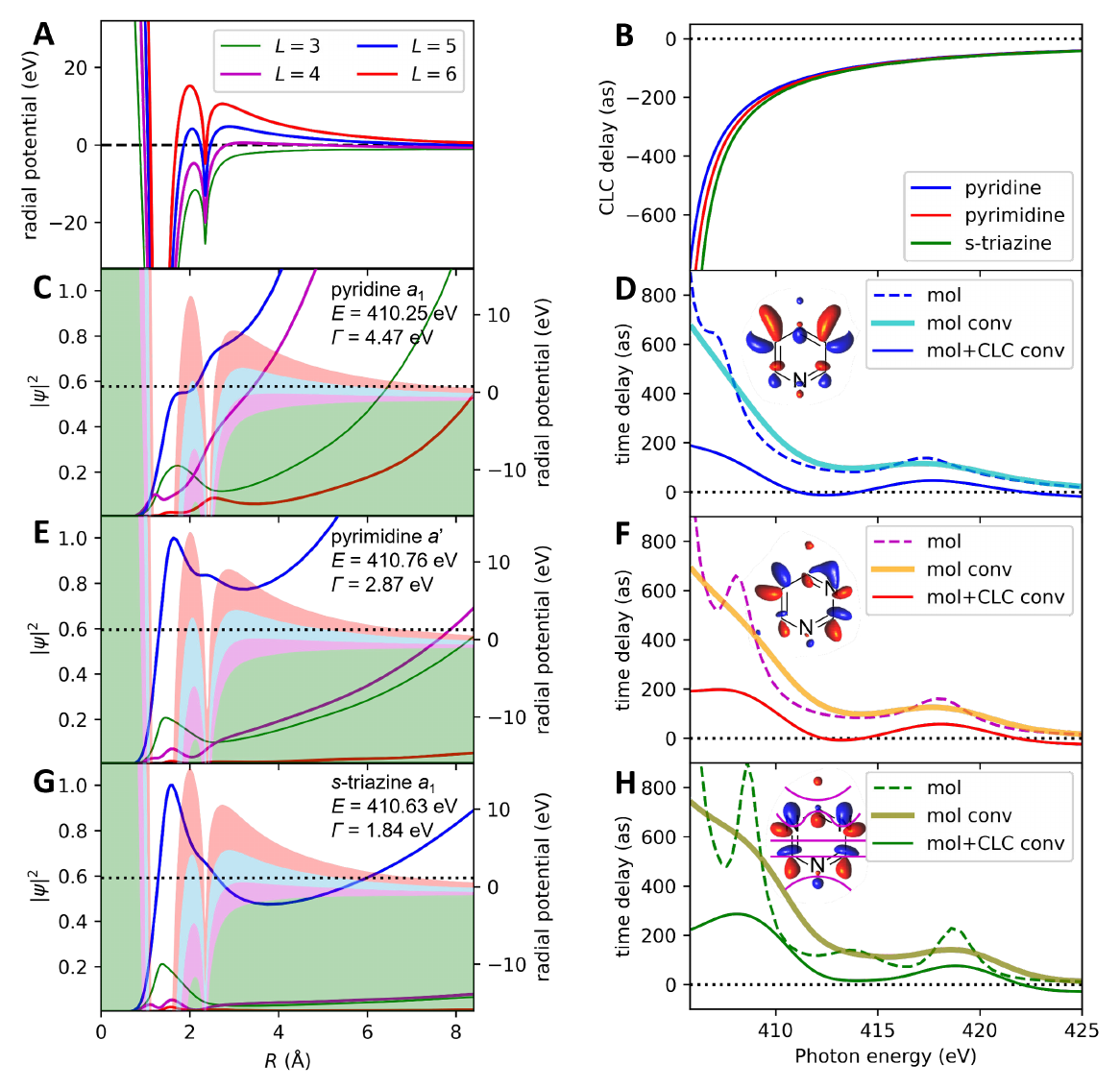}
\caption{
\textcolor{black}{
\textbf{Mechanisms of photoelectron trapping in the azabenzene molecules} (A) Sketch of radial potentials of $s$-triazine averaging over the polar angles with the angular-momentum barriers of different $L$. (B) CLC delays for the three molecules. (C,E,G) Resonant wavefunctions (Kapur-Peierls states) corresponding to the shape resonance centered at the energies indicated in the insets, decomposed into partial waves indicated by the same color code in (A). Different components under the same $L$ are summed to obtain the total probability density. The shaded areas correspond to the radial potentials of the corresponding molecules with the angular-momentum barriers of $L=3,4,5,6$ along the direction defined in (A) with the corresponding color code, and the dotted lines correspond to the asymptotic energies for the resonant states. (D,F,H) Molecular photoionization delays as obtained from the CP calculations (dashed lines), total delays obtained by adding the CLC delays to the molecular delays (thin full lines) and convolution of the total delays with a 3.8-eV full width at half maximum (FWHM) Gaussian line profile (thick full lines). The resonant wavefunction at about 410 eV for each molecule is illustrated as the inset in the corresponding panel. The magenta curves in (H) qualitatively sketch the location of the nodal surfaces of the resonant wavefunction for $s$-triazine.}}
\label{fig:new5}
\end{figure*}

Since our measurements extend to within $\sim$2~eV of the N-1s ionization threshold, they are particularly sensitive to the CLC delays. It has previously been shown that CLC delays in attosecond streaking experiments have the same origin of the continuum-continuum (CC) delays known from reconstruction of attosecond beating by interference of two-photon transitions (RABBIT) measurements \cite{pazourek15a}. 
The most commonly used expressions for the CC delays are reported in Ref. \cite{dahlstrom12a}, which is based on the hydrogen-like atoms, and the radial wavefunctions are approximated considering the asymptotic phase shift (denoted as ``P'') or with additional consideration of the long-range amplitude correction (denoted as ``P+A''). The approximation of the hydrogenic radial wavefunction causes errors compared to the time-dependent Schr{\"o}dinger equation (TDSE), particularly in the low-kinetic-energy region \cite{dahlstrom12a}, which is relevant in our work. Recently, Ji {\it et al.} \cite{ji2024analytical} reported using the analytical formula with hydrogenic radial wavefunctions (confluent hypergeometric functions) \cite{karule1990integrals} instead of their asymptotic forms to calculate the CC transition of the hydrogen-like atoms, which shows excellent agreement with the TDSE in the whole kinetic-energy range. Here, we have adapted the same approach for calculating the CLC delay with a slight modification, where for the 
CC transition the increase or decrease of the kinetic energy is a continuous quantity instead of being equal to the energy of one IR photon, and the time delay is defined based on its derivative. The details can be found in Section 4.2 of the SM. This new approach yields very good agreement to the TDSE values reported in Ref. \cite{kheifets_ionization_2022} (see Fig. S24A). Therefore, we used those results in Figs. \ref{fig:3} and \ref{fig:new5}.
We also investigated the role of the Auger-Meitner decay and the post-collision interaction (PCI) between the Auger-Meitner electron and the photoelectron. We find that the correction of the Coulomb delay from a singly- to a doubly-charged cation is nearly compensated by the delay that is caused by the PCI effect (Fig.~S25). Since these corrections are small and do not affect the relative delays between the molecular species that we studied, we refer the interested reader to the SM, and neglect these corrections for the remainder of the present discussion.

\section{Discussion}
These results open the novel opportunity to analyze the mechanisms that govern the trapping of the photoelectron on the attosecond time scale. For this purpose, we analyze the properties of the trapped photoelectron wave function and the potential that traps it. Figure~\ref{fig:new4} compares the trapping of the outgoing photoelectron wave in the resonances at around 410 eV of pyridine and $s$-triazine. The atomic cores provide attractive potentials in their vicinities, while the centrifugal energy creates a repulsive region centered on the aromatic ring. Their addition results in a potential barrier (black curves in Fig.~\ref{fig:new4}) which traps the resonant wavefunction. The calculations provide a lifetime of $\sim$150~as for pyridine and $\sim$360~as for $s$-triazine at this shape resonance. The stark difference in lifetime is reflected in the time-delay difference in the same energy region. The photoelectron wave initially originates from the 1s shells of the nitrogen atoms and then scatters in the molecular potential, which results in comparable amplitudes of the resonant wave function on all atoms of the molecular ring. But what causes the increase of the photoionization delays and resonance lifetimes with increasing number of nitrogen atoms and increasing symmetry?

To answer this question, Figure~\ref{fig:new5}A shows a sketch of effective radial potentials which indicate the height of the angular-momentum barrier for different values of the angular momentum, $L$, and Figs.~\ref{fig:new5}C,E,G show the absolute squares of resonant one-electron wavefunctions, resolved into partial waves of angular momentum $L$ and irreducible representations of the respective point groups (specified in each panel). Figure~\ref{fig:new5}B shows the CLC delays and Figs.~\ref{fig:new5}D,F,H show the calculated delays for all three molecules. We will focus the remainder of this discussion on the shape resonance located at about 410~eV in each molecule.

In all three molecules, the partial-wave components with $L=5$ dominate in probability density, which is the consequence of the high barrier in the radial potential (panel A) arising from the centrifugal-potential contribution $\hbar^2L(L+1)/(2 m_e R^2)$, where $m_e$ is the mass of the electron and $R$ its distance from the center of charge. The effective potentials of different partial waves are indicated by the shaded areas with different colors in the corresponding panels. It is clear that the main part of the wavefunction is confined in the quasi-bound region in the vicinity of the N and C atoms, while a tail tunnels through the barrier.
In the case of pyridine (Fig.~\ref{fig:new5}A), the dominant $L=5$ partial wave is least-well trapped compared to the other molecules, which is clear from comparing the modulations of the probability density at the barriers. This identifies tunneling through an angular-momentum barrier as the first mechanism governing electron trapping in the azabenzene series and shows that the trapping strength of the dominant partial wave increases in the order pyridine, pyrimidine, $s$-triazine. This shows that substituting a C-H group with the more electronegative N atom increases the trapping strength and therefore the lifetime of the resonance. We note that the divergence of the Kapur-Peierls state for large $R$ is a consequence of their mathematical nature, specifically their complex-valued momentum, see SM Section 3.2.

Figure~\ref{fig:new5} also reveals a second mechanism at play. In addition to the changing probability density ratio between the wavefunctions in the quasi-bound and tunneling regions, we also observe the manifestation of molecular symmetry. The reason that the $L=5$ partial wave prevails in the resonant wavefunctions can be qualitatively understood from the inset panel in Fig. \ref{fig:new5}H. 
The resonant wavefunction displays five nodal surfaces, as qualitatively illustrated in the inset of Fig. \ref{fig:new5}H, corresponding to a dominant $L=5$ partial-wave character. However, the wavefunction also contains contributions from the other partial waves. As the symmetry of the molecular scaffold of the neutral molecule decreases from $s$-triazine (D$_{\rm 3h}$) to pyrimidine and pyridine (C$_{\rm 2v}$), the relative amplitudes of $L<5$ to $L=5$ partial waves increases as a consequence of lower symmetry. The increasing relative amplitudes of the untrapped $L<5$ partial waves cause a faster decay of the resonant state.

To summarize, these two mechanisms explain the trend in the resonant photoelectron trapping times in the azabenzene series. This trend is most clearly visualized in Figs.~\ref{fig:new5}D,F,H, which show that the photoionization delays in the vicinity of the discussed shape resonance indeed increase in the sequence pyridine, pyrimidine, $s$-triazine.

\section{Conclusion}
In conclusion, we have used the novel attosecond capabilities of LCLS to measure core-level photoionization delays of aromatic molecules. These measurements were made possible by the combination of attosecond X-ray pulses with the angular streaking technique. Absolute N-1s photoionization delays were accessed by using the much faster and therefore prompt C-1s photoelectrons as a time reference. These self-referenced measurements were realized on a fine energy grid and revealed the presence of local maxima at 3-15~eV above the N-1s threshold. Comparison to advanced core-level photoionization calculations identified two dominant shape resonances in each molecule and showed that photoelectron trapping occurs in the plane of the aromatic molecules. The two dominant mechanisms governing the photoelectron trapping were identified as (i) changes in the barrier height of the molecular potential that lead to an increase of the electron trapping with increasing electronegativity of the functional groups and (ii) the increasing symmetry of the molecular scaffold reduces the coupling between the well-trapped high-$L$ and the less-trapped lower-$L$ states. Both effects individually and collectively increase the trapping time in $s$-triazine compared to pyridine.
These results demonstrate the promise of performing attosecond core-level photoemission experiments at free-electron-laser facilities for the following reasons. The studied molecules are indeed key building blocks of biologically active molecules, as well as molecules employed in optoelectronics, which opens the perspective of extending attosecond chronoscopy to these systems. The key advantage of core-level photoionization delays over valence-shell photoionization delays, is that the photoelectron spectra always remain comparably simple. This will enable the application of the methods described in this work to complex molecules, molecular assemblies and polymers, such as organic semiconductors, which offers the perspective of studying attosecond electron scattering and trapping in functional materials.


\section{Methods}
\subsection*{Experimental methodology}
The attosecond soft X-ray pulses were produced using the photocathode shaping method~\cite{zhang2020experimental}. 
The mean pulse energy is $39~\mu$J and the mean FWHM bandwidth is $3.8~$eV. 
From our previous measurements, we can estimate a FWHM temporal pulse duration of $460~$as for the X-ray pulses. 

The attosecond soft X-ray pulses described above were focused to a spot-size of $\sim30~\mu$m in diameter using a pair of Kirkpatrick-Baez focusing mirrors~\cite{seaberg_x-ray_2022}. 
The $1300$-nm streaking pulse with a duration of $\sim$40~fs was produced by a commercial optical parametric amplifier~(TOPAS-HE, Light Conversion) which is pumped by a $\sim30$~fs, $9$~mJ, $800$-nm laser source.
The signal beam was sent through a superachromatic quarter-wave plate~(Thorlabs SAQWP05M-1700) to produce a circularly polarized laser field and was focused with a $f=600$~mm CaF$_2$ lens before being reflected from a silver-coated mirror with a $2$-mm-diameter hole to co-propagate the laser and X-ray fields to the interaction point in the c-VMI spectrometer chamber. 

High pressure~(6.4~bar) helium gas was bubbled through a vessel heated to 80-140$^{\circ}$C to deliver the samples into the c-VMI spectrometer chamber, and the photoelectrons ionized from the samples were collected using an electrostatic lens.
A variable-line spaced (VLS) grating X-ray photon spectrometer~\cite{obaid2018lcls, hettrick1988resolving} was employed downstream of the interaction point to obtain the spectra for individual X-ray shots. 

\subsection*{Data analysis}
The relative photoionization delay between the carbon and nitrogen K-shell photoemission features, $\Delta \tau$, is related to the difference in streaking angle, $\Delta \vartheta$, between the momentum shift of the nitrogen and carbon photoemission features,
$\Delta \tau = T_L \times \Delta \vartheta/2\pi$, 
where $T_L=4.33~$fs is the period of the IR streaking laser. 
We use two methods to extract the differential streaking angle $\Delta \vartheta$ from the dataset. 
The first method uses a partial covariance approach to isolate the correlated changes of streaking-induced variations in the photoelectron momentum distribution, while the second procedure uses a sorting method which tags the arrival time of the attosecond X-ray pulse relative to the IR-field. The two approaches are further illustrated and compared in the supplementary information.

\subsection*{Theoretical methodology}
The N $(1{\rm s})^{-1}$  photoionization of the three azabenzenes considered here were studied using two methods, the basis-set complex Kohn method \cite{Orel1990a} and \texttt{ePolyScat} \cite{Gianturco1994a,Natalense1999a}. 
The one-electron basis set used to obtain the occupied molecular orbitals was a correlation consistent valence triple zeta \cite{Dunning1989a} and the geometry used minimized the Hartree-Fock energy. 
The \texttt{ePolyScat} calculations used a static-exchange potential with equivalent-core method. An additional CP potential was introduced to approximate the interaction between the emitted electron and the residual ion of the molecule \cite{Perdew1981}. The calculated values with and without the CP potential are compared in the SM. 
The multichannel calculation from the coupled N $(1{\rm s})^{-1}$ holes using the complex-Kohn method \cite{Marante2020a} and the one-channel calculation using \texttt{ePolyScat} from the localized N $(1{\rm s})^{-1}$ hole show consistency, as demonstrated in the SM.

\section*{Data availability} 
All data will be made available on the public ETH data repository upon acceptance of the manuscript.

\section*{Acknowledgments}
Use of the Linac Coherent Light Source (LCLS), SLAC National Accelerator Laboratory, is supported by the U.S. Department of Energy, Office of Science, Office of Basic Energy Sciences under Contract No. DE-AC02-76SF00515. 
The work of J.-B.J., X.G., D.H., M.H. (ETH Z{\"u}rich), S.H. and H.J.W. was supported by ETH Zürich and project 200021\_172946 of the Swiss National Science Foundation. The work of M.H. (ETH Z{\"u}rich) was additionally supported by the European Union’s Horizon 2020 research and innovation program under the Marie Skłodowska-Curie grant agreement No 801459 - FP-RESOMUS.
The effort from T.D., J.T.O. A.L.W., J.W., T.J.A.W. and J.P.C. was supported by DOE, BES, Chemical Sciences, Geosciences, and Biosciences Division (CSGB).
Z.G., D.C., J.D., P.L.F., R.R.R. and A.M. acknowledge support from the Accelerator and Detector Research Program of the Department of Energy, Basic Energy Sciences division. 
Z.G., P.L.F. and R.R.R. also acknowledge support from Robert Siemann Fellowship of Stanford University. The effort by R.R.L. and C.W.M. at LBNL was supported by the U.S. DOE BES, CSGB under Contract No. DE-AC02-05CH11231.  Calculations at LBNL made use of the resources of the National Energy Research Scientific Computing Center, a DOE Office of Science User
Facility, and the Lawrencium computational cluster resource provided by the IT Division at the LBNL.
C.S.T. was supported in part by the Berkeley Lab Undergraduate Faculty Fellowship (BLUFF) Program, managed by Workforce Development \& Education at Berkeley Lab.

\section*{Author contributions}  
J.-B.J., T.D., K.U., A.M., J.P.C. and H.J.W. conceived the experiment. All authors contributed to the realization of the experiment and discussed the results. J.-B.J., Z.G. and T.D. developed the data-analysis methodology. J.-B.J. and Z.G. analyzed the data. R.R.L., C.S.T. and C.W.M. performed the calculations of molecular photoionization delays. J.-B.J., Z.G. and H.J.W. wrote the initial draft of the manuscript, which was reviewed and edited by all authors.

\section*{Competing interests} 
The authors declare no competing interests. 

\bibliography{references.bib, attobib.bib, theory.bib}

\begin{thebibliography}{49}%
\makeatletter
\providecommand \@ifxundefined [1]{%
 \@ifx{#1\undefined}
}%
\providecommand \@ifnum [1]{%
 \ifnum #1\expandafter \@firstoftwo
 \else \expandafter \@secondoftwo
 \fi
}%
\providecommand \@ifx [1]{%
 \ifx #1\expandafter \@firstoftwo
 \else \expandafter \@secondoftwo
 \fi
}%
\providecommand \natexlab [1]{#1}%
\providecommand \enquote  [1]{``#1''}%
\providecommand \bibnamefont  [1]{#1}%
\providecommand \bibfnamefont [1]{#1}%
\providecommand \citenamefont [1]{#1}%
\providecommand \href@noop [0]{\@secondoftwo}%
\providecommand \href [0]{\begingroup \@sanitize@url \@href}%
\providecommand \@href[1]{\@@startlink{#1}\@@href}%
\providecommand \@@href[1]{\endgroup#1\@@endlink}%
\providecommand \@sanitize@url [0]{\catcode `\\12\catcode `\$12\catcode `\&12\catcode `\#12\catcode `\^12\catcode `\_12\catcode `\%12\relax}%
\providecommand \@@startlink[1]{}%
\providecommand \@@endlink[0]{}%
\providecommand \url  [0]{\begingroup\@sanitize@url \@url }%
\providecommand \@url [1]{\endgroup\@href {#1}{\urlprefix }}%
\providecommand \urlprefix  [0]{URL }%
\providecommand \Eprint [0]{\href }%
\providecommand \doibase [0]{https://doi.org/}%
\providecommand \selectlanguage [0]{\@gobble}%
\providecommand \bibinfo  [0]{\@secondoftwo}%
\providecommand \bibfield  [0]{\@secondoftwo}%
\providecommand \translation [1]{[#1]}%
\providecommand \BibitemOpen [0]{}%
\providecommand \bibitemStop [0]{}%
\providecommand \bibitemNoStop [0]{.\EOS\space}%
\providecommand \EOS [0]{\spacefactor3000\relax}%
\providecommand \BibitemShut  [1]{\csname bibitem#1\endcsname}%
\let\auto@bib@innerbib\@empty
\bibitem [{\citenamefont {Schultze}\ \emph {et~al.}(2010)\citenamefont {Schultze}, \citenamefont {Fiess}, \citenamefont {Karpowicz}, \citenamefont {Gagnon}, \citenamefont {Korbman}, \citenamefont {Hofstetter}, \citenamefont {Neppl}, \citenamefont {Cavalieri}, \citenamefont {Komninos}, \citenamefont {Mercouris}, \citenamefont {Nicolaides}, \citenamefont {Pazourek}, \citenamefont {N\"{a}gele}, \citenamefont {Feist}, \citenamefont {Burgd\"{o}rfer}, \citenamefont {Azzeer}, \citenamefont {Ernstorfer}, \citenamefont {Kienberger}, \citenamefont {Kleineberg}, \citenamefont {Goulielmakis}, \citenamefont {Krausz},\ and\ \citenamefont {Yakovlev}}]{schultze10a}%
  \BibitemOpen
  \bibfield  {author} {\bibinfo {author} {\bibfnamefont {M.}~\bibnamefont {Schultze}}, \bibinfo {author} {\bibfnamefont {M.}~\bibnamefont {Fiess}}, \bibinfo {author} {\bibfnamefont {N.}~\bibnamefont {Karpowicz}}, \bibinfo {author} {\bibfnamefont {J.}~\bibnamefont {Gagnon}}, \bibinfo {author} {\bibfnamefont {M.}~\bibnamefont {Korbman}}, \bibinfo {author} {\bibfnamefont {M.}~\bibnamefont {Hofstetter}}, \bibinfo {author} {\bibfnamefont {S.}~\bibnamefont {Neppl}}, \bibinfo {author} {\bibfnamefont {A.~L.}\ \bibnamefont {Cavalieri}}, \bibinfo {author} {\bibfnamefont {Y.}~\bibnamefont {Komninos}}, \bibinfo {author} {\bibfnamefont {T.}~\bibnamefont {Mercouris}}, \bibinfo {author} {\bibfnamefont {C.~A.}\ \bibnamefont {Nicolaides}}, \bibinfo {author} {\bibfnamefont {R.}~\bibnamefont {Pazourek}}, \bibinfo {author} {\bibfnamefont {S.}~\bibnamefont {N\"{a}gele}}, \bibinfo {author} {\bibfnamefont {J.}~\bibnamefont {Feist}}, \bibinfo {author} {\bibfnamefont {J.}~\bibnamefont {Burgd\"{o}rfer}}, \bibinfo {author}
  {\bibfnamefont {A.~M.}\ \bibnamefont {Azzeer}}, \bibinfo {author} {\bibfnamefont {R.}~\bibnamefont {Ernstorfer}}, \bibinfo {author} {\bibfnamefont {R.}~\bibnamefont {Kienberger}}, \bibinfo {author} {\bibfnamefont {U.}~\bibnamefont {Kleineberg}}, \bibinfo {author} {\bibfnamefont {E.}~\bibnamefont {Goulielmakis}}, \bibinfo {author} {\bibfnamefont {F.}~\bibnamefont {Krausz}},\ and\ \bibinfo {author} {\bibfnamefont {V.~S.}\ \bibnamefont {Yakovlev}},\ }\href {http://www.sciencemag.org/cgi/content/abstract/328/5986/1658} {\bibfield  {journal} {\bibinfo  {journal} {Science}\ }\textbf {\bibinfo {volume} {328}},\ \bibinfo {pages} {1658} (\bibinfo {year} {2010})}\BibitemShut {NoStop}%
\bibitem [{\citenamefont {Kl\"under}\ \emph {et~al.}(2011)\citenamefont {Kl\"under}, \citenamefont {Dahlstr\"om}, \citenamefont {Gisselbrecht}, \citenamefont {Fordell}, \citenamefont {Swoboda}, \citenamefont {Gu\'enot}, \citenamefont {Johnsson}, \citenamefont {Caillat}, \citenamefont {Mauritsson}, \citenamefont {Maquet}, \citenamefont {Ta\"{i}eb},\ and\ \citenamefont {L'Huillier}}]{kluender11a}%
  \BibitemOpen
  \bibfield  {author} {\bibinfo {author} {\bibfnamefont {K.}~\bibnamefont {Kl\"under}}, \bibinfo {author} {\bibfnamefont {J.~M.}\ \bibnamefont {Dahlstr\"om}}, \bibinfo {author} {\bibfnamefont {M.}~\bibnamefont {Gisselbrecht}}, \bibinfo {author} {\bibfnamefont {T.}~\bibnamefont {Fordell}}, \bibinfo {author} {\bibfnamefont {M.}~\bibnamefont {Swoboda}}, \bibinfo {author} {\bibfnamefont {D.}~\bibnamefont {Gu\'enot}}, \bibinfo {author} {\bibfnamefont {P.}~\bibnamefont {Johnsson}}, \bibinfo {author} {\bibfnamefont {J.}~\bibnamefont {Caillat}}, \bibinfo {author} {\bibfnamefont {J.}~\bibnamefont {Mauritsson}}, \bibinfo {author} {\bibfnamefont {A.}~\bibnamefont {Maquet}}, \bibinfo {author} {\bibnamefont {Ta\"{i}eb}},\ and\ \bibinfo {author} {\bibfnamefont {A.}~\bibnamefont {L'Huillier}},\ }\href {https://doi.org/10.1103/PhysRevLett.106.143002} {\bibfield  {journal} {\bibinfo  {journal} {Phys. Rev. Lett.}\ }\textbf {\bibinfo {volume} {106}},\ \bibinfo {pages} {143002} (\bibinfo {year} {2011})}\BibitemShut {NoStop}%
\bibitem [{\citenamefont {Pazourek}\ \emph {et~al.}(2015)\citenamefont {Pazourek}, \citenamefont {Nagele},\ and\ \citenamefont {Burgd\"orfer}}]{pazourek15a}%
  \BibitemOpen
  \bibfield  {author} {\bibinfo {author} {\bibfnamefont {R.}~\bibnamefont {Pazourek}}, \bibinfo {author} {\bibfnamefont {S.}~\bibnamefont {Nagele}},\ and\ \bibinfo {author} {\bibfnamefont {J.}~\bibnamefont {Burgd\"orfer}},\ }\href {https://doi.org/10.1103/RevModPhys.87.765} {\bibfield  {journal} {\bibinfo  {journal} {Rev. Mod. Phys.}\ }\textbf {\bibinfo {volume} {87}},\ \bibinfo {pages} {765} (\bibinfo {year} {2015})}\BibitemShut {NoStop}%
\bibitem [{\citenamefont {Huppert}\ \emph {et~al.}(2016)\citenamefont {Huppert}, \citenamefont {Jordan}, \citenamefont {Baykusheva}, \citenamefont {{von Conta}},\ and\ \citenamefont {W\"orner}}]{huppert16a}%
  \BibitemOpen
  \bibfield  {author} {\bibinfo {author} {\bibfnamefont {M.}~\bibnamefont {Huppert}}, \bibinfo {author} {\bibfnamefont {I.}~\bibnamefont {Jordan}}, \bibinfo {author} {\bibfnamefont {D.}~\bibnamefont {Baykusheva}}, \bibinfo {author} {\bibfnamefont {A.}~\bibnamefont {{von Conta}}},\ and\ \bibinfo {author} {\bibfnamefont {H.~J.}\ \bibnamefont {W\"orner}},\ }\href@noop {} {\bibfield  {journal} {\bibinfo  {journal} {Phys. Rev. Lett.}\ }\textbf {\bibinfo {volume} {117}} (\bibinfo {year} {2016})}\BibitemShut {NoStop}%
\bibitem [{\citenamefont {Vos}\ \emph {et~al.}(2018)\citenamefont {Vos}, \citenamefont {Cattaneo}, \citenamefont {Patchkovskii}, \citenamefont {Zimmermann}, \citenamefont {Cirelli}, \citenamefont {Lucchini}, \citenamefont {Kheifets}, \citenamefont {Landsman},\ and\ \citenamefont {Keller}}]{vos18a}%
  \BibitemOpen
  \bibfield  {author} {\bibinfo {author} {\bibfnamefont {J.}~\bibnamefont {Vos}}, \bibinfo {author} {\bibfnamefont {L.}~\bibnamefont {Cattaneo}}, \bibinfo {author} {\bibfnamefont {S.}~\bibnamefont {Patchkovskii}}, \bibinfo {author} {\bibfnamefont {T.}~\bibnamefont {Zimmermann}}, \bibinfo {author} {\bibfnamefont {C.}~\bibnamefont {Cirelli}}, \bibinfo {author} {\bibfnamefont {M.}~\bibnamefont {Lucchini}}, \bibinfo {author} {\bibfnamefont {A.}~\bibnamefont {Kheifets}}, \bibinfo {author} {\bibfnamefont {A.~S.}\ \bibnamefont {Landsman}},\ and\ \bibinfo {author} {\bibfnamefont {U.}~\bibnamefont {Keller}},\ }\href {https://doi.org/10.1126/science.aao4731} {\bibfield  {journal} {\bibinfo  {journal} {Science}\ }\textbf {\bibinfo {volume} {360}},\ \bibinfo {pages} {1326} (\bibinfo {year} {2018})}\BibitemShut {NoStop}%
\bibitem [{\citenamefont {Biswas}\ \emph {et~al.}(2020)\citenamefont {Biswas}, \citenamefont {F{\"o}rg}, \citenamefont {Sch{\"o}tz}, \citenamefont {Schweinberger}, \citenamefont {Ortmann}, \citenamefont {Zimmermann}, \citenamefont {Pi}, \citenamefont {Baykusheva}, \citenamefont {Masood}, \citenamefont {Liontos}, \citenamefont {M.}, \citenamefont {G.}, \citenamefont {Alharbi}, \citenamefont {Alharbi}, \citenamefont {Azzeer}, \citenamefont {Hartmann}, \citenamefont {W\"orner}, \citenamefont {Landsman},\ and\ \citenamefont {Kling}}]{biswas20a}%
  \BibitemOpen
  \bibfield  {author} {\bibinfo {author} {\bibfnamefont {S.}~\bibnamefont {Biswas}}, \bibinfo {author} {\bibfnamefont {B.}~\bibnamefont {F{\"o}rg}}, \bibinfo {author} {\bibfnamefont {J.}~\bibnamefont {Sch{\"o}tz}}, \bibinfo {author} {\bibfnamefont {W.}~\bibnamefont {Schweinberger}}, \bibinfo {author} {\bibfnamefont {L.}~\bibnamefont {Ortmann}}, \bibinfo {author} {\bibfnamefont {T.}~\bibnamefont {Zimmermann}}, \bibinfo {author} {\bibfnamefont {L.-W.}\ \bibnamefont {Pi}}, \bibinfo {author} {\bibfnamefont {D.}~\bibnamefont {Baykusheva}}, \bibinfo {author} {\bibfnamefont {H.}~\bibnamefont {Masood}}, \bibinfo {author} {\bibfnamefont {I.}~\bibnamefont {Liontos}}, \bibinfo {author} {\bibfnamefont {K.~A.}\ \bibnamefont {M.}}, \bibinfo {author} {\bibfnamefont {K.~N.}\ \bibnamefont {G.}}, \bibinfo {author} {\bibfnamefont {A.~F.}\ \bibnamefont {Alharbi}}, \bibinfo {author} {\bibfnamefont {M.}~\bibnamefont {Alharbi}}, \bibinfo {author} {\bibfnamefont {A.~M.}\ \bibnamefont {Azzeer}}, \bibinfo {author} {\bibfnamefont
  {G.}~\bibnamefont {Hartmann}}, \bibinfo {author} {\bibfnamefont {H.~J.}\ \bibnamefont {W\"orner}}, \bibinfo {author} {\bibfnamefont {A.~S.}\ \bibnamefont {Landsman}},\ and\ \bibinfo {author} {\bibfnamefont {M.~F.}\ \bibnamefont {Kling}},\ }\bibfield  {booktitle} {\emph {\bibinfo {booktitle} {2019 Conference on Lasers and Electro-Optics Europe \& European Quantum Electronics Conference (CLEO/Europe-EQEC)}},\ }\href@noop {} {\bibfield  {journal} {\bibinfo  {journal} {Nat. Phys.}\ }\textbf {\bibinfo {volume} {16}},\ \bibinfo {pages} {778} (\bibinfo {year} {2020})}\BibitemShut {NoStop}%
\bibitem [{\citenamefont {Kamalov}\ \emph {et~al.}(2020)\citenamefont {Kamalov}, \citenamefont {Wang}, \citenamefont {Bucksbaum}, \citenamefont {Haxton},\ and\ \citenamefont {Cryan}}]{kamalov20a}%
  \BibitemOpen
  \bibfield  {author} {\bibinfo {author} {\bibfnamefont {A.}~\bibnamefont {Kamalov}}, \bibinfo {author} {\bibfnamefont {A.~L.}\ \bibnamefont {Wang}}, \bibinfo {author} {\bibfnamefont {P.~H.}\ \bibnamefont {Bucksbaum}}, \bibinfo {author} {\bibfnamefont {D.~J.}\ \bibnamefont {Haxton}},\ and\ \bibinfo {author} {\bibfnamefont {J.~P.}\ \bibnamefont {Cryan}},\ }\href@noop {} {\bibfield  {journal} {\bibinfo  {journal} {Phys. Rev. A}\ }\textbf {\bibinfo {volume} {102}},\ \bibinfo {pages} {023118} (\bibinfo {year} {2020})}\BibitemShut {NoStop}%
\bibitem [{\citenamefont {Nandi}\ \emph {et~al.}(2020)\citenamefont {Nandi}, \citenamefont {Pl{\'e}siat}, \citenamefont {Zhong}, \citenamefont {Palacios}, \citenamefont {Busto}, \citenamefont {Isinger}, \citenamefont {Neori{\v{c}}i{\'c}}, \citenamefont {Arnold}, \citenamefont {Squibb}, \citenamefont {Feifel} \emph {et~al.}}]{nandi20a}%
  \BibitemOpen
  \bibfield  {author} {\bibinfo {author} {\bibfnamefont {S.}~\bibnamefont {Nandi}}, \bibinfo {author} {\bibfnamefont {E.}~\bibnamefont {Pl{\'e}siat}}, \bibinfo {author} {\bibfnamefont {S.}~\bibnamefont {Zhong}}, \bibinfo {author} {\bibfnamefont {A.}~\bibnamefont {Palacios}}, \bibinfo {author} {\bibfnamefont {D.}~\bibnamefont {Busto}}, \bibinfo {author} {\bibfnamefont {M.}~\bibnamefont {Isinger}}, \bibinfo {author} {\bibfnamefont {L.}~\bibnamefont {Neori{\v{c}}i{\'c}}}, \bibinfo {author} {\bibfnamefont {C.}~\bibnamefont {Arnold}}, \bibinfo {author} {\bibfnamefont {R.}~\bibnamefont {Squibb}}, \bibinfo {author} {\bibfnamefont {R.}~\bibnamefont {Feifel}}, \emph {et~al.},\ }\href@noop {} {\bibfield  {journal} {\bibinfo  {journal} {Sci. Adv.}\ }\textbf {\bibinfo {volume} {6}},\ \bibinfo {pages} {eaba7762} (\bibinfo {year} {2020})}\BibitemShut {NoStop}%
\bibitem [{\citenamefont {Heck}\ \emph {et~al.}(2021)\citenamefont {Heck}, \citenamefont {Baykusheva}, \citenamefont {Han}, \citenamefont {Ji}, \citenamefont {Perry}, \citenamefont {Gong},\ and\ \citenamefont {W{\"o}rner}}]{heck21a}%
  \BibitemOpen
  \bibfield  {author} {\bibinfo {author} {\bibfnamefont {S.}~\bibnamefont {Heck}}, \bibinfo {author} {\bibfnamefont {D.}~\bibnamefont {Baykusheva}}, \bibinfo {author} {\bibfnamefont {M.}~\bibnamefont {Han}}, \bibinfo {author} {\bibfnamefont {J.-B.}\ \bibnamefont {Ji}}, \bibinfo {author} {\bibfnamefont {C.}~\bibnamefont {Perry}}, \bibinfo {author} {\bibfnamefont {X.}~\bibnamefont {Gong}},\ and\ \bibinfo {author} {\bibfnamefont {H.~J.}\ \bibnamefont {W{\"o}rner}},\ }\href@noop {} {\bibfield  {journal} {\bibinfo  {journal} {Sci. Adv.}\ }\textbf {\bibinfo {volume} {7}},\ \bibinfo {pages} {eabj8121} (\bibinfo {year} {2021})}\BibitemShut {NoStop}%
\bibitem [{\citenamefont {Gong}\ \emph {et~al.}(2022)\citenamefont {Gong}, \citenamefont {Heck}, \citenamefont {Jelovina}, \citenamefont {Perry}, \citenamefont {Zinchenko}, \citenamefont {Lucchese},\ and\ \citenamefont {W{\"o}rner}}]{gong22a}%
  \BibitemOpen
  \bibfield  {author} {\bibinfo {author} {\bibfnamefont {X.}~\bibnamefont {Gong}}, \bibinfo {author} {\bibfnamefont {S.}~\bibnamefont {Heck}}, \bibinfo {author} {\bibfnamefont {D.}~\bibnamefont {Jelovina}}, \bibinfo {author} {\bibfnamefont {C.}~\bibnamefont {Perry}}, \bibinfo {author} {\bibfnamefont {K.}~\bibnamefont {Zinchenko}}, \bibinfo {author} {\bibfnamefont {R.}~\bibnamefont {Lucchese}},\ and\ \bibinfo {author} {\bibfnamefont {H.~J.}\ \bibnamefont {W{\"o}rner}},\ }\href@noop {} {\bibfield  {journal} {\bibinfo  {journal} {Nature}\ }\textbf {\bibinfo {volume} {609}},\ \bibinfo {pages} {507} (\bibinfo {year} {2022})}\BibitemShut {NoStop}%
\bibitem [{\citenamefont {Heck}\ \emph {et~al.}(2022)\citenamefont {Heck}, \citenamefont {Han}, \citenamefont {Jelovina}, \citenamefont {Ji}, \citenamefont {Perry}, \citenamefont {Gong}, \citenamefont {Lucchese}, \citenamefont {Ueda},\ and\ \citenamefont {W{\"o}rner}}]{heck22a}%
  \BibitemOpen
  \bibfield  {author} {\bibinfo {author} {\bibfnamefont {S.}~\bibnamefont {Heck}}, \bibinfo {author} {\bibfnamefont {M.}~\bibnamefont {Han}}, \bibinfo {author} {\bibfnamefont {D.}~\bibnamefont {Jelovina}}, \bibinfo {author} {\bibfnamefont {J.-B.}\ \bibnamefont {Ji}}, \bibinfo {author} {\bibfnamefont {C.}~\bibnamefont {Perry}}, \bibinfo {author} {\bibfnamefont {X.}~\bibnamefont {Gong}}, \bibinfo {author} {\bibfnamefont {R.}~\bibnamefont {Lucchese}}, \bibinfo {author} {\bibfnamefont {K.}~\bibnamefont {Ueda}},\ and\ \bibinfo {author} {\bibfnamefont {H.~J.}\ \bibnamefont {W{\"o}rner}},\ }\href@noop {} {\bibfield  {journal} {\bibinfo  {journal} {Phys. Rev. Lett.}\ }\textbf {\bibinfo {volume} {129}},\ \bibinfo {pages} {133002} (\bibinfo {year} {2022})}\BibitemShut {NoStop}%
\bibitem [{\citenamefont {Jordan}\ \emph {et~al.}(2020)\citenamefont {Jordan}, \citenamefont {Huppert}, \citenamefont {Rattenbacher}, \citenamefont {Peper}, \citenamefont {Jelovina}, \citenamefont {Perry}, \citenamefont {von Conta}, \citenamefont {Schild},\ and\ \citenamefont {W{\"o}rner}}]{jordan20a}%
  \BibitemOpen
  \bibfield  {author} {\bibinfo {author} {\bibfnamefont {I.}~\bibnamefont {Jordan}}, \bibinfo {author} {\bibfnamefont {M.}~\bibnamefont {Huppert}}, \bibinfo {author} {\bibfnamefont {D.}~\bibnamefont {Rattenbacher}}, \bibinfo {author} {\bibfnamefont {M.}~\bibnamefont {Peper}}, \bibinfo {author} {\bibfnamefont {D.}~\bibnamefont {Jelovina}}, \bibinfo {author} {\bibfnamefont {C.}~\bibnamefont {Perry}}, \bibinfo {author} {\bibfnamefont {A.}~\bibnamefont {von Conta}}, \bibinfo {author} {\bibfnamefont {A.}~\bibnamefont {Schild}},\ and\ \bibinfo {author} {\bibfnamefont {H.~J.}\ \bibnamefont {W{\"o}rner}},\ }\href {https://doi.org/10.1126/science.abb0979} {\bibfield  {journal} {\bibinfo  {journal} {Science}\ }\textbf {\bibinfo {volume} {369}},\ \bibinfo {pages} {974} (\bibinfo {year} {2020})}\BibitemShut {NoStop}%
\bibitem [{\citenamefont {Cavalieri}\ \emph {et~al.}(2007)\citenamefont {Cavalieri}, \citenamefont {Muller}, \citenamefont {Uphues}, \citenamefont {Yakovlev}, \citenamefont {Baltuska}, \citenamefont {Horvath}, \citenamefont {Schmidt}, \citenamefont {Blumel}, \citenamefont {Holzwarth}, \citenamefont {Hendel}, \citenamefont {Drescher}, \citenamefont {Kleineberg}, \citenamefont {Echenique}, \citenamefont {Kienberger}, \citenamefont {Krausz},\ and\ \citenamefont {Heinzmann}}]{cavalieri07a}%
  \BibitemOpen
  \bibfield  {author} {\bibinfo {author} {\bibfnamefont {A.~L.}\ \bibnamefont {Cavalieri}}, \bibinfo {author} {\bibfnamefont {N.}~\bibnamefont {Muller}}, \bibinfo {author} {\bibfnamefont {T.}~\bibnamefont {Uphues}}, \bibinfo {author} {\bibfnamefont {V.~S.}\ \bibnamefont {Yakovlev}}, \bibinfo {author} {\bibfnamefont {A.}~\bibnamefont {Baltuska}}, \bibinfo {author} {\bibfnamefont {B.}~\bibnamefont {Horvath}}, \bibinfo {author} {\bibfnamefont {B.}~\bibnamefont {Schmidt}}, \bibinfo {author} {\bibfnamefont {L.}~\bibnamefont {Blumel}}, \bibinfo {author} {\bibfnamefont {R.}~\bibnamefont {Holzwarth}}, \bibinfo {author} {\bibfnamefont {S.}~\bibnamefont {Hendel}}, \bibinfo {author} {\bibfnamefont {M.}~\bibnamefont {Drescher}}, \bibinfo {author} {\bibfnamefont {U.}~\bibnamefont {Kleineberg}}, \bibinfo {author} {\bibfnamefont {P.~M.}\ \bibnamefont {Echenique}}, \bibinfo {author} {\bibfnamefont {R.}~\bibnamefont {Kienberger}}, \bibinfo {author} {\bibfnamefont {F.}~\bibnamefont {Krausz}},\ and\ \bibinfo {author}
  {\bibfnamefont {U.}~\bibnamefont {Heinzmann}},\ }\href {http://dx.doi.org/10.1038/nature06229} {\bibfield  {journal} {\bibinfo  {journal} {Nature}\ }\textbf {\bibinfo {volume} {449}},\ \bibinfo {pages} {1029} (\bibinfo {year} {2007})}\BibitemShut {NoStop}%
\bibitem [{\citenamefont {Neppl}\ \emph {et~al.}(2015)\citenamefont {Neppl}, \citenamefont {Ernstorfer}, \citenamefont {Cavalieri}, \citenamefont {Lemell}, \citenamefont {Wachter}, \citenamefont {Magerl}, \citenamefont {Bothschafter}, \citenamefont {Jobst}, \citenamefont {Hofstetter}, \citenamefont {Kleineberg}, \citenamefont {Barth}, \citenamefont {Menzel}, \citenamefont {Burgdorfer}, \citenamefont {Feulner}, \citenamefont {Krausz},\ and\ \citenamefont {Kienberger}}]{neppl15a}%
  \BibitemOpen
  \bibfield  {author} {\bibinfo {author} {\bibfnamefont {S.}~\bibnamefont {Neppl}}, \bibinfo {author} {\bibfnamefont {R.}~\bibnamefont {Ernstorfer}}, \bibinfo {author} {\bibfnamefont {A.~L.}\ \bibnamefont {Cavalieri}}, \bibinfo {author} {\bibfnamefont {C.}~\bibnamefont {Lemell}}, \bibinfo {author} {\bibfnamefont {G.}~\bibnamefont {Wachter}}, \bibinfo {author} {\bibfnamefont {E.}~\bibnamefont {Magerl}}, \bibinfo {author} {\bibfnamefont {E.~M.}\ \bibnamefont {Bothschafter}}, \bibinfo {author} {\bibfnamefont {M.}~\bibnamefont {Jobst}}, \bibinfo {author} {\bibfnamefont {M.}~\bibnamefont {Hofstetter}}, \bibinfo {author} {\bibfnamefont {U.}~\bibnamefont {Kleineberg}}, \bibinfo {author} {\bibfnamefont {J.~V.}\ \bibnamefont {Barth}}, \bibinfo {author} {\bibfnamefont {D.}~\bibnamefont {Menzel}}, \bibinfo {author} {\bibfnamefont {J.}~\bibnamefont {Burgdorfer}}, \bibinfo {author} {\bibfnamefont {P.}~\bibnamefont {Feulner}}, \bibinfo {author} {\bibfnamefont {F.}~\bibnamefont {Krausz}},\ and\ \bibinfo {author}
  {\bibfnamefont {R.}~\bibnamefont {Kienberger}},\ }\href {http://dx.doi.org/10.1038/nature14094} {\bibfield  {journal} {\bibinfo  {journal} {Nature}\ }\textbf {\bibinfo {volume} {517}},\ \bibinfo {pages} {342} (\bibinfo {year} {2015})}\BibitemShut {NoStop}%
\bibitem [{\citenamefont {Tao}\ \emph {et~al.}(2016)\citenamefont {Tao}, \citenamefont {Chen}, \citenamefont {Szilv\'asi}, \citenamefont {Keller}, \citenamefont {Mavrikakis}, \citenamefont {Kapteyn},\ and\ \citenamefont {Murnane}}]{tao16a}%
  \BibitemOpen
  \bibfield  {author} {\bibinfo {author} {\bibfnamefont {Z.}~\bibnamefont {Tao}}, \bibinfo {author} {\bibfnamefont {C.}~\bibnamefont {Chen}}, \bibinfo {author} {\bibfnamefont {T.}~\bibnamefont {Szilv\'asi}}, \bibinfo {author} {\bibfnamefont {M.}~\bibnamefont {Keller}}, \bibinfo {author} {\bibfnamefont {M.}~\bibnamefont {Mavrikakis}}, \bibinfo {author} {\bibfnamefont {H.}~\bibnamefont {Kapteyn}},\ and\ \bibinfo {author} {\bibfnamefont {M.}~\bibnamefont {Murnane}},\ }\href@noop {} {\bibfield  {journal} {\bibinfo  {journal} {Science}\ }\textbf {\bibinfo {volume} {353}},\ \bibinfo {pages} {62} (\bibinfo {year} {2016})}\BibitemShut {NoStop}%
\bibitem [{\citenamefont {Siek}\ \emph {et~al.}(2017)\citenamefont {Siek}, \citenamefont {Neb}, \citenamefont {Bartz}, \citenamefont {Hensen}, \citenamefont {Str{\"u}ber}, \citenamefont {Fiechter}, \citenamefont {Torrent-Sucarrat}, \citenamefont {Silkin}, \citenamefont {Krasovskii}, \citenamefont {Kabachnik} \emph {et~al.}}]{siek17a}%
  \BibitemOpen
  \bibfield  {author} {\bibinfo {author} {\bibfnamefont {F.}~\bibnamefont {Siek}}, \bibinfo {author} {\bibfnamefont {S.}~\bibnamefont {Neb}}, \bibinfo {author} {\bibfnamefont {P.}~\bibnamefont {Bartz}}, \bibinfo {author} {\bibfnamefont {M.}~\bibnamefont {Hensen}}, \bibinfo {author} {\bibfnamefont {C.}~\bibnamefont {Str{\"u}ber}}, \bibinfo {author} {\bibfnamefont {S.}~\bibnamefont {Fiechter}}, \bibinfo {author} {\bibfnamefont {M.}~\bibnamefont {Torrent-Sucarrat}}, \bibinfo {author} {\bibfnamefont {V.~M.}\ \bibnamefont {Silkin}}, \bibinfo {author} {\bibfnamefont {E.~E.}\ \bibnamefont {Krasovskii}}, \bibinfo {author} {\bibfnamefont {N.~M.}\ \bibnamefont {Kabachnik}}, \emph {et~al.},\ }\href@noop {} {\bibfield  {journal} {\bibinfo  {journal} {Science}\ }\textbf {\bibinfo {volume} {357}},\ \bibinfo {pages} {1274} (\bibinfo {year} {2017})}\BibitemShut {NoStop}%
\bibitem [{\citenamefont {Duris}\ \emph {et~al.}(2020)\citenamefont {Duris}, \citenamefont {Li}, \citenamefont {Driver}, \citenamefont {Champenois}, \citenamefont {MacArthur}, \citenamefont {Lutman}, \citenamefont {Zhang}, \citenamefont {Rosenberger}, \citenamefont {Aldrich}, \citenamefont {Coffee} \emph {et~al.}}]{duris20a}%
  \BibitemOpen
  \bibfield  {author} {\bibinfo {author} {\bibfnamefont {J.}~\bibnamefont {Duris}}, \bibinfo {author} {\bibfnamefont {S.}~\bibnamefont {Li}}, \bibinfo {author} {\bibfnamefont {T.}~\bibnamefont {Driver}}, \bibinfo {author} {\bibfnamefont {E.~G.}\ \bibnamefont {Champenois}}, \bibinfo {author} {\bibfnamefont {J.~P.}\ \bibnamefont {MacArthur}}, \bibinfo {author} {\bibfnamefont {A.~A.}\ \bibnamefont {Lutman}}, \bibinfo {author} {\bibfnamefont {Z.}~\bibnamefont {Zhang}}, \bibinfo {author} {\bibfnamefont {P.}~\bibnamefont {Rosenberger}}, \bibinfo {author} {\bibfnamefont {J.~W.}\ \bibnamefont {Aldrich}}, \bibinfo {author} {\bibfnamefont {R.}~\bibnamefont {Coffee}}, \emph {et~al.},\ }\href@noop {} {\bibfield  {journal} {\bibinfo  {journal} {Nat. Photonics}\ }\textbf {\bibinfo {volume} {14}},\ \bibinfo {pages} {30} (\bibinfo {year} {2020})}\BibitemShut {NoStop}%
\bibitem [{\citenamefont {Li}\ \emph {et~al.}(2018)\citenamefont {Li}, \citenamefont {Guo}, \citenamefont {Coffee}, \citenamefont {Hegazy}, \citenamefont {Huang}, \citenamefont {Natan}, \citenamefont {Osipov}, \citenamefont {Ray}, \citenamefont {Marinelli},\ and\ \citenamefont {Cryan}}]{li18a}%
  \BibitemOpen
  \bibfield  {author} {\bibinfo {author} {\bibfnamefont {S.}~\bibnamefont {Li}}, \bibinfo {author} {\bibfnamefont {Z.}~\bibnamefont {Guo}}, \bibinfo {author} {\bibfnamefont {R.~N.}\ \bibnamefont {Coffee}}, \bibinfo {author} {\bibfnamefont {K.}~\bibnamefont {Hegazy}}, \bibinfo {author} {\bibfnamefont {Z.}~\bibnamefont {Huang}}, \bibinfo {author} {\bibfnamefont {A.}~\bibnamefont {Natan}}, \bibinfo {author} {\bibfnamefont {T.}~\bibnamefont {Osipov}}, \bibinfo {author} {\bibfnamefont {D.}~\bibnamefont {Ray}}, \bibinfo {author} {\bibfnamefont {A.}~\bibnamefont {Marinelli}},\ and\ \bibinfo {author} {\bibfnamefont {J.~P.}\ \bibnamefont {Cryan}},\ }\href@noop {} {\bibfield  {journal} {\bibinfo  {journal} {Opt. Express}\ }\textbf {\bibinfo {volume} {26}},\ \bibinfo {pages} {4531} (\bibinfo {year} {2018})}\BibitemShut {NoStop}%
\bibitem [{\citenamefont {Li}\ \emph {et~al.}(2022)\citenamefont {Li}, \citenamefont {Driver}, \citenamefont {Rosenberger}, \citenamefont {Champenois}, \citenamefont {Duris}, \citenamefont {Al-Haddad}, \citenamefont {Averbukh}, \citenamefont {Barnard}, \citenamefont {Berrah}, \citenamefont {Bostedt} \emph {et~al.}}]{li22a}%
  \BibitemOpen
  \bibfield  {author} {\bibinfo {author} {\bibfnamefont {S.}~\bibnamefont {Li}}, \bibinfo {author} {\bibfnamefont {T.}~\bibnamefont {Driver}}, \bibinfo {author} {\bibfnamefont {P.}~\bibnamefont {Rosenberger}}, \bibinfo {author} {\bibfnamefont {E.~G.}\ \bibnamefont {Champenois}}, \bibinfo {author} {\bibfnamefont {J.}~\bibnamefont {Duris}}, \bibinfo {author} {\bibfnamefont {A.}~\bibnamefont {Al-Haddad}}, \bibinfo {author} {\bibfnamefont {V.}~\bibnamefont {Averbukh}}, \bibinfo {author} {\bibfnamefont {J.~C.}\ \bibnamefont {Barnard}}, \bibinfo {author} {\bibfnamefont {N.}~\bibnamefont {Berrah}}, \bibinfo {author} {\bibfnamefont {C.}~\bibnamefont {Bostedt}}, \emph {et~al.},\ }\href@noop {} {\bibfield  {journal} {\bibinfo  {journal} {Science}\ }\textbf {\bibinfo {volume} {375}},\ \bibinfo {pages} {285} (\bibinfo {year} {2022})}\BibitemShut {NoStop}%
\bibitem [{\citenamefont {Driver}\ \emph {et~al.}(2024)\citenamefont {Driver}, \citenamefont {Mountney}, \citenamefont {Wang}, \citenamefont {Ortmann}, \citenamefont {Al-Haddad}, \citenamefont {Berrah}, \citenamefont {Bostedt}, \citenamefont {Champenois}, \citenamefont {DiMauro}, \citenamefont {Duris} \emph {et~al.}}]{driver24a}%
  \BibitemOpen
  \bibfield  {author} {\bibinfo {author} {\bibfnamefont {T.}~\bibnamefont {Driver}}, \bibinfo {author} {\bibfnamefont {M.}~\bibnamefont {Mountney}}, \bibinfo {author} {\bibfnamefont {J.}~\bibnamefont {Wang}}, \bibinfo {author} {\bibfnamefont {L.}~\bibnamefont {Ortmann}}, \bibinfo {author} {\bibfnamefont {A.}~\bibnamefont {Al-Haddad}}, \bibinfo {author} {\bibfnamefont {N.}~\bibnamefont {Berrah}}, \bibinfo {author} {\bibfnamefont {C.}~\bibnamefont {Bostedt}}, \bibinfo {author} {\bibfnamefont {E.~G.}\ \bibnamefont {Champenois}}, \bibinfo {author} {\bibfnamefont {L.~F.}\ \bibnamefont {DiMauro}}, \bibinfo {author} {\bibfnamefont {J.}~\bibnamefont {Duris}}, \emph {et~al.},\ }\href@noop {} {\bibfield  {journal} {\bibinfo  {journal} {Nature}\ }\textbf {\bibinfo {volume} {632}},\ \bibinfo {pages} {762} (\bibinfo {year} {2024})}\BibitemShut {NoStop}%
\bibitem [{\citenamefont {Ling}\ \emph {et~al.}(2021)\citenamefont {Ling}, \citenamefont {Hao}, \citenamefont {Liang}, \citenamefont {Zhang}, \citenamefont {Liu},\ and\ \citenamefont {Wang}}]{ling21a}%
  \BibitemOpen
  \bibfield  {author} {\bibinfo {author} {\bibfnamefont {Y.}~\bibnamefont {Ling}}, \bibinfo {author} {\bibfnamefont {Z.-Y.}\ \bibnamefont {Hao}}, \bibinfo {author} {\bibfnamefont {D.}~\bibnamefont {Liang}}, \bibinfo {author} {\bibfnamefont {C.-L.}\ \bibnamefont {Zhang}}, \bibinfo {author} {\bibfnamefont {Y.-F.}\ \bibnamefont {Liu}},\ and\ \bibinfo {author} {\bibfnamefont {Y.}~\bibnamefont {Wang}},\ }\href@noop {} {\bibfield  {journal} {\bibinfo  {journal} {Drug. Des. Devel. Ther.}\ ,\ \bibinfo {pages} {4289}} (\bibinfo {year} {2021})}\BibitemShut {NoStop}%
\bibitem [{\citenamefont {Selvam}\ \emph {et~al.}(2015)\citenamefont {Selvam}, \citenamefont {James}, \citenamefont {Dniandev},\ and\ \citenamefont {Valzita}}]{selvam15a}%
  \BibitemOpen
  \bibfield  {author} {\bibinfo {author} {\bibfnamefont {T.~P.}\ \bibnamefont {Selvam}}, \bibinfo {author} {\bibfnamefont {C.~R.}\ \bibnamefont {James}}, \bibinfo {author} {\bibfnamefont {P.~V.}\ \bibnamefont {Dniandev}},\ and\ \bibinfo {author} {\bibfnamefont {S.~K.}\ \bibnamefont {Valzita}},\ }\href@noop {} {\bibfield  {journal} {\bibinfo  {journal} {Research in Pharmacy}\ }\textbf {\bibinfo {volume} {2}} (\bibinfo {year} {2015})}\BibitemShut {NoStop}%
\bibitem [{\citenamefont {Sharma}\ \emph {et~al.}(2021)\citenamefont {Sharma}, \citenamefont {Sheyi}, \citenamefont {de~la Torre}, \citenamefont {El-Faham},\ and\ \citenamefont {Albericio}}]{sharma21a}%
  \BibitemOpen
  \bibfield  {author} {\bibinfo {author} {\bibfnamefont {A.}~\bibnamefont {Sharma}}, \bibinfo {author} {\bibfnamefont {R.}~\bibnamefont {Sheyi}}, \bibinfo {author} {\bibfnamefont {B.~G.}\ \bibnamefont {de~la Torre}}, \bibinfo {author} {\bibfnamefont {A.}~\bibnamefont {El-Faham}},\ and\ \bibinfo {author} {\bibfnamefont {F.}~\bibnamefont {Albericio}},\ }\href@noop {} {\bibfield  {journal} {\bibinfo  {journal} {Molecules}\ }\textbf {\bibinfo {volume} {26}},\ \bibinfo {pages} {864} (\bibinfo {year} {2021})}\BibitemShut {NoStop}%
\bibitem [{\citenamefont {Moliton}(2010)}]{moliton10a}%
  \BibitemOpen
  \bibfield  {author} {\bibinfo {author} {\bibfnamefont {A.}~\bibnamefont {Moliton}},\ }\href@noop {} {\emph {\bibinfo {title} {Optoelectronics of molecules and polymers}}},\ Vol.\ \bibinfo {volume} {104}\ (\bibinfo  {publisher} {Springer},\ \bibinfo {address} {New York},\ \bibinfo {year} {2010})\BibitemShut {NoStop}%
\bibitem [{\citenamefont {Walter}\ \emph {et~al.}(2022)\citenamefont {Walter}, \citenamefont {Osipov}, \citenamefont {Lin}, \citenamefont {Cryan}, \citenamefont {Driver}, \citenamefont {Kamalov}, \citenamefont {Marinelli}, \citenamefont {Robinson}, \citenamefont {Seaberg}, \citenamefont {Wolf}, \citenamefont {Aldrich}, \citenamefont {Brown}, \citenamefont {Champenois}, \citenamefont {Cheng}, \citenamefont {Cocco}, \citenamefont {Conder}, \citenamefont {Curiel}, \citenamefont {Egger}, \citenamefont {Glownia}, \citenamefont {Heimann}, \citenamefont {Holmes}, \citenamefont {Johnson}, \citenamefont {Lee}, \citenamefont {Li}, \citenamefont {Moeller}, \citenamefont {Morton}, \citenamefont {Ng}, \citenamefont {Ninh}, \citenamefont {O'Neal}, \citenamefont {Obaid}, \citenamefont {Pai}, \citenamefont {Schlotter}, \citenamefont {Shepard}, \citenamefont {Shivaram}, \citenamefont {Stefan}, \citenamefont {Van}, \citenamefont {Wang}, \citenamefont {Wang}, \citenamefont {Yin}, \citenamefont {Yunus}, \citenamefont {Fritz},
  \citenamefont {James},\ and\ \citenamefont {Castagna}}]{Walter22a}%
  \BibitemOpen
  \bibfield  {author} {\bibinfo {author} {\bibfnamefont {P.}~\bibnamefont {Walter}}, \bibinfo {author} {\bibfnamefont {T.}~\bibnamefont {Osipov}}, \bibinfo {author} {\bibfnamefont {M.-F.}\ \bibnamefont {Lin}}, \bibinfo {author} {\bibfnamefont {J.}~\bibnamefont {Cryan}}, \bibinfo {author} {\bibfnamefont {T.}~\bibnamefont {Driver}}, \bibinfo {author} {\bibfnamefont {A.}~\bibnamefont {Kamalov}}, \bibinfo {author} {\bibfnamefont {A.}~\bibnamefont {Marinelli}}, \bibinfo {author} {\bibfnamefont {J.}~\bibnamefont {Robinson}}, \bibinfo {author} {\bibfnamefont {M.~H.}\ \bibnamefont {Seaberg}}, \bibinfo {author} {\bibfnamefont {T.~J.~A.}\ \bibnamefont {Wolf}}, \bibinfo {author} {\bibfnamefont {J.}~\bibnamefont {Aldrich}}, \bibinfo {author} {\bibfnamefont {N.}~\bibnamefont {Brown}}, \bibinfo {author} {\bibfnamefont {E.~G.}\ \bibnamefont {Champenois}}, \bibinfo {author} {\bibfnamefont {X.}~\bibnamefont {Cheng}}, \bibinfo {author} {\bibfnamefont {D.}~\bibnamefont {Cocco}}, \bibinfo {author} {\bibfnamefont
  {A.}~\bibnamefont {Conder}}, \bibinfo {author} {\bibfnamefont {I.}~\bibnamefont {Curiel}}, \bibinfo {author} {\bibfnamefont {A.}~\bibnamefont {Egger}}, \bibinfo {author} {\bibfnamefont {J.~M.}\ \bibnamefont {Glownia}}, \bibinfo {author} {\bibfnamefont {P.}~\bibnamefont {Heimann}}, \bibinfo {author} {\bibfnamefont {M.}~\bibnamefont {Holmes}}, \bibinfo {author} {\bibfnamefont {T.}~\bibnamefont {Johnson}}, \bibinfo {author} {\bibfnamefont {L.}~\bibnamefont {Lee}}, \bibinfo {author} {\bibfnamefont {X.}~\bibnamefont {Li}}, \bibinfo {author} {\bibfnamefont {S.}~\bibnamefont {Moeller}}, \bibinfo {author} {\bibfnamefont {D.~S.}\ \bibnamefont {Morton}}, \bibinfo {author} {\bibfnamefont {M.~L.}\ \bibnamefont {Ng}}, \bibinfo {author} {\bibfnamefont {K.}~\bibnamefont {Ninh}}, \bibinfo {author} {\bibfnamefont {J.~T.}\ \bibnamefont {O'Neal}}, \bibinfo {author} {\bibfnamefont {R.}~\bibnamefont {Obaid}}, \bibinfo {author} {\bibfnamefont {A.}~\bibnamefont {Pai}}, \bibinfo {author} {\bibfnamefont {W.}~\bibnamefont
  {Schlotter}}, \bibinfo {author} {\bibfnamefont {J.}~\bibnamefont {Shepard}}, \bibinfo {author} {\bibfnamefont {N.}~\bibnamefont {Shivaram}}, \bibinfo {author} {\bibfnamefont {P.}~\bibnamefont {Stefan}}, \bibinfo {author} {\bibfnamefont {X.}~\bibnamefont {Van}}, \bibinfo {author} {\bibfnamefont {A.~L.}\ \bibnamefont {Wang}}, \bibinfo {author} {\bibfnamefont {H.}~\bibnamefont {Wang}}, \bibinfo {author} {\bibfnamefont {J.}~\bibnamefont {Yin}}, \bibinfo {author} {\bibfnamefont {S.}~\bibnamefont {Yunus}}, \bibinfo {author} {\bibfnamefont {D.}~\bibnamefont {Fritz}}, \bibinfo {author} {\bibfnamefont {J.}~\bibnamefont {James}},\ and\ \bibinfo {author} {\bibfnamefont {J.-C.}\ \bibnamefont {Castagna}},\ }\href {https://doi.org/10.1107/S1600577522004283} {\bibfield  {journal} {\bibinfo  {journal} {J. Synchrotron Radiat.}\ }\textbf {\bibinfo {volume} {29}},\ \bibinfo {pages} {957} (\bibinfo {year} {2022})}\BibitemShut {NoStop}%
\bibitem [{\citenamefont {Zhao}\ \emph {et~al.}(2005)\citenamefont {Zhao}, \citenamefont {Chang}, \citenamefont {Tong},\ and\ \citenamefont {Lin}}]{zhao05a}%
  \BibitemOpen
  \bibfield  {author} {\bibinfo {author} {\bibfnamefont {Z.}~\bibnamefont {Zhao}}, \bibinfo {author} {\bibfnamefont {Z.}~\bibnamefont {Chang}}, \bibinfo {author} {\bibfnamefont {X.}~\bibnamefont {Tong}},\ and\ \bibinfo {author} {\bibfnamefont {C.}~\bibnamefont {Lin}},\ }\href@noop {} {\bibfield  {journal} {\bibinfo  {journal} {Opt. Express}\ }\textbf {\bibinfo {volume} {13}},\ \bibinfo {pages} {1966} (\bibinfo {year} {2005})}\BibitemShut {NoStop}%
\bibitem [{\citenamefont {Eckle}\ \emph {et~al.}(2008)\citenamefont {Eckle}, \citenamefont {Smolarski}, \citenamefont {Schlup}, \citenamefont {Biegert}, \citenamefont {Staudte}, \citenamefont {Sch{\"o}ffler}, \citenamefont {Muller}, \citenamefont {D{\"o}rner},\ and\ \citenamefont {Keller}}]{eckle08b}%
  \BibitemOpen
  \bibfield  {author} {\bibinfo {author} {\bibfnamefont {P.}~\bibnamefont {Eckle}}, \bibinfo {author} {\bibfnamefont {M.}~\bibnamefont {Smolarski}}, \bibinfo {author} {\bibfnamefont {P.}~\bibnamefont {Schlup}}, \bibinfo {author} {\bibfnamefont {J.}~\bibnamefont {Biegert}}, \bibinfo {author} {\bibfnamefont {A.}~\bibnamefont {Staudte}}, \bibinfo {author} {\bibfnamefont {M.}~\bibnamefont {Sch{\"o}ffler}}, \bibinfo {author} {\bibfnamefont {H.~G.}\ \bibnamefont {Muller}}, \bibinfo {author} {\bibfnamefont {R.}~\bibnamefont {D{\"o}rner}},\ and\ \bibinfo {author} {\bibfnamefont {U.}~\bibnamefont {Keller}},\ }\href@noop {} {\bibfield  {journal} {\bibinfo  {journal} {Nat. Phys.}\ }\textbf {\bibinfo {volume} {4}},\ \bibinfo {pages} {565} (\bibinfo {year} {2008})}\BibitemShut {NoStop}%
\bibitem [{\citenamefont {Kazansky}\ \emph {et~al.}(2016)\citenamefont {Kazansky}, \citenamefont {Bozhevolnov}, \citenamefont {Sazhina},\ and\ \citenamefont {Kabachnik}}]{kazansky16a}%
  \BibitemOpen
  \bibfield  {author} {\bibinfo {author} {\bibfnamefont {A.}~\bibnamefont {Kazansky}}, \bibinfo {author} {\bibfnamefont {A.}~\bibnamefont {Bozhevolnov}}, \bibinfo {author} {\bibfnamefont {I.}~\bibnamefont {Sazhina}},\ and\ \bibinfo {author} {\bibfnamefont {N.}~\bibnamefont {Kabachnik}},\ }\href@noop {} {\bibfield  {journal} {\bibinfo  {journal} {Phys. Rev. A}\ }\textbf {\bibinfo {volume} {93}},\ \bibinfo {pages} {013407} (\bibinfo {year} {2016})}\BibitemShut {NoStop}%
\bibitem [{\citenamefont {Kazansky}\ \emph {et~al.}(2019)\citenamefont {Kazansky}, \citenamefont {Sazhina},\ and\ \citenamefont {Kabachnik}}]{kazansky19a}%
  \BibitemOpen
  \bibfield  {author} {\bibinfo {author} {\bibfnamefont {A.}~\bibnamefont {Kazansky}}, \bibinfo {author} {\bibfnamefont {I.}~\bibnamefont {Sazhina}},\ and\ \bibinfo {author} {\bibfnamefont {N.}~\bibnamefont {Kabachnik}},\ }\href@noop {} {\bibfield  {journal} {\bibinfo  {journal} {Opt. Express}\ }\textbf {\bibinfo {volume} {27}},\ \bibinfo {pages} {12939} (\bibinfo {year} {2019})}\BibitemShut {NoStop}%
\bibitem [{\citenamefont {Hartmann}\ \emph {et~al.}(2018)\citenamefont {Hartmann}, \citenamefont {Hartmann}, \citenamefont {Heider}, \citenamefont {Wagner}, \citenamefont {Ilchen}, \citenamefont {Buck}, \citenamefont {Lindahl}, \citenamefont {Benko}, \citenamefont {Gr{\"u}nert}, \citenamefont {Krzywinski} \emph {et~al.}}]{hartmann18a}%
  \BibitemOpen
  \bibfield  {author} {\bibinfo {author} {\bibfnamefont {N.}~\bibnamefont {Hartmann}}, \bibinfo {author} {\bibfnamefont {G.}~\bibnamefont {Hartmann}}, \bibinfo {author} {\bibfnamefont {R.}~\bibnamefont {Heider}}, \bibinfo {author} {\bibfnamefont {M.}~\bibnamefont {Wagner}}, \bibinfo {author} {\bibfnamefont {M.}~\bibnamefont {Ilchen}}, \bibinfo {author} {\bibfnamefont {J.}~\bibnamefont {Buck}}, \bibinfo {author} {\bibfnamefont {A.}~\bibnamefont {Lindahl}}, \bibinfo {author} {\bibfnamefont {C.}~\bibnamefont {Benko}}, \bibinfo {author} {\bibfnamefont {J.}~\bibnamefont {Gr{\"u}nert}}, \bibinfo {author} {\bibfnamefont {J.}~\bibnamefont {Krzywinski}}, \emph {et~al.},\ }\href@noop {} {\bibfield  {journal} {\bibinfo  {journal} {Nat. Photonics}\ }\textbf {\bibinfo {volume} {12}},\ \bibinfo {pages} {215} (\bibinfo {year} {2018})}\BibitemShut {NoStop}%
\bibitem [{\citenamefont {Kheifets}\ \emph {et~al.}(2022)\citenamefont {Kheifets}, \citenamefont {Wielian}, \citenamefont {Serov}, \citenamefont {Ivanov}, \citenamefont {Wang}, \citenamefont {Marinelli},\ and\ \citenamefont {Cryan}}]{kheifets_ionization_2022}%
  \BibitemOpen
  \bibfield  {author} {\bibinfo {author} {\bibfnamefont {A.~S.}\ \bibnamefont {Kheifets}}, \bibinfo {author} {\bibfnamefont {R.}~\bibnamefont {Wielian}}, \bibinfo {author} {\bibfnamefont {V.~V.}\ \bibnamefont {Serov}}, \bibinfo {author} {\bibfnamefont {I.~A.}\ \bibnamefont {Ivanov}}, \bibinfo {author} {\bibfnamefont {A.~L.}\ \bibnamefont {Wang}}, \bibinfo {author} {\bibfnamefont {A.}~\bibnamefont {Marinelli}},\ and\ \bibinfo {author} {\bibfnamefont {J.~P.}\ \bibnamefont {Cryan}},\ }\href {https://doi.org/10.1103/PhysRevA.106.033106} {\bibfield  {journal} {\bibinfo  {journal} {Phys. Rev. A}\ }\textbf {\bibinfo {volume} {106}},\ \bibinfo {pages} {033106} (\bibinfo {year} {2022})}\BibitemShut {NoStop}%
\bibitem [{\citenamefont {Serov}\ and\ \citenamefont {Kheifets}(2023)}]{serov23a}%
  \BibitemOpen
  \bibfield  {author} {\bibinfo {author} {\bibfnamefont {V.~V.}\ \bibnamefont {Serov}}\ and\ \bibinfo {author} {\bibfnamefont {A.~S.}\ \bibnamefont {Kheifets}},\ }\href@noop {} {\bibfield  {journal} {\bibinfo  {journal} {J. Phys. B: At. Mol. Opt. Phys.}\ }\textbf {\bibinfo {volume} {56}},\ \bibinfo {pages} {025601} (\bibinfo {year} {2023})}\BibitemShut {NoStop}%
\bibitem [{\citenamefont {Glownia}\ \emph {et~al.}(2010)\citenamefont {Glownia}, \citenamefont {Cryan}, \citenamefont {Andreasson}, \citenamefont {Belkacem}, \citenamefont {Berrah}, \citenamefont {Blaga}, \citenamefont {Bostedt}, \citenamefont {Bozek}, \citenamefont {DiMauro}, \citenamefont {Fang} \emph {et~al.}}]{glownia2010time}%
  \BibitemOpen
  \bibfield  {author} {\bibinfo {author} {\bibfnamefont {J.~M.}\ \bibnamefont {Glownia}}, \bibinfo {author} {\bibfnamefont {J.}~\bibnamefont {Cryan}}, \bibinfo {author} {\bibfnamefont {J.}~\bibnamefont {Andreasson}}, \bibinfo {author} {\bibfnamefont {A.}~\bibnamefont {Belkacem}}, \bibinfo {author} {\bibfnamefont {N.}~\bibnamefont {Berrah}}, \bibinfo {author} {\bibfnamefont {C.}~\bibnamefont {Blaga}}, \bibinfo {author} {\bibfnamefont {C.}~\bibnamefont {Bostedt}}, \bibinfo {author} {\bibfnamefont {J.}~\bibnamefont {Bozek}}, \bibinfo {author} {\bibfnamefont {L.}~\bibnamefont {DiMauro}}, \bibinfo {author} {\bibfnamefont {L.}~\bibnamefont {Fang}}, \emph {et~al.},\ }\href@noop {} {\bibfield  {journal} {\bibinfo  {journal} {Opt. express}\ }\textbf {\bibinfo {volume} {18}},\ \bibinfo {pages} {17620} (\bibinfo {year} {2010})}\BibitemShut {NoStop}%
\bibitem [{\citenamefont {Wang}\ \emph {et~al.}(2024)\citenamefont {Wang}, \citenamefont {Guo}, \citenamefont {Isele}, \citenamefont {Bucksbaum}, \citenamefont {Marinelli}, \citenamefont {Cryan},\ and\ \citenamefont {Driver}}]{wang2024covariance}%
  \BibitemOpen
  \bibfield  {author} {\bibinfo {author} {\bibfnamefont {J.}~\bibnamefont {Wang}}, \bibinfo {author} {\bibfnamefont {Z.}~\bibnamefont {Guo}}, \bibinfo {author} {\bibfnamefont {E.}~\bibnamefont {Isele}}, \bibinfo {author} {\bibfnamefont {P.~H.}\ \bibnamefont {Bucksbaum}}, \bibinfo {author} {\bibfnamefont {A.}~\bibnamefont {Marinelli}}, \bibinfo {author} {\bibfnamefont {J.~P.}\ \bibnamefont {Cryan}},\ and\ \bibinfo {author} {\bibfnamefont {T.}~\bibnamefont {Driver}},\ }\href@noop {} {\bibfield  {journal} {\bibinfo  {journal} {arXiv preprint arXiv:2411.01729}\ } (\bibinfo {year} {2024})}\BibitemShut {NoStop}%
\bibitem [{\citenamefont {Vall-Ilosera}\ \emph {et~al.}(2008)\citenamefont {Vall-Ilosera}, \citenamefont {Gao}, \citenamefont {Kivimaki}, \citenamefont {Coreno}, \citenamefont {Ruiz}, \citenamefont {de~Simone}, \citenamefont {{\AA}gren},\ and\ \citenamefont {Rachlew}}]{vall08a}%
  \BibitemOpen
  \bibfield  {author} {\bibinfo {author} {\bibfnamefont {G.}~\bibnamefont {Vall-Ilosera}}, \bibinfo {author} {\bibfnamefont {B.}~\bibnamefont {Gao}}, \bibinfo {author} {\bibfnamefont {A.}~\bibnamefont {Kivimaki}}, \bibinfo {author} {\bibfnamefont {M.}~\bibnamefont {Coreno}}, \bibinfo {author} {\bibfnamefont {J.~A.}\ \bibnamefont {Ruiz}}, \bibinfo {author} {\bibfnamefont {M.}~\bibnamefont {de~Simone}}, \bibinfo {author} {\bibfnamefont {H.}~\bibnamefont {{\AA}gren}},\ and\ \bibinfo {author} {\bibfnamefont {E.}~\bibnamefont {Rachlew}},\ }\href@noop {} {\bibfield  {journal} {\bibinfo  {journal} {J. Chem. Phys.}\ }\textbf {\bibinfo {volume} {128}} (\bibinfo {year} {2008})}\BibitemShut {NoStop}%
\bibitem [{\citenamefont {Piancastelli}(1999)}]{piancastelli99a}%
  \BibitemOpen
  \bibfield  {author} {\bibinfo {author} {\bibfnamefont {M.}~\bibnamefont {Piancastelli}},\ }\href@noop {} {\bibfield  {journal} {\bibinfo  {journal} {J. Electron. Spectrosc. Relat. Phenom.}\ }\textbf {\bibinfo {volume} {100}},\ \bibinfo {pages} {167} (\bibinfo {year} {1999})}\BibitemShut {NoStop}%
\bibitem [{\citenamefont {Dahlstr\"om}\ \emph {et~al.}(2012)\citenamefont {Dahlstr\"om}, \citenamefont {L'Huillier},\ and\ \citenamefont {Maquet}}]{dahlstrom12a}%
  \BibitemOpen
  \bibfield  {author} {\bibinfo {author} {\bibfnamefont {J.~M.}\ \bibnamefont {Dahlstr\"om}}, \bibinfo {author} {\bibfnamefont {A.}~\bibnamefont {L'Huillier}},\ and\ \bibinfo {author} {\bibfnamefont {A.}~\bibnamefont {Maquet}},\ }\href {http://stacks.iop.org/0953-4075/45/i=18/a=183001} {\bibfield  {journal} {\bibinfo  {journal} {J. Phys. B: At. Mol. Opt. Phys.}\ }\textbf {\bibinfo {volume} {45}},\ \bibinfo {pages} {183001} (\bibinfo {year} {2012})}\BibitemShut {NoStop}%
\bibitem [{\citenamefont {Ji}\ \emph {et~al.}(2024)\citenamefont {Ji}, \citenamefont {Ueda}, \citenamefont {Han},\ and\ \citenamefont {W{\"o}rner}}]{ji2024analytical}%
  \BibitemOpen
  \bibfield  {author} {\bibinfo {author} {\bibfnamefont {J.-B.}\ \bibnamefont {Ji}}, \bibinfo {author} {\bibfnamefont {K.}~\bibnamefont {Ueda}}, \bibinfo {author} {\bibfnamefont {M.}~\bibnamefont {Han}},\ and\ \bibinfo {author} {\bibfnamefont {H.~J.}\ \bibnamefont {W{\"o}rner}},\ }\href@noop {} {\bibfield  {journal} {\bibinfo  {journal} {J. Phys. B: At. Mol. Opt. Phys.}\ }\textbf {\bibinfo {volume} {57}},\ \bibinfo {pages} {235601} (\bibinfo {year} {2024})}\BibitemShut {NoStop}%
\bibitem [{\citenamefont {Karule}(1990)}]{karule1990integrals}%
  \BibitemOpen
  \bibfield  {author} {\bibinfo {author} {\bibfnamefont {E.}~\bibnamefont {Karule}},\ }\href@noop {} {\bibfield  {journal} {\bibinfo  {journal} {Journal of Physics A: Mathematical and General}\ }\textbf {\bibinfo {volume} {23}},\ \bibinfo {pages} {1969} (\bibinfo {year} {1990})}\BibitemShut {NoStop}%
\bibitem [{\citenamefont {Zhang}\ \emph {et~al.}(2020)\citenamefont {Zhang}, \citenamefont {Duris}, \citenamefont {MacArthur}, \citenamefont {Zholents}, \citenamefont {Huang},\ and\ \citenamefont {Marinelli}}]{zhang2020experimental}%
  \BibitemOpen
  \bibfield  {author} {\bibinfo {author} {\bibfnamefont {Z.}~\bibnamefont {Zhang}}, \bibinfo {author} {\bibfnamefont {J.}~\bibnamefont {Duris}}, \bibinfo {author} {\bibfnamefont {J.~P.}\ \bibnamefont {MacArthur}}, \bibinfo {author} {\bibfnamefont {A.}~\bibnamefont {Zholents}}, \bibinfo {author} {\bibfnamefont {Z.}~\bibnamefont {Huang}},\ and\ \bibinfo {author} {\bibfnamefont {A.}~\bibnamefont {Marinelli}},\ }\href@noop {} {\bibfield  {journal} {\bibinfo  {journal} {New J. Phys.}\ }\textbf {\bibinfo {volume} {22}},\ \bibinfo {pages} {083030} (\bibinfo {year} {2020})}\BibitemShut {NoStop}%
\bibitem [{\citenamefont {Seaberg}\ \emph {et~al.}(2022)\citenamefont {Seaberg}, \citenamefont {Lee}, \citenamefont {Morton}, \citenamefont {Cheng}, \citenamefont {Cryan}, \citenamefont {Curiel}, \citenamefont {Dix}, \citenamefont {Driver}, \citenamefont {Fox}, \citenamefont {Hardin}, \citenamefont {Kamalov}, \citenamefont {Li}, \citenamefont {Li}, \citenamefont {Lin}, \citenamefont {Liu}, \citenamefont {Montagne}, \citenamefont {Obaid}, \citenamefont {Sakdinawat}, \citenamefont {Stefan}, \citenamefont {Whitney}, \citenamefont {Wolf}, \citenamefont {Zhang}, \citenamefont {Fritz}, \citenamefont {Walter}, \citenamefont {Cocco},\ and\ \citenamefont {Ng}}]{seaberg_x-ray_2022}%
  \BibitemOpen
  \bibfield  {author} {\bibinfo {author} {\bibfnamefont {M.}~\bibnamefont {Seaberg}}, \bibinfo {author} {\bibfnamefont {L.}~\bibnamefont {Lee}}, \bibinfo {author} {\bibfnamefont {D.}~\bibnamefont {Morton}}, \bibinfo {author} {\bibfnamefont {X.}~\bibnamefont {Cheng}}, \bibinfo {author} {\bibfnamefont {J.}~\bibnamefont {Cryan}}, \bibinfo {author} {\bibfnamefont {G.~I.}\ \bibnamefont {Curiel}}, \bibinfo {author} {\bibfnamefont {B.}~\bibnamefont {Dix}}, \bibinfo {author} {\bibfnamefont {T.}~\bibnamefont {Driver}}, \bibinfo {author} {\bibfnamefont {K.}~\bibnamefont {Fox}}, \bibinfo {author} {\bibfnamefont {C.}~\bibnamefont {Hardin}}, \bibinfo {author} {\bibfnamefont {A.}~\bibnamefont {Kamalov}}, \bibinfo {author} {\bibfnamefont {K.}~\bibnamefont {Li}}, \bibinfo {author} {\bibfnamefont {X.}~\bibnamefont {Li}}, \bibinfo {author} {\bibfnamefont {M.-F.}\ \bibnamefont {Lin}}, \bibinfo {author} {\bibfnamefont {Y.}~\bibnamefont {Liu}}, \bibinfo {author} {\bibfnamefont {T.}~\bibnamefont {Montagne}}, \bibinfo {author}
  {\bibfnamefont {R.}~\bibnamefont {Obaid}}, \bibinfo {author} {\bibfnamefont {A.}~\bibnamefont {Sakdinawat}}, \bibinfo {author} {\bibfnamefont {P.}~\bibnamefont {Stefan}}, \bibinfo {author} {\bibfnamefont {R.}~\bibnamefont {Whitney}}, \bibinfo {author} {\bibfnamefont {T.}~\bibnamefont {Wolf}}, \bibinfo {author} {\bibfnamefont {L.}~\bibnamefont {Zhang}}, \bibinfo {author} {\bibfnamefont {D.}~\bibnamefont {Fritz}}, \bibinfo {author} {\bibfnamefont {P.}~\bibnamefont {Walter}}, \bibinfo {author} {\bibfnamefont {D.}~\bibnamefont {Cocco}},\ and\ \bibinfo {author} {\bibfnamefont {M.~L.}\ \bibnamefont {Ng}},\ }\href {https://doi.org/10.1080/08940886.2022.2066416} {\bibfield  {journal} {\bibinfo  {journal} {Synchrotron Radiat. News}\ }\textbf {\bibinfo {volume} {35}},\ \bibinfo {pages} {20} (\bibinfo {year} {2022})}\BibitemShut {NoStop}%
\bibitem [{\citenamefont {Obaid}\ \emph {et~al.}(2018)\citenamefont {Obaid}, \citenamefont {Buth}, \citenamefont {Dakovski}, \citenamefont {Beerwerth}, \citenamefont {Holmes}, \citenamefont {Aldrich}, \citenamefont {Lin}, \citenamefont {Minitti}, \citenamefont {Osipov}, \citenamefont {Schlotter} \emph {et~al.}}]{obaid2018lcls}%
  \BibitemOpen
  \bibfield  {author} {\bibinfo {author} {\bibfnamefont {R.}~\bibnamefont {Obaid}}, \bibinfo {author} {\bibfnamefont {C.}~\bibnamefont {Buth}}, \bibinfo {author} {\bibfnamefont {G.~L.}\ \bibnamefont {Dakovski}}, \bibinfo {author} {\bibfnamefont {R.}~\bibnamefont {Beerwerth}}, \bibinfo {author} {\bibfnamefont {M.}~\bibnamefont {Holmes}}, \bibinfo {author} {\bibfnamefont {J.}~\bibnamefont {Aldrich}}, \bibinfo {author} {\bibfnamefont {M.-F.}\ \bibnamefont {Lin}}, \bibinfo {author} {\bibfnamefont {M.}~\bibnamefont {Minitti}}, \bibinfo {author} {\bibfnamefont {T.}~\bibnamefont {Osipov}}, \bibinfo {author} {\bibfnamefont {W.}~\bibnamefont {Schlotter}}, \emph {et~al.},\ }\href@noop {} {\bibfield  {journal} {\bibinfo  {journal} {J. Phys. B: At. Mol. Opt. Phys.}\ }\textbf {\bibinfo {volume} {51}},\ \bibinfo {pages} {034003} (\bibinfo {year} {2018})}\BibitemShut {NoStop}%
\bibitem [{\citenamefont {Hettrick}\ \emph {et~al.}(1988)\citenamefont {Hettrick}, \citenamefont {Underwood}, \citenamefont {Batson},\ and\ \citenamefont {Eckart}}]{hettrick1988resolving}%
  \BibitemOpen
  \bibfield  {author} {\bibinfo {author} {\bibfnamefont {M.~C.}\ \bibnamefont {Hettrick}}, \bibinfo {author} {\bibfnamefont {J.~H.}\ \bibnamefont {Underwood}}, \bibinfo {author} {\bibfnamefont {P.~J.}\ \bibnamefont {Batson}},\ and\ \bibinfo {author} {\bibfnamefont {M.~J.}\ \bibnamefont {Eckart}},\ }\href@noop {} {\bibfield  {journal} {\bibinfo  {journal} {Appl. Opt.}\ }\textbf {\bibinfo {volume} {27}},\ \bibinfo {pages} {200} (\bibinfo {year} {1988})}\BibitemShut {NoStop}%
\bibitem [{\citenamefont {Orel}\ \emph {et~al.}(1990)\citenamefont {Orel}, \citenamefont {Rescigno},\ and\ \citenamefont {Lengsfield~III}}]{Orel1990a}%
  \BibitemOpen
  \bibfield  {author} {\bibinfo {author} {\bibfnamefont {A.~E.}\ \bibnamefont {Orel}}, \bibinfo {author} {\bibfnamefont {T.~N.}\ \bibnamefont {Rescigno}},\ and\ \bibinfo {author} {\bibfnamefont {B.~H.}\ \bibnamefont {Lengsfield~III}},\ }\href {https://doi.org/10.1103/PhysRevA.42.5292} {\bibfield  {journal} {\bibinfo  {journal} {Phys. Rev. A}\ }\textbf {\bibinfo {volume} {42}},\ \bibinfo {pages} {5292} (\bibinfo {year} {1990})}\BibitemShut {NoStop}%
\bibitem [{\citenamefont {Gianturco}\ \emph {et~al.}(1994)\citenamefont {Gianturco}, \citenamefont {Lucchese},\ and\ \citenamefont {Sanna}}]{Gianturco1994a}%
  \BibitemOpen
  \bibfield  {author} {\bibinfo {author} {\bibfnamefont {F.~A.}\ \bibnamefont {Gianturco}}, \bibinfo {author} {\bibfnamefont {R.~R.}\ \bibnamefont {Lucchese}},\ and\ \bibinfo {author} {\bibfnamefont {N.}~\bibnamefont {Sanna}},\ }\href {https://doi.org/10.1063/1.467237} {\bibfield  {journal} {\bibinfo  {journal} {J. Chem. Phys.}\ }\textbf {\bibinfo {volume} {100}},\ \bibinfo {pages} {6464} (\bibinfo {year} {1994})}\BibitemShut {NoStop}%
\bibitem [{\citenamefont {Natalense}\ and\ \citenamefont {Lucchese}(1999)}]{Natalense1999a}%
  \BibitemOpen
  \bibfield  {author} {\bibinfo {author} {\bibfnamefont {A.~P.~P.}\ \bibnamefont {Natalense}}\ and\ \bibinfo {author} {\bibfnamefont {R.~R.}\ \bibnamefont {Lucchese}},\ }\href {https://doi.org/10.1063/1.479794} {\bibfield  {journal} {\bibinfo  {journal} {J. Chem. Phys.}\ }\textbf {\bibinfo {volume} {111}},\ \bibinfo {pages} {5344} (\bibinfo {year} {1999})}\BibitemShut {NoStop}%
\bibitem [{\citenamefont {Dunning}(1989)}]{Dunning1989a}%
  \BibitemOpen
  \bibfield  {author} {\bibinfo {author} {\bibfnamefont {J.}~\bibnamefont {Dunning}, \bibfnamefont {Thom~H.}},\ }\href {https://doi.org/10.1063/1.456153} {\bibfield  {journal} {\bibinfo  {journal} {J. Chem. Phys.}\ }\textbf {\bibinfo {volume} {90}},\ \bibinfo {pages} {1007} (\bibinfo {year} {1989})}\BibitemShut {NoStop}%
\bibitem [{\citenamefont {Perdew}\ and\ \citenamefont {Zunger}(1981)}]{Perdew1981}%
  \BibitemOpen
  \bibfield  {author} {\bibinfo {author} {\bibfnamefont {J.~P.}\ \bibnamefont {Perdew}}\ and\ \bibinfo {author} {\bibfnamefont {A.}~\bibnamefont {Zunger}},\ }\href@noop {} {\bibfield  {journal} {\bibinfo  {journal} {Phys. Rev. B}\ }\textbf {\bibinfo {volume} {23}},\ \bibinfo {pages} {5048} (\bibinfo {year} {1981})}\BibitemShut {NoStop}%
\bibitem [{\citenamefont {Marante}\ \emph {et~al.}(2020)\citenamefont {Marante}, \citenamefont {Greenman}, \citenamefont {Trevisan}, \citenamefont {Rescigno}, \citenamefont {McCurdy},\ and\ \citenamefont {Lucchese}}]{Marante2020a}%
  \BibitemOpen
  \bibfield  {author} {\bibinfo {author} {\bibfnamefont {C.~A.}\ \bibnamefont {Marante}}, \bibinfo {author} {\bibfnamefont {L.}~\bibnamefont {Greenman}}, \bibinfo {author} {\bibfnamefont {C.~S.}\ \bibnamefont {Trevisan}}, \bibinfo {author} {\bibfnamefont {T.~N.}\ \bibnamefont {Rescigno}}, \bibinfo {author} {\bibfnamefont {C.~W.}\ \bibnamefont {McCurdy}},\ and\ \bibinfo {author} {\bibfnamefont {R.~R.}\ \bibnamefont {Lucchese}},\ }\href {https://doi.org/10.1103/PhysRevA.102.012815} {\bibfield  {journal} {\bibinfo  {journal} {Phys. Rev. A}\ }\textbf {\bibinfo {volume} {102}},\ \bibinfo {pages} {012815} (\bibinfo {year} {2020})}\BibitemShut {NoStop}%
\end{thebibliography}%


\begin{thebibliography}{27}%
\makeatletter
\providecommand \@ifxundefined [1]{%
 \@ifx{#1\undefined}
}%
\providecommand \@ifnum [1]{%
 \ifnum #1\expandafter \@firstoftwo
 \else \expandafter \@secondoftwo
 \fi
}%
\providecommand \@ifx [1]{%
 \ifx #1\expandafter \@firstoftwo
 \else \expandafter \@secondoftwo
 \fi
}%
\providecommand \natexlab [1]{#1}%
\providecommand \enquote  [1]{``#1''}%
\providecommand \bibnamefont  [1]{#1}%
\providecommand \bibfnamefont [1]{#1}%
\providecommand \citenamefont [1]{#1}%
\providecommand \href@noop [0]{\@secondoftwo}%
\providecommand \href [0]{\begingroup \@sanitize@url \@href}%
\providecommand \@href[1]{\@@startlink{#1}\@@href}%
\providecommand \@@href[1]{\endgroup#1\@@endlink}%
\providecommand \@sanitize@url [0]{\catcode `\\12\catcode `\$12\catcode `\&12\catcode `\#12\catcode `\^12\catcode `\_12\catcode `\%12\relax}%
\providecommand \@@startlink[1]{}%
\providecommand \@@endlink[0]{}%
\providecommand \url  [0]{\begingroup\@sanitize@url \@url }%
\providecommand \@url [1]{\endgroup\@href {#1}{\urlprefix }}%
\providecommand \urlprefix  [0]{URL }%
\providecommand \Eprint [0]{\href }%
\providecommand \doibase [0]{https://doi.org/}%
\providecommand \selectlanguage [0]{\@gobble}%
\providecommand \bibinfo  [0]{\@secondoftwo}%
\providecommand \bibfield  [0]{\@secondoftwo}%
\providecommand \translation [1]{[#1]}%
\providecommand \BibitemOpen [0]{}%
\providecommand \bibitemStop [0]{}%
\providecommand \bibitemNoStop [0]{.\EOS\space}%
\providecommand \EOS [0]{\spacefactor3000\relax}%
\providecommand \BibitemShut  [1]{\csname bibitem#1\endcsname}%
\let\auto@bib@innerbib\@empty
\bibitem [{\citenamefont {Zhang}\ \emph {et~al.}(2020)\citenamefont {Zhang}, \citenamefont {Duris}, \citenamefont {MacArthur}, \citenamefont {Zholents}, \citenamefont {Huang},\ and\ \citenamefont {Marinelli}}]{zhang2020experimental}%
  \BibitemOpen
  \bibfield  {author} {\bibinfo {author} {\bibfnamefont {Z.}~\bibnamefont {Zhang}}, \bibinfo {author} {\bibfnamefont {J.}~\bibnamefont {Duris}}, \bibinfo {author} {\bibfnamefont {J.~P.}\ \bibnamefont {MacArthur}}, \bibinfo {author} {\bibfnamefont {A.}~\bibnamefont {Zholents}}, \bibinfo {author} {\bibfnamefont {Z.}~\bibnamefont {Huang}},\ and\ \bibinfo {author} {\bibfnamefont {A.}~\bibnamefont {Marinelli}},\ }\href@noop {} {\bibfield  {journal} {\bibinfo  {journal} {New J. Phys.}\ }\textbf {\bibinfo {volume} {22}},\ \bibinfo {pages} {083030} (\bibinfo {year} {2020})}\BibitemShut {NoStop}%
\bibitem [{\citenamefont {Duris}\ \emph {et~al.}(2020)\citenamefont {Duris}, \citenamefont {Li}, \citenamefont {Driver}, \citenamefont {Champenois}, \citenamefont {MacArthur}, \citenamefont {Lutman}, \citenamefont {Zhang}, \citenamefont {Rosenberger}, \citenamefont {Aldrich}, \citenamefont {Coffee} \emph {et~al.}}]{duris20a}%
  \BibitemOpen
  \bibfield  {author} {\bibinfo {author} {\bibfnamefont {J.}~\bibnamefont {Duris}}, \bibinfo {author} {\bibfnamefont {S.}~\bibnamefont {Li}}, \bibinfo {author} {\bibfnamefont {T.}~\bibnamefont {Driver}}, \bibinfo {author} {\bibfnamefont {E.~G.}\ \bibnamefont {Champenois}}, \bibinfo {author} {\bibfnamefont {J.~P.}\ \bibnamefont {MacArthur}}, \bibinfo {author} {\bibfnamefont {A.~A.}\ \bibnamefont {Lutman}}, \bibinfo {author} {\bibfnamefont {Z.}~\bibnamefont {Zhang}}, \bibinfo {author} {\bibfnamefont {P.}~\bibnamefont {Rosenberger}}, \bibinfo {author} {\bibfnamefont {J.~W.}\ \bibnamefont {Aldrich}}, \bibinfo {author} {\bibfnamefont {R.}~\bibnamefont {Coffee}}, \emph {et~al.},\ }\href@noop {} {\bibfield  {journal} {\bibinfo  {journal} {Nat. Photonics}\ }\textbf {\bibinfo {volume} {14}},\ \bibinfo {pages} {30} (\bibinfo {year} {2020})}\BibitemShut {NoStop}%
\bibitem [{\citenamefont {Seaberg}\ \emph {et~al.}(2022)\citenamefont {Seaberg}, \citenamefont {Lee}, \citenamefont {Morton}, \citenamefont {Cheng}, \citenamefont {Cryan}, \citenamefont {Curiel}, \citenamefont {Dix}, \citenamefont {Driver}, \citenamefont {Fox}, \citenamefont {Hardin}, \citenamefont {Kamalov}, \citenamefont {Li}, \citenamefont {Li}, \citenamefont {Lin}, \citenamefont {Liu}, \citenamefont {Montagne}, \citenamefont {Obaid}, \citenamefont {Sakdinawat}, \citenamefont {Stefan}, \citenamefont {Whitney}, \citenamefont {Wolf}, \citenamefont {Zhang}, \citenamefont {Fritz}, \citenamefont {Walter}, \citenamefont {Cocco},\ and\ \citenamefont {Ng}}]{seaberg_X-ray_2022}%
  \BibitemOpen
  \bibfield  {author} {\bibinfo {author} {\bibfnamefont {M.}~\bibnamefont {Seaberg}}, \bibinfo {author} {\bibfnamefont {L.}~\bibnamefont {Lee}}, \bibinfo {author} {\bibfnamefont {D.}~\bibnamefont {Morton}}, \bibinfo {author} {\bibfnamefont {X.}~\bibnamefont {Cheng}}, \bibinfo {author} {\bibfnamefont {J.}~\bibnamefont {Cryan}}, \bibinfo {author} {\bibfnamefont {G.~I.}\ \bibnamefont {Curiel}}, \bibinfo {author} {\bibfnamefont {B.}~\bibnamefont {Dix}}, \bibinfo {author} {\bibfnamefont {T.}~\bibnamefont {Driver}}, \bibinfo {author} {\bibfnamefont {K.}~\bibnamefont {Fox}}, \bibinfo {author} {\bibfnamefont {C.}~\bibnamefont {Hardin}}, \bibinfo {author} {\bibfnamefont {A.}~\bibnamefont {Kamalov}}, \bibinfo {author} {\bibfnamefont {K.}~\bibnamefont {Li}}, \bibinfo {author} {\bibfnamefont {X.}~\bibnamefont {Li}}, \bibinfo {author} {\bibfnamefont {M.-F.}\ \bibnamefont {Lin}}, \bibinfo {author} {\bibfnamefont {Y.}~\bibnamefont {Liu}}, \bibinfo {author} {\bibfnamefont {T.}~\bibnamefont {Montagne}}, \bibinfo {author}
  {\bibfnamefont {R.}~\bibnamefont {Obaid}}, \bibinfo {author} {\bibfnamefont {A.}~\bibnamefont {Sakdinawat}}, \bibinfo {author} {\bibfnamefont {P.}~\bibnamefont {Stefan}}, \bibinfo {author} {\bibfnamefont {R.}~\bibnamefont {Whitney}}, \bibinfo {author} {\bibfnamefont {T.}~\bibnamefont {Wolf}}, \bibinfo {author} {\bibfnamefont {L.}~\bibnamefont {Zhang}}, \bibinfo {author} {\bibfnamefont {D.}~\bibnamefont {Fritz}}, \bibinfo {author} {\bibfnamefont {P.}~\bibnamefont {Walter}}, \bibinfo {author} {\bibfnamefont {D.}~\bibnamefont {Cocco}},\ and\ \bibinfo {author} {\bibfnamefont {M.~L.}\ \bibnamefont {Ng}},\ }\href {https://doi.org/10.1080/08940886.2022.2066416} {\bibfield  {journal} {\bibinfo  {journal} {Synchrotron Radiat. News}\ }\textbf {\bibinfo {volume} {35}},\ \bibinfo {pages} {20} (\bibinfo {year} {2022})}\BibitemShut {NoStop}%
\bibitem [{\citenamefont {Li}\ \emph {et~al.}(2018)\citenamefont {Li}, \citenamefont {Champenois}, \citenamefont {Coffee}, \citenamefont {Guo}, \citenamefont {Hegazy}, \citenamefont {Kamalov}, \citenamefont {Natan}, \citenamefont {O’Neal}, \citenamefont {Osipov}, \citenamefont {Owens}, \citenamefont {Ray}, \citenamefont {Rich}, \citenamefont {Walter}, \citenamefont {Marinelli},\ and\ \citenamefont {Cryan}}]{li_co-axial_2018}%
  \BibitemOpen
  \bibfield  {author} {\bibinfo {author} {\bibfnamefont {S.}~\bibnamefont {Li}}, \bibinfo {author} {\bibfnamefont {E.~G.}\ \bibnamefont {Champenois}}, \bibinfo {author} {\bibfnamefont {R.}~\bibnamefont {Coffee}}, \bibinfo {author} {\bibfnamefont {Z.}~\bibnamefont {Guo}}, \bibinfo {author} {\bibfnamefont {K.}~\bibnamefont {Hegazy}}, \bibinfo {author} {\bibfnamefont {A.}~\bibnamefont {Kamalov}}, \bibinfo {author} {\bibfnamefont {A.}~\bibnamefont {Natan}}, \bibinfo {author} {\bibfnamefont {J.}~\bibnamefont {O’Neal}}, \bibinfo {author} {\bibfnamefont {T.}~\bibnamefont {Osipov}}, \bibinfo {author} {\bibfnamefont {M.}~\bibnamefont {Owens}}, \bibinfo {author} {\bibfnamefont {D.}~\bibnamefont {Ray}}, \bibinfo {author} {\bibfnamefont {D.}~\bibnamefont {Rich}}, \bibinfo {author} {\bibfnamefont {P.}~\bibnamefont {Walter}}, \bibinfo {author} {\bibfnamefont {A.}~\bibnamefont {Marinelli}},\ and\ \bibinfo {author} {\bibfnamefont {J.~P.}\ \bibnamefont {Cryan}},\ }\href {https://doi.org/10.1063/1.5046192} {\bibfield
  {journal} {\bibinfo  {journal} {AIP Adv.}\ }\textbf {\bibinfo {volume} {8}},\ \bibinfo {pages} {115308} (\bibinfo {year} {2018})}\BibitemShut {NoStop}%
\bibitem [{\citenamefont {Obaid}\ \emph {et~al.}(2018)\citenamefont {Obaid}, \citenamefont {Buth}, \citenamefont {Dakovski}, \citenamefont {Beerwerth}, \citenamefont {Holmes}, \citenamefont {Aldrich}, \citenamefont {Lin}, \citenamefont {Minitti}, \citenamefont {Osipov}, \citenamefont {Schlotter}, \citenamefont {Cederbaum}, \citenamefont {Fritzsche},\ and\ \citenamefont {Berrah}}]{obaid_lcls_2018}%
  \BibitemOpen
  \bibfield  {author} {\bibinfo {author} {\bibfnamefont {R.}~\bibnamefont {Obaid}}, \bibinfo {author} {\bibfnamefont {C.}~\bibnamefont {Buth}}, \bibinfo {author} {\bibfnamefont {G.~L.}\ \bibnamefont {Dakovski}}, \bibinfo {author} {\bibfnamefont {R.}~\bibnamefont {Beerwerth}}, \bibinfo {author} {\bibfnamefont {M.}~\bibnamefont {Holmes}}, \bibinfo {author} {\bibfnamefont {J.}~\bibnamefont {Aldrich}}, \bibinfo {author} {\bibfnamefont {M.-F.}\ \bibnamefont {Lin}}, \bibinfo {author} {\bibfnamefont {M.}~\bibnamefont {Minitti}}, \bibinfo {author} {\bibfnamefont {T.}~\bibnamefont {Osipov}}, \bibinfo {author} {\bibfnamefont {W.}~\bibnamefont {Schlotter}}, \bibinfo {author} {\bibfnamefont {L.~S.}\ \bibnamefont {Cederbaum}}, \bibinfo {author} {\bibfnamefont {S.}~\bibnamefont {Fritzsche}},\ and\ \bibinfo {author} {\bibfnamefont {N.}~\bibnamefont {Berrah}},\ }\href {https://doi.org/10.1088/1361-6455/aaa189} {\bibfield  {journal} {\bibinfo  {journal} {J. Phys. B: At. Mol. Opt. Phys.}\ }\textbf {\bibinfo {volume} {51}},\
  \bibinfo {pages} {034003} (\bibinfo {year} {2018})}\BibitemShut {NoStop}%
\bibitem [{\citenamefont {Hettrick}\ \emph {et~al.}(1988)\citenamefont {Hettrick}, \citenamefont {Underwood}, \citenamefont {Batson},\ and\ \citenamefont {Eckart}}]{hettrick1988resolving}%
  \BibitemOpen
  \bibfield  {author} {\bibinfo {author} {\bibfnamefont {M.~C.}\ \bibnamefont {Hettrick}}, \bibinfo {author} {\bibfnamefont {J.~H.}\ \bibnamefont {Underwood}}, \bibinfo {author} {\bibfnamefont {P.~J.}\ \bibnamefont {Batson}},\ and\ \bibinfo {author} {\bibfnamefont {M.~J.}\ \bibnamefont {Eckart}},\ }\href@noop {} {\bibfield  {journal} {\bibinfo  {journal} {Appl. Opt.}\ }\textbf {\bibinfo {volume} {27}},\ \bibinfo {pages} {200} (\bibinfo {year} {1988})}\BibitemShut {NoStop}%
\bibitem [{\citenamefont {Guo}\ \emph {et~al.}(2024)\citenamefont {Guo}, \citenamefont {Driver}, \citenamefont {Beauvarlet}, \citenamefont {Cesar}, \citenamefont {Duris}, \citenamefont {Franz}, \citenamefont {Alexander}, \citenamefont {Bohler}, \citenamefont {Bostedt}, \citenamefont {Averbukh}, \citenamefont {Cheng}, \citenamefont {DiMauro}, \citenamefont {Doumy}, \citenamefont {Forbes}, \citenamefont {Gessner}, \citenamefont {Glownia}, \citenamefont {Isele}, \citenamefont {Kamalov}, \citenamefont {Larsen}, \citenamefont {Li}, \citenamefont {Li}, \citenamefont {Lin}, \citenamefont {McCracken}, \citenamefont {Obaid}, \citenamefont {O’Neal}, \citenamefont {Robles}, \citenamefont {Rolles}, \citenamefont {Ruberti}, \citenamefont {Rudenko}, \citenamefont {Slaughter}, \citenamefont {Sudar}, \citenamefont {Thierstein}, \citenamefont {Tuthill}, \citenamefont {Ueda}, \citenamefont {Wang}, \citenamefont {Wang}, \citenamefont {Wang}, \citenamefont {Weber}, \citenamefont {Wolf}, \citenamefont {Young}, \citenamefont
  {Zhang}, \citenamefont {Bucksbaum}, \citenamefont {Marangos}, \citenamefont {Kling}, \citenamefont {Huang}, \citenamefont {Walter}, \citenamefont {Inhester}, \citenamefont {Berrah}, \citenamefont {Cryan},\ and\ \citenamefont {Marinelli}}]{guo_experimental_2024}%
  \BibitemOpen
  \bibfield  {author} {\bibinfo {author} {\bibfnamefont {Z.}~\bibnamefont {Guo}}, \bibinfo {author} {\bibfnamefont {T.}~\bibnamefont {Driver}}, \bibinfo {author} {\bibfnamefont {S.}~\bibnamefont {Beauvarlet}}, \bibinfo {author} {\bibfnamefont {D.}~\bibnamefont {Cesar}}, \bibinfo {author} {\bibfnamefont {J.}~\bibnamefont {Duris}}, \bibinfo {author} {\bibfnamefont {P.~L.}\ \bibnamefont {Franz}}, \bibinfo {author} {\bibfnamefont {O.}~\bibnamefont {Alexander}}, \bibinfo {author} {\bibfnamefont {D.}~\bibnamefont {Bohler}}, \bibinfo {author} {\bibfnamefont {C.}~\bibnamefont {Bostedt}}, \bibinfo {author} {\bibfnamefont {V.}~\bibnamefont {Averbukh}}, \bibinfo {author} {\bibfnamefont {X.}~\bibnamefont {Cheng}}, \bibinfo {author} {\bibfnamefont {L.~F.}\ \bibnamefont {DiMauro}}, \bibinfo {author} {\bibfnamefont {G.}~\bibnamefont {Doumy}}, \bibinfo {author} {\bibfnamefont {R.}~\bibnamefont {Forbes}}, \bibinfo {author} {\bibfnamefont {O.}~\bibnamefont {Gessner}}, \bibinfo {author} {\bibfnamefont {J.~M.}\ \bibnamefont
  {Glownia}}, \bibinfo {author} {\bibfnamefont {E.}~\bibnamefont {Isele}}, \bibinfo {author} {\bibfnamefont {A.}~\bibnamefont {Kamalov}}, \bibinfo {author} {\bibfnamefont {K.~A.}\ \bibnamefont {Larsen}}, \bibinfo {author} {\bibfnamefont {S.}~\bibnamefont {Li}}, \bibinfo {author} {\bibfnamefont {X.}~\bibnamefont {Li}}, \bibinfo {author} {\bibfnamefont {M.-F.}\ \bibnamefont {Lin}}, \bibinfo {author} {\bibfnamefont {G.~A.}\ \bibnamefont {McCracken}}, \bibinfo {author} {\bibfnamefont {R.}~\bibnamefont {Obaid}}, \bibinfo {author} {\bibfnamefont {J.~T.}\ \bibnamefont {O’Neal}}, \bibinfo {author} {\bibfnamefont {R.~R.}\ \bibnamefont {Robles}}, \bibinfo {author} {\bibfnamefont {D.}~\bibnamefont {Rolles}}, \bibinfo {author} {\bibfnamefont {M.}~\bibnamefont {Ruberti}}, \bibinfo {author} {\bibfnamefont {A.}~\bibnamefont {Rudenko}}, \bibinfo {author} {\bibfnamefont {D.~S.}\ \bibnamefont {Slaughter}}, \bibinfo {author} {\bibfnamefont {N.~S.}\ \bibnamefont {Sudar}}, \bibinfo {author} {\bibfnamefont {E.}~\bibnamefont
  {Thierstein}}, \bibinfo {author} {\bibfnamefont {D.}~\bibnamefont {Tuthill}}, \bibinfo {author} {\bibfnamefont {K.}~\bibnamefont {Ueda}}, \bibinfo {author} {\bibfnamefont {E.}~\bibnamefont {Wang}}, \bibinfo {author} {\bibfnamefont {A.~L.}\ \bibnamefont {Wang}}, \bibinfo {author} {\bibfnamefont {J.}~\bibnamefont {Wang}}, \bibinfo {author} {\bibfnamefont {T.}~\bibnamefont {Weber}}, \bibinfo {author} {\bibfnamefont {T.~J.~A.}\ \bibnamefont {Wolf}}, \bibinfo {author} {\bibfnamefont {L.}~\bibnamefont {Young}}, \bibinfo {author} {\bibfnamefont {Z.}~\bibnamefont {Zhang}}, \bibinfo {author} {\bibfnamefont {P.~H.}\ \bibnamefont {Bucksbaum}}, \bibinfo {author} {\bibfnamefont {J.~P.}\ \bibnamefont {Marangos}}, \bibinfo {author} {\bibfnamefont {M.~F.}\ \bibnamefont {Kling}}, \bibinfo {author} {\bibfnamefont {Z.}~\bibnamefont {Huang}}, \bibinfo {author} {\bibfnamefont {P.}~\bibnamefont {Walter}}, \bibinfo {author} {\bibfnamefont {L.}~\bibnamefont {Inhester}}, \bibinfo {author} {\bibfnamefont {N.}~\bibnamefont {Berrah}},
  \bibinfo {author} {\bibfnamefont {J.~P.}\ \bibnamefont {Cryan}},\ and\ \bibinfo {author} {\bibfnamefont {A.}~\bibnamefont {Marinelli}},\ }\href {https://doi.org/10.1038/s41566-024-01419-w} {\bibfield  {journal} {\bibinfo  {journal} {Nat. Photonics}\ }\textbf {\bibinfo {volume} {18}},\ \bibinfo {pages} {691} (\bibinfo {year} {2024})}\BibitemShut {NoStop}%
\bibitem [{\citenamefont {Orel}\ \emph {et~al.}(1990)\citenamefont {Orel}, \citenamefont {Rescigno},\ and\ \citenamefont {Lengsfield~III}}]{Orel1990a}%
  \BibitemOpen
  \bibfield  {author} {\bibinfo {author} {\bibfnamefont {A.~E.}\ \bibnamefont {Orel}}, \bibinfo {author} {\bibfnamefont {T.~N.}\ \bibnamefont {Rescigno}},\ and\ \bibinfo {author} {\bibfnamefont {B.~H.}\ \bibnamefont {Lengsfield~III}},\ }\href {https://doi.org/10.1103/PhysRevA.42.5292} {\bibfield  {journal} {\bibinfo  {journal} {Phys. Rev. A}\ }\textbf {\bibinfo {volume} {42}},\ \bibinfo {pages} {5292} (\bibinfo {year} {1990})}\BibitemShut {NoStop}%
\bibitem [{\citenamefont {Gianturco}\ \emph {et~al.}(1994)\citenamefont {Gianturco}, \citenamefont {Lucchese},\ and\ \citenamefont {Sanna}}]{Gianturco1994a}%
  \BibitemOpen
  \bibfield  {author} {\bibinfo {author} {\bibfnamefont {F.~A.}\ \bibnamefont {Gianturco}}, \bibinfo {author} {\bibfnamefont {R.~R.}\ \bibnamefont {Lucchese}},\ and\ \bibinfo {author} {\bibfnamefont {N.}~\bibnamefont {Sanna}},\ }\href {https://doi.org/10.1063/1.467237} {\bibfield  {journal} {\bibinfo  {journal} {J. Chem. Phys.}\ }\textbf {\bibinfo {volume} {100}},\ \bibinfo {pages} {6464} (\bibinfo {year} {1994})}\BibitemShut {NoStop}%
\bibitem [{\citenamefont {Natalense}\ and\ \citenamefont {Lucchese}(1999)}]{Natalense1999a}%
  \BibitemOpen
  \bibfield  {author} {\bibinfo {author} {\bibfnamefont {A.~P.~P.}\ \bibnamefont {Natalense}}\ and\ \bibinfo {author} {\bibfnamefont {R.~R.}\ \bibnamefont {Lucchese}},\ }\href {https://doi.org/10.1063/1.479794} {\bibfield  {journal} {\bibinfo  {journal} {J. Chem. Phys.}\ }\textbf {\bibinfo {volume} {111}},\ \bibinfo {pages} {5344} (\bibinfo {year} {1999})}\BibitemShut {NoStop}%
\bibitem [{\citenamefont {Perdew}\ and\ \citenamefont {Zunger}(1981)}]{Perdew1981}%
  \BibitemOpen
  \bibfield  {author} {\bibinfo {author} {\bibfnamefont {J.~P.}\ \bibnamefont {Perdew}}\ and\ \bibinfo {author} {\bibfnamefont {A.}~\bibnamefont {Zunger}},\ }\href@noop {} {\bibfield  {journal} {\bibinfo  {journal} {Phys. Rev. B}\ }\textbf {\bibinfo {volume} {23}},\ \bibinfo {pages} {5048} (\bibinfo {year} {1981})}\BibitemShut {NoStop}%
\bibitem [{\citenamefont {Dunning}(1989)}]{Dunning1989a}%
  \BibitemOpen
  \bibfield  {author} {\bibinfo {author} {\bibfnamefont {J.}~\bibnamefont {Dunning}, \bibfnamefont {Thom~H.}},\ }\href {https://doi.org/10.1063/1.456153} {\bibfield  {journal} {\bibinfo  {journal} {J. Chem. Phys.}\ }\textbf {\bibinfo {volume} {90}},\ \bibinfo {pages} {1007} (\bibinfo {year} {1989})}\BibitemShut {NoStop}%
\bibitem [{\citenamefont {Marante}\ \emph {et~al.}(2020)\citenamefont {Marante}, \citenamefont {Greenman}, \citenamefont {Trevisan}, \citenamefont {Rescigno}, \citenamefont {McCurdy},\ and\ \citenamefont {Lucchese}}]{Marante2020a}%
  \BibitemOpen
  \bibfield  {author} {\bibinfo {author} {\bibfnamefont {C.~A.}\ \bibnamefont {Marante}}, \bibinfo {author} {\bibfnamefont {L.}~\bibnamefont {Greenman}}, \bibinfo {author} {\bibfnamefont {C.~S.}\ \bibnamefont {Trevisan}}, \bibinfo {author} {\bibfnamefont {T.~N.}\ \bibnamefont {Rescigno}}, \bibinfo {author} {\bibfnamefont {C.~W.}\ \bibnamefont {McCurdy}},\ and\ \bibinfo {author} {\bibfnamefont {R.~R.}\ \bibnamefont {Lucchese}},\ }\href {https://doi.org/10.1103/PhysRevA.102.012815} {\bibfield  {journal} {\bibinfo  {journal} {Phys. Rev. A}\ }\textbf {\bibinfo {volume} {102}},\ \bibinfo {pages} {012815} (\bibinfo {year} {2020})}\BibitemShut {NoStop}%
\bibitem [{\citenamefont {Hoshino}\ \emph {et~al.}(2018)\citenamefont {Hoshino}, \citenamefont {Kato}, \citenamefont {Kuze}, \citenamefont {Tanaka}, \citenamefont {Fukuzawa}, \citenamefont {Ueda},\ and\ \citenamefont {Lucchese}}]{Hoshino2018a}%
  \BibitemOpen
  \bibfield  {author} {\bibinfo {author} {\bibfnamefont {M.}~\bibnamefont {Hoshino}}, \bibinfo {author} {\bibfnamefont {H.}~\bibnamefont {Kato}}, \bibinfo {author} {\bibfnamefont {N.}~\bibnamefont {Kuze}}, \bibinfo {author} {\bibfnamefont {H.}~\bibnamefont {Tanaka}}, \bibinfo {author} {\bibfnamefont {H.}~\bibnamefont {Fukuzawa}}, \bibinfo {author} {\bibfnamefont {K.}~\bibnamefont {Ueda}},\ and\ \bibinfo {author} {\bibfnamefont {R.~R.}\ \bibnamefont {Lucchese}},\ }\href {https://doi.org/10.1088/1361-6455/aaaeb3} {\bibfield  {journal} {\bibinfo  {journal} {J. Phys. B: At. Mol. Opt. Phys.}\ }\textbf {\bibinfo {volume} {51}},\ \bibinfo {pages} {065402} (\bibinfo {year} {2018})}\BibitemShut {NoStop}%
\bibitem [{\citenamefont {Vall-Ilosera}\ \emph {et~al.}(2008)\citenamefont {Vall-Ilosera}, \citenamefont {Gao}, \citenamefont {Kivimaki}, \citenamefont {Coreno}, \citenamefont {Ruiz}, \citenamefont {de~Simone}, \citenamefont {{\AA}gren},\ and\ \citenamefont {Rachlew}}]{vall08a}%
  \BibitemOpen
  \bibfield  {author} {\bibinfo {author} {\bibfnamefont {G.}~\bibnamefont {Vall-Ilosera}}, \bibinfo {author} {\bibfnamefont {B.}~\bibnamefont {Gao}}, \bibinfo {author} {\bibfnamefont {A.}~\bibnamefont {Kivimaki}}, \bibinfo {author} {\bibfnamefont {M.}~\bibnamefont {Coreno}}, \bibinfo {author} {\bibfnamefont {J.~A.}\ \bibnamefont {Ruiz}}, \bibinfo {author} {\bibfnamefont {M.}~\bibnamefont {de~Simone}}, \bibinfo {author} {\bibfnamefont {H.}~\bibnamefont {{\AA}gren}},\ and\ \bibinfo {author} {\bibfnamefont {E.}~\bibnamefont {Rachlew}},\ }\href@noop {} {\bibfield  {journal} {\bibinfo  {journal} {J. Chem. Phys.}\ }\textbf {\bibinfo {volume} {128}} (\bibinfo {year} {2008})}\BibitemShut {NoStop}%
\bibitem [{\citenamefont {Lucchese}\ and\ \citenamefont {Gianturco}(1996)}]{Lucchese1996a}%
  \BibitemOpen
  \bibfield  {author} {\bibinfo {author} {\bibfnamefont {R.~R.}\ \bibnamefont {Lucchese}}\ and\ \bibinfo {author} {\bibfnamefont {F.~A.}\ \bibnamefont {Gianturco}},\ }\href@noop {} {\bibfield  {journal} {\bibinfo  {journal} {Int. Rev. Phys. Chem.}\ }\textbf {\bibinfo {volume} {15}},\ \bibinfo {pages} {429} (\bibinfo {year} {1996})}\BibitemShut {NoStop}%
\bibitem [{\citenamefont {Lane}\ and\ \citenamefont {Robson}(1966)}]{Lane1966}%
  \BibitemOpen
  \bibfield  {author} {\bibinfo {author} {\bibfnamefont {A.~M.}\ \bibnamefont {Lane}}\ and\ \bibinfo {author} {\bibfnamefont {D.}~\bibnamefont {Robson}},\ }\href {https://doi.org/10.1103/PhysRev.151.774} {\bibfield  {journal} {\bibinfo  {journal} {Phys. Rev.}\ }\textbf {\bibinfo {volume} {151}},\ \bibinfo {pages} {774} (\bibinfo {year} {1966})}\BibitemShut {NoStop}%
\bibitem [{\citenamefont {Meyer}\ and\ \citenamefont {Walter}(1982)}]{Meyer1982}%
  \BibitemOpen
  \bibfield  {author} {\bibinfo {author} {\bibfnamefont {H.~D.}\ \bibnamefont {Meyer}}\ and\ \bibinfo {author} {\bibfnamefont {O.}~\bibnamefont {Walter}},\ }\href {https://doi.org/10.1088/0022-3700/15/20/013} {\bibfield  {journal} {\bibinfo  {journal} {J. Phys. B: At. Mol. Phys.}\ }\textbf {\bibinfo {volume} {15}},\ \bibinfo {pages} {3647} (\bibinfo {year} {1982})}\BibitemShut {NoStop}%
\bibitem [{\citenamefont {Kukulin}\ \emph {et~al.}(1989)\citenamefont {Kukulin}, \citenamefont {Krasnopol’sky},\ and\ \citenamefont {Horáček}}]{Kukulin1989}%
  \BibitemOpen
  \bibfield  {author} {\bibinfo {author} {\bibfnamefont {V.~I.}\ \bibnamefont {Kukulin}}, \bibinfo {author} {\bibfnamefont {V.~M.}\ \bibnamefont {Krasnopol’sky}},\ and\ \bibinfo {author} {\bibfnamefont {J.}~\bibnamefont {Horáček}},\ }\href@noop {} {\emph {\bibinfo {title} {Theory of Resonances: Principles and Applications}}}\ (\bibinfo  {publisher} {Springer Netherlands},\ \bibinfo {address} {Dordrecht},\ \bibinfo {year} {1989})\BibitemShut {NoStop}%
\bibitem [{\citenamefont {Pazourek}\ \emph {et~al.}(2015)\citenamefont {Pazourek}, \citenamefont {Nagele},\ and\ \citenamefont {Burgd\"orfer}}]{pazourek15a}%
  \BibitemOpen
  \bibfield  {author} {\bibinfo {author} {\bibfnamefont {R.}~\bibnamefont {Pazourek}}, \bibinfo {author} {\bibfnamefont {S.}~\bibnamefont {Nagele}},\ and\ \bibinfo {author} {\bibfnamefont {J.}~\bibnamefont {Burgd\"orfer}},\ }\href {https://doi.org/10.1103/RevModPhys.87.765} {\bibfield  {journal} {\bibinfo  {journal} {Rev. Mod. Phys.}\ }\textbf {\bibinfo {volume} {87}},\ \bibinfo {pages} {765} (\bibinfo {year} {2015})}\BibitemShut {NoStop}%
\bibitem [{\citenamefont {Dahlstr\"om}\ \emph {et~al.}(2012)\citenamefont {Dahlstr\"om}, \citenamefont {L'Huillier},\ and\ \citenamefont {Maquet}}]{dahlstrom12a}%
  \BibitemOpen
  \bibfield  {author} {\bibinfo {author} {\bibfnamefont {J.~M.}\ \bibnamefont {Dahlstr\"om}}, \bibinfo {author} {\bibfnamefont {A.}~\bibnamefont {L'Huillier}},\ and\ \bibinfo {author} {\bibfnamefont {A.}~\bibnamefont {Maquet}},\ }\href {http://stacks.iop.org/0953-4075/45/i=18/a=183001} {\bibfield  {journal} {\bibinfo  {journal} {J. Phys. B: At. Mol. Opt. Phys.}\ }\textbf {\bibinfo {volume} {45}},\ \bibinfo {pages} {183001} (\bibinfo {year} {2012})}\BibitemShut {NoStop}%
\bibitem [{\citenamefont {Dahlstr{\"{o}}m}\ \emph {et~al.}(2013)\citenamefont {Dahlstr{\"{o}}m}, \citenamefont {Gu{\'{e}}not}, \citenamefont {Kl{\"{u}}nder}, \citenamefont {Gisselbrecht}, \citenamefont {Mauritsson}, \citenamefont {L'Huillier}, \citenamefont {Maquet},\ and\ \citenamefont {Ta{\"{i}}eb}}]{Dahlstrom2013}%
  \BibitemOpen
  \bibfield  {author} {\bibinfo {author} {\bibfnamefont {J.~M.}\ \bibnamefont {Dahlstr{\"{o}}m}}, \bibinfo {author} {\bibfnamefont {D.}~\bibnamefont {Gu{\'{e}}not}}, \bibinfo {author} {\bibfnamefont {K.}~\bibnamefont {Kl{\"{u}}nder}}, \bibinfo {author} {\bibfnamefont {M.}~\bibnamefont {Gisselbrecht}}, \bibinfo {author} {\bibfnamefont {J.}~\bibnamefont {Mauritsson}}, \bibinfo {author} {\bibfnamefont {A.}~\bibnamefont {L'Huillier}}, \bibinfo {author} {\bibfnamefont {A.}~\bibnamefont {Maquet}},\ and\ \bibinfo {author} {\bibfnamefont {R.}~\bibnamefont {Ta{\"{i}}eb}},\ }\href {https://doi.org/10.1016/j.chemphys.2012.01.017} {\bibfield  {journal} {\bibinfo  {journal} {Chem. Phys.}\ }\textbf {\bibinfo {volume} {414}},\ \bibinfo {pages} {53} (\bibinfo {year} {2013})},\ \Eprint {https://arxiv.org/abs/arXiv:1112.4144v2} {arXiv:arXiv:1112.4144v2} \BibitemShut {NoStop}%
\bibitem [{\citenamefont {Serov}\ \emph {et~al.}(2015)\citenamefont {Serov}, \citenamefont {Derbov},\ and\ \citenamefont {Sergeeva}}]{serov_interpretation_2015}%
  \BibitemOpen
  \bibfield  {author} {\bibinfo {author} {\bibfnamefont {V.~V.}\ \bibnamefont {Serov}}, \bibinfo {author} {\bibfnamefont {V.~L.}\ \bibnamefont {Derbov}},\ and\ \bibinfo {author} {\bibfnamefont {T.~A.}\ \bibnamefont {Sergeeva}},\ }in\ \href {https://doi.org/10.1007/978-94-017-9481-7_14} {{\selectlanguage {en}\emph {\bibinfo {booktitle} {Advanced {Lasers}: {Laser} {Physics} and {Technology} for {Applied} and {Fundamental} {Science}}}}},\ \bibinfo {series and number} {Springer {Series} in {Optical} {Sciences}},\ \bibinfo {editor} {edited by\ \bibinfo {editor} {\bibfnamefont {O.}~\bibnamefont {Shulika}}\ and\ \bibinfo {editor} {\bibfnamefont {I.}~\bibnamefont {Sukhoivanov}}}\ (\bibinfo  {publisher} {Springer Netherlands},\ \bibinfo {address} {Dordrecht},\ \bibinfo {year} {2015})\ pp.\ \bibinfo {pages} {213--230}\BibitemShut {NoStop}%
\bibitem [{\citenamefont {Ji}\ \emph {et~al.}(2024)\citenamefont {Ji}, \citenamefont {Ueda}, \citenamefont {Han},\ and\ \citenamefont {W{\"o}rner}}]{ji2024analytical}%
  \BibitemOpen
  \bibfield  {author} {\bibinfo {author} {\bibfnamefont {J.-B.}\ \bibnamefont {Ji}}, \bibinfo {author} {\bibfnamefont {K.}~\bibnamefont {Ueda}}, \bibinfo {author} {\bibfnamefont {M.}~\bibnamefont {Han}},\ and\ \bibinfo {author} {\bibfnamefont {H.~J.}\ \bibnamefont {W{\"o}rner}},\ }\href@noop {} {\bibfield  {journal} {\bibinfo  {journal} {J. Phys. B: At. Mol. Opt. Phys.}\ }\textbf {\bibinfo {volume} {57}},\ \bibinfo {pages} {235601} (\bibinfo {year} {2024})}\BibitemShut {NoStop}%
\bibitem [{\citenamefont {Kheifets}\ \emph {et~al.}(2022)\citenamefont {Kheifets}, \citenamefont {Wielian}, \citenamefont {Serov}, \citenamefont {Ivanov}, \citenamefont {Wang}, \citenamefont {Marinelli},\ and\ \citenamefont {Cryan}}]{kheifets_ionization_2022}%
  \BibitemOpen
  \bibfield  {author} {\bibinfo {author} {\bibfnamefont {A.~S.}\ \bibnamefont {Kheifets}}, \bibinfo {author} {\bibfnamefont {R.}~\bibnamefont {Wielian}}, \bibinfo {author} {\bibfnamefont {V.~V.}\ \bibnamefont {Serov}}, \bibinfo {author} {\bibfnamefont {I.~A.}\ \bibnamefont {Ivanov}}, \bibinfo {author} {\bibfnamefont {A.~L.}\ \bibnamefont {Wang}}, \bibinfo {author} {\bibfnamefont {A.}~\bibnamefont {Marinelli}},\ and\ \bibinfo {author} {\bibfnamefont {J.~P.}\ \bibnamefont {Cryan}},\ }\href {https://doi.org/10.1103/PhysRevA.106.033106} {\bibfield  {journal} {\bibinfo  {journal} {Phys. Rev. A}\ }\textbf {\bibinfo {volume} {106}},\ \bibinfo {pages} {033106} (\bibinfo {year} {2022})}\BibitemShut {NoStop}%
\bibitem [{\citenamefont {Nicolas}\ and\ \citenamefont {Miron}(2012)}]{nicolas_lifetime_2012}%
  \BibitemOpen
  \bibfield  {author} {\bibinfo {author} {\bibfnamefont {C.}~\bibnamefont {Nicolas}}\ and\ \bibinfo {author} {\bibfnamefont {C.}~\bibnamefont {Miron}},\ }\href {https://doi.org/10.1016/j.elspec.2012.05.008} {\bibfield  {journal} {\bibinfo  {journal} {J. Electron. Spectrosc. Relat. Phenom.}\ }\textbf {\bibinfo {volume} {185}},\ \bibinfo {pages} {267} (\bibinfo {year} {2012})}\BibitemShut {NoStop}%
\bibitem [{\citenamefont {Russek}\ and\ \citenamefont {Mehlhorn}(1986)}]{russek_post-collision_1986}%
  \BibitemOpen
  \bibfield  {author} {\bibinfo {author} {\bibfnamefont {A.}~\bibnamefont {Russek}}\ and\ \bibinfo {author} {\bibfnamefont {W.}~\bibnamefont {Mehlhorn}},\ }\href {https://doi.org/10.1088/0022-3700/19/6/013} {\bibfield  {journal} {\bibinfo  {journal} {J. Phys. B: At. Mol. Phys.}\ }\textbf {\bibinfo {volume} {19}},\ \bibinfo {pages} {911} (\bibinfo {year} {1986})}\BibitemShut {NoStop}%
\end{thebibliography}%

\end{document}


\title{Supplemental Material for Attosecond X-ray Core-level Chronoscopy of Aromatic Molecules} 

\newcommand{\AffilETH}{Laboratory of Physical Chemistry, ETH Z\"urich, Zurich, Switzerland}
\newcommand{\AffilSLAC}{SLAC National Accelerator Laboratory, Menlo Park, CA, USA}
\newcommand{\AffilPULSE}{Stanford PULSE Institute, SLAC National Accelerator Laboratory, Menlo Park, CA, USA}
\newcommand{\AffilLUXS}{Laboratory for Ultrafast X-ray Sciences, EPFL, Lausanne, Switzerland}
\newcommand{\AffilCSU}{California State University, Maritime Academy, Vallejo, CA, USA}
\newcommand{\AffilZJU}{State Key Laboratory of Extreme Photonics and Instrumentation, College of Optical Science and Engineering, Zhejiang University, Hangzhou, China}
\newcommand{\AffilKSU}{J. R. Macdonald Laboratory, Department of Physics, Kansas State University, Manhattan, KS, USA}
\newcommand{\AffilLBL}{Lawrence Berkeley National Laboratory, Berkeley, CA, USA}
\newcommand{\AffilUCD}{Department of Chemistry, University of California, Davis, CA, USA}
\newcommand{\AffilTHK}{Department of Chemistry, Tohoku University, Sendai, Japan}
\newcommand{\AffilSHT}{School of Physical Science and Technology, ShanghaiTech University, Shanghai, China}


\author{Jia-Bao Ji}\thanks{These authors contributed equally.}
\affiliation{\AffilETH}

\author{Zhaoheng Guo}\thanks{These authors contributed equally.}
\affiliation{\AffilSLAC} 
\affiliation{\AffilPULSE} 
\affiliation{\AffilLUXS}

\author{Taran Driver} 
\affiliation{\AffilSLAC} 
\affiliation{\AffilPULSE} 

\author{Cynthia S. Trevisan} 
\affiliation{\AffilCSU}

\author{David Cesar} \affiliation{\AffilSLAC}
\author{Xinxin Cheng} \affiliation{\AffilSLAC}
\author{Joseph Duris} \affiliation{\AffilSLAC}
\author{Paris L. Franz} \affiliation{\AffilSLAC} \affiliation{\AffilPULSE}
\author{James Glownia} \affiliation{\AffilSLAC}
\author{Xiaochun Gong} \affiliation{\AffilETH} \affiliation{\AffilZJU}
\author{Daniel Hammerland} \affiliation{\AffilETH}
\author{Meng Han} \affiliation{\AffilETH} \affiliation{\AffilKSU}
\author{Saijoscha Heck} \affiliation{\AffilETH}
\author{Matthias Hoffmann} \affiliation{\AffilSLAC}
\author{Andrei Kamalov} \affiliation{\AffilSLAC}
\author{Kirk A. Larsen} \affiliation{\AffilSLAC} \affiliation{\AffilPULSE}
\author{Xiang Li} \affiliation{\AffilSLAC}
\author{Ming-Fu Lin} \affiliation{\AffilSLAC}
\author{Yuchen Liu} \affiliation{\AffilLBL} \affiliation{\AffilUCD}
\author{C. William McCurdy} \affiliation{\AffilLBL} \affiliation{\AffilUCD}
\author{Razib Obaid} \affiliation{\AffilSLAC}
\author{Jordan T. O'Neal} \affiliation{\AffilPULSE}
\author{Thomas N. Rescigno} \affiliation{\AffilLBL}
\author{River R. Robles} \affiliation{\AffilSLAC} \affiliation{\AffilPULSE}
\author{Nicholas Sudar} \affiliation{\AffilSLAC} \affiliation{\AffilPULSE}
\author{Peter Walter} \affiliation{\AffilSLAC}
\author{Anna L. Wang} \affiliation{\AffilSLAC} \affiliation{\AffilPULSE}
\author{Jun Wang} \affiliation{\AffilSLAC} \affiliation{\AffilPULSE}
\author{Thomas J. A. Wolf} \affiliation{\AffilSLAC} \affiliation{\AffilPULSE}
\author{Zhen Zhang} \affiliation{\AffilSLAC}
\author{Kiyoshi Ueda} \affiliation{\AffilETH} \affiliation{\AffilTHK} \affiliation{\AffilSHT}

\author{Robert R. Lucchese} 
\email{rlucchese@lbl.gov} 
\affiliation{\AffilLBL}

\author{Agostino Marinelli}\email{marinelli@slac.stanford.edu}
\affiliation{\AffilSLAC}
\affiliation{\AffilPULSE} 

\author{James P. Cryan}\email{jcryan@stanford.edu}
\affiliation{\AffilSLAC}
\affiliation{\AffilPULSE} 

\author{Hans Jakob W\"orner}\email{hwoerner@ethz.ch}
\affiliation{\AffilETH}

\maketitle

\clearpage
\clearpage

\section{Experimental Methodology}
\subsection{Attosecond soft X-ray pulses from LCLS}

\begin{figure}[h!]
    \centering
    \includegraphics[width=\textwidth]{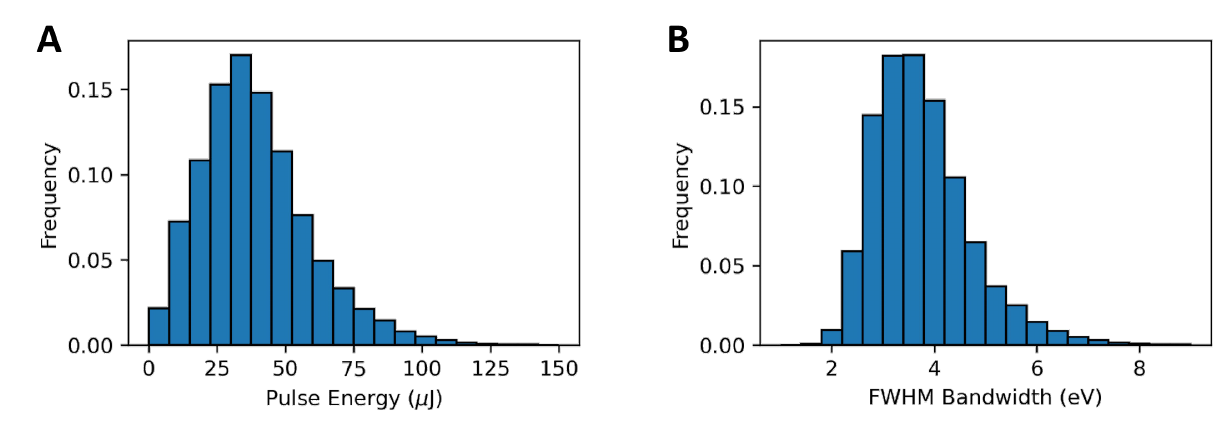}
    \caption{Distributions of pulse energies (A) and FWHM bandwidths (B) of single isolated soft X-ray XFEL pulses. We limit our analysis of FWHM bandwidth to shots with top $50\%$ of pulse energies to increase the signal-to-noise ratio.}
    \label{fig:s_GMD_and_Bandwidth}
\end{figure}

To generate attosecond pulses with an X-ray free-electron laser (XFEL), we create a high current spike in the relativistic electron bunch using temporal shaping of the photocathode laser pulse~\cite{zhang2020experimental}. 
This electron bunch produces an isolated attosecond soft X-ray pulse after passing through the undulator line of the Linac Coherent Light Source~(LCLS). 
To shape the temporal profile of the photocathode laser, we used an interferometer to create a pulse shape with a temporal notch at the center of the pulse. 
Such a laser profile produces an electron bunch with a modulated current profile, i.e. there is a local minimum of current profile in the electron bunch core. 
This current modulation was amplified to produce a high current spike in the downstream accelerator via the microbunching instability. 
The electron bunch was accelerated to $5~$GeV before entering the undulator, and has a peak current of $\sim10~$kA and $\sim1.5$~fs full width at half maximum~(FWHM) duration.
The positive energy chirp of the electron beam was on the order of $1\%$ of the electron bunch central energy.
The undulator beam line is tapered to match this energy chirp, which results in a gigawatt-level single isolated attosecond pulse~\cite{duris20a}. 
Such a large undulator taper ($\sim 0.15\%$ per module) was much larger than the acceptance of the XFEL, which therefore suppressed the background XFEL radiation outside of the main current spike. 
Figure~\ref{fig:s_GMD_and_Bandwidth} shows the distributions of XFEL pulse energies and FWHM bandwidths in a dataset with $416~$eV central X-ray photon energy. 
The mean pulse energy is $39~\mu$J and the mean FWHM bandwidth is $3.8~$eV. 
From our previous measurements, we can estimate a FWHM temporal pulse duration of $460~$as for the X-ray pulses.  

The XFEL central photon energy could be continuously adjusted by tuning the undulator gap to adjust the undulator parameter, $K$. We scanned the XFEL central photon energy from just below the nitrogen K-edge up to $418~$eV to study the energy-dependent photoionization delays between nitrogen and carbon K-shell photoemission features.
We jumped to $\sim450~$eV to take a reference measurement, where both photoelectrons have a high kinetic energy, and we expect a negligible photoemission delay. 

\subsection{Attosecond angular streaking}

Our measurements were performed at the TMO end station of LCLS.
The attosecond X-ray pulses described above were focused to a spot-size of $\sim30~\mu$m in diameter using a pair of Kirkpatrick-Baez focusing mirrors~\cite{seaberg_X-ray_2022}.
The X-ray beam was focused into the co-axial velocity map imaging~(c-VMI) spectrometer~\cite{li_co-axial_2018} which collected photoelectrons ionized from the target molecules.
At the interaction point of the c-VMI, the X-ray beam was overlapped with a focused $1300$~nm, circularly polarized ``streaking" field. 

The $1300$-nm streaking pulse with a duration of $\sim$40~fs was produced by a commercial optical parametric amplifier~(TOPAS-HE, Light Conversion) which is pumped by a $\sim30$~fs, $9$~mJ, $800$-nm laser source.
The signal and idler beams were separated by a dichroic mirror and the signal beam was sent through a superachromatic quarter-wave plate~(Thorlabs SAQWP05M-1700) to produce a circularly polarized laser field.
The 1300-nm signal beam was then focused with a $f=600$~mm CaF$_2$ lens before being reflected from a silver-coated mirror with a $2$-mm-diameter hole to co-propagate the laser and X-ray fields to the interaction point. 
The focus of the IR laser is shifted $1-2$~mm downstream of the interaction point to mitigate the effects of the Gouy phase of the IR-field across the focal volume. 

The sample molecules were introduced via a skimmed, molecular beam source. 
High pressure~(6.4~bar) helium gas is bubbled through a vessel heated to 80-140$^{\circ}$C containing one of the sample molecules.
The sample is injected into the vacuum chamber by a pulsed valve, which is separated from the main chamber by a $2$~mm diameter skimmer. 

The electrostatic lens of the c-VMI spectrometer was set for an optimal resolution on the high-energy electrons photoionized from the carbon K-shell following the procedure outlined in Ref.~\cite{li_co-axial_2018}.
The experimental voltages are outlined in Table~\ref{sm:table:cVMI}.
\begin{table}[]
    \centering
    \begin{tabular}{c|c}
     Electrode Number & Voltage (V)  \\
     \hline
     1 & -1073 \\ 2& -1522 \\ 3& -1973\\ 4 (Repeller) & -1973\\ 5 (Extractor)& -1403\\ 6& -1300\\ 7&-1243\\ 8&-1100 \\ 9& -950\\ 10& -803 \\ 11& -657\\ 12& -510\\ 13& -364\\ 14& -290\\ MCP Front& 0  
    \end{tabular}
    \caption{Voltage settings for co-axial velocity map imaging~(c-VMI)~\cite{li_co-axial_2018} spectrometer used to collect X-ray ionized photoelectrons. }
    \label{sm:table:cVMI}
\end{table}

Downstream of the interaction point, the TMO beamline features an X-ray photon spectrometer featuring a variable-line spaced (VLS) grating~\cite{obaid_lcls_2018, hettrick1988resolving}.
This spectrometer measures the incident X-ray spectrum for each X-ray shot. 
The calibration of VLS pixel to X-ray photon energy is:
\begin{equation}
\text{X-ray Photon Energy}~[\text{eV}] = (6.71 \pm 0.06)\times10^{-2}~[\text{eV/pixel}] \times \text{VLS Pixel} + (363.2 \pm 0.1)~[\text{eV}].\label{eqs:VLS_calibration}
\end{equation}
This information is used to sort the data as described below.

\clearpage

\section{Data analysis}

\subsection{Image Preprocessing}

To account for errors in the imaging system we process the raw image data. 
We apply a transformation to the raw image from the c-VMI detector so that both the nitrogen K-shell and carbon K-shell photoemission features have the expected circular momentum distribution in the absence of the streaking laser field.   
The momentum distribution is then further processed to extract the photoemission delay~(Sec.~\ref{sec_s:delay_analysis_methods}) and the photoelectron asymmetry parameters~(Sec.~\ref{sec_s:beta_2_from_cov_analysis}).

\subsection{Determination of Photoionization Delays}\label{sec_s:delay_analysis_methods}

The relative photoionization delay between the carbon and nitrogen K-shell photoemission features, $\Delta \tau$, is related to the difference in streaking angle, $\Delta \vartheta$, between the momentum shift of the nitrogen and carbon photoemission features,
\begin{equation}
    \Delta \tau = T_L \times \Delta \vartheta/2\pi,
\end{equation}
where $T_L=4.33~$fs is the period of the IR streaking laser. 

We use two methods to extract the differential streaking angle $\Delta \vartheta$ from the dataset. 
The first method~(Sec.~\ref{sec_s:partial_cov_analysis}), uses a partial covariance approach to isolate the correlated changes of streaking-induced variations in the photoelectron momentum distribution. 
A second procedure~(Sec.~\ref{sec_s:com_delay_analysis}), uses a sorting method which tags the arrival time of the attosecond X-ray pulse relative to the IR-field. 
Following the sorting, a fast Fourier Transform across this sorting direction is then applied to the momentum distribution. 
A global phase shift $\Delta \vartheta$ between the nitrogen and carbon K-shell photoelectrons is extracted from the transformed data, as further described below.

\subsubsection{Partial Covariance Analysis}\label{sec_s:partial_cov_analysis}

In the partial covariance analysis, we directly determine $\Delta \vartheta$ by studying the correlated changes of streaking-induced signal variations in two photoemission features, circumventing the need of sorting the XFEL arrival time on a single-shot basis. 
Given a relative photoionization delay $\Delta \tau = T_L \times \Delta\vartheta/2\pi$, the streaking-induced signal change in the C-1s photoline at detector angle $\theta$ has the best joint variability with the streaking-induced signal change in the N-1s photoline at detector angle $\theta+\Delta \vartheta$. 
Therefore, the information of $\Delta\vartheta$ is encoded in the covariance  between streaking-induced signal changes in two photoemission features. 
Here the covariance between two variables $A$ and $B$ is defined as:
\begin{equation}
\operatorname{Cov}[A, B] = \langle AB\rangle-\langle A \rangle \langle B \rangle,
\end{equation}
where $\langle \cdot \rangle$ denotes the statistical average.

\begin{figure}[h!]
    \centering
    \includegraphics[width=\textwidth]{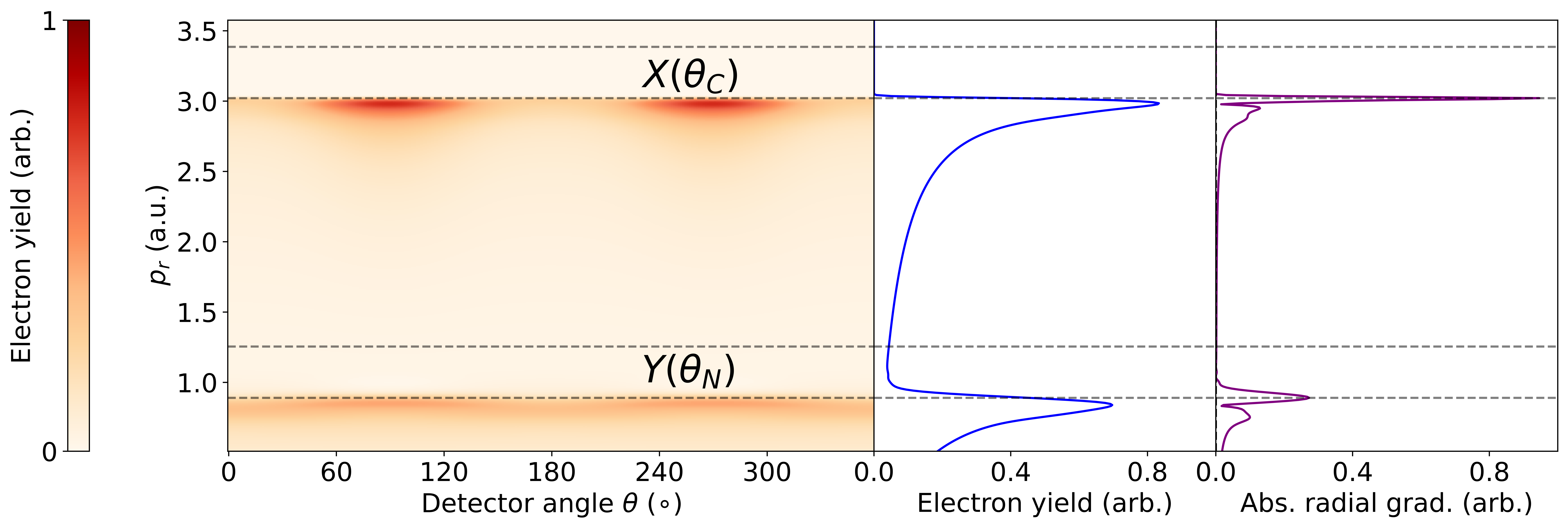}
    \caption{Regions of interest (ROIs) used in the partial covariance analysis of photoionization delays.
    The false-color map in the left panel shows the polar-rebinned average unstreaked c-VMI image of pyrimidine data, ionized by $417~$eV XFEL pulses. 
    The blue curve in the middle panel shows the radial electron yield integrated over all detector angles. 
    The purple curve in the right panel shows the absolute value of the radial gradient of data shown in the middle panel. 
    Two pairs of dashed lines present lower and upper bounds of collecting signal in C-1s ($X(\theta_C)$) and N-1s ($Y(\theta_N)$) photoelectrons for the partial covariance analysis of photoionization delays.}
    \label{fig:s_ROI_cov_analysis}
\end{figure}

The mathematical model behind the covariance analysis is discussed in the Supplementary Information of Ref.~\cite{guo_experimental_2024}. 
Here we give a brief introduction to the data analysis pipeline for this publication. 
Given a dataset of c-VMI data, we calculate two 1-D traces, namely $X(\theta_C)$ for C-1s photoelectrons and $Y(\theta_N)$ for N-1s photoelectrons, on every single-shot streaked c-VMI image. 
The region of interest (ROI) for integrating each 1-D trace is taken on the high-energy flank of the corresponding unstreaked photoemission feature. 
Essentially, the lower bound of the ROI is strictly determined by the maximal gradient of the radial electron yield in the average unstreaked c-VMI image, as shown in Fig.~\ref{fig:s_ROI_cov_analysis}. 
The selection of the upper bound of the ROI is not strict. 
In our data analysis, we chose the upper bound of the ROI to be high enough to cover a sufficient range of the streaked signal, without including too much of the X-ray artifacts. The latter originate from X-ray radiation that is diffracted on slits along the beam path and scattered onto the microchannel plate (MCP) detector. Since these artifacts change with the X-ray-photon energy, they can result in unwanted joint variability in the covariance analysis.
Two 1-D traces $X(\theta_C)$ and $Y(\theta_N)$ contain the information of streaking-induced signal changes in the C-1s and N-1s photoemission features. 

We aim to study the covariance between streaking-induced signal changes in C-1s and N-1s photoemission features. 
However, besides streaking-induced momentum shifts, fluctuations in XFEL pulse properties can also induce correlated signal changes, notably in the high-energy flanks of two photoemission features. 
For example, the electron yield in both photoemission features changes together with the XFEL pulse energy. 
To this end, we employ the rotation symmetry of the unstreaked c-VMI image and utilize the fact that the streaking-induced momentum shift breaks the rotational symmetry. 
We define the anti-symmetric traces of $X(\theta_C)$ and $Y(\theta_N)$ as:
\begin{equation}
\begin{split}
\tilde{X}(\theta_C) &= \frac{1}{2}\left( X(\theta_C)-X(\theta_C+\pi)\right),\\
\tilde{Y}(\theta_N) &= \frac{1}{2}\left( Y(\theta_N)-Y(\theta_N+\pi)\right).
\end{split}\label{eq_s:anti_symmetrization}
\end{equation}
The covariance $\operatorname{Cov}[\tilde{X}(\theta_C), \tilde{Y}(\theta_N)]$ mainly describes the joint variability between streaking-induced signal changes in two photoemission feature. 
In our data analysis, we notice that there are some X-ray artifacts near the central hole of the c-VMI. 
These X-ray artifacts do not shift with the streaking laser. 
However, they are X-ray pulse-energy dependent and they are also not symmetric in geometry. 
The anti-symmetrization defined in Eqs.~\ref{eq_s:anti_symmetrization} can not get rid of these X-ray artifacts. 
To better remove the effect of these X-ray artifacts on our covariance analysis, we further use the following partial covariance to study the photoionization delay,
\begin{equation}
\begin{split}
\operatorname{pCov}[\tilde{X}(\theta_C), \tilde{Y}(\theta_N); I(\omega)] 
&= \operatorname{Cov}[\tilde{X}(\theta_C), \tilde{Y}(\theta_N)]- \\
&\operatorname{Cov}[\tilde{X}(\theta_C), I(\omega)] \operatorname{Cov}^{-1}[I(\omega), I(\omega)] \operatorname{Cov}[I(\omega), \tilde{Y}(\theta_N)] ,
\end{split}\label{eq_s:pCov}
\end{equation}
where $I(\omega)$ is the single-shot X-ray spectrum measured on VLS. 
The sum of $I(\omega)$ is a measurement of the single-shot XFEL pulse energy. 
The partial covariance coefficient defined in Eq.~\ref{eq_s:pCov} resolves the linear dependence of the X-ray artifacts on the X-ray pulse energy. 
As a result, the partial covariance coefficient defined in Eq.~\ref{eq_s:pCov} gets rid of most contributions from the X-ray artifacts, and is dominated by the information on the joint variability in streaking-induced signal changes. 
We use the 2-D partial covariance map $\operatorname{pCov}[\tilde{X}(\theta_C), \tilde{Y}(\theta_N); I(\omega)]$ as a function of $(\theta_C, \theta_N)$ to analyze photoionization delays. 

\begin{figure}[h!]
    \centering
    \includegraphics[width=1\textwidth]{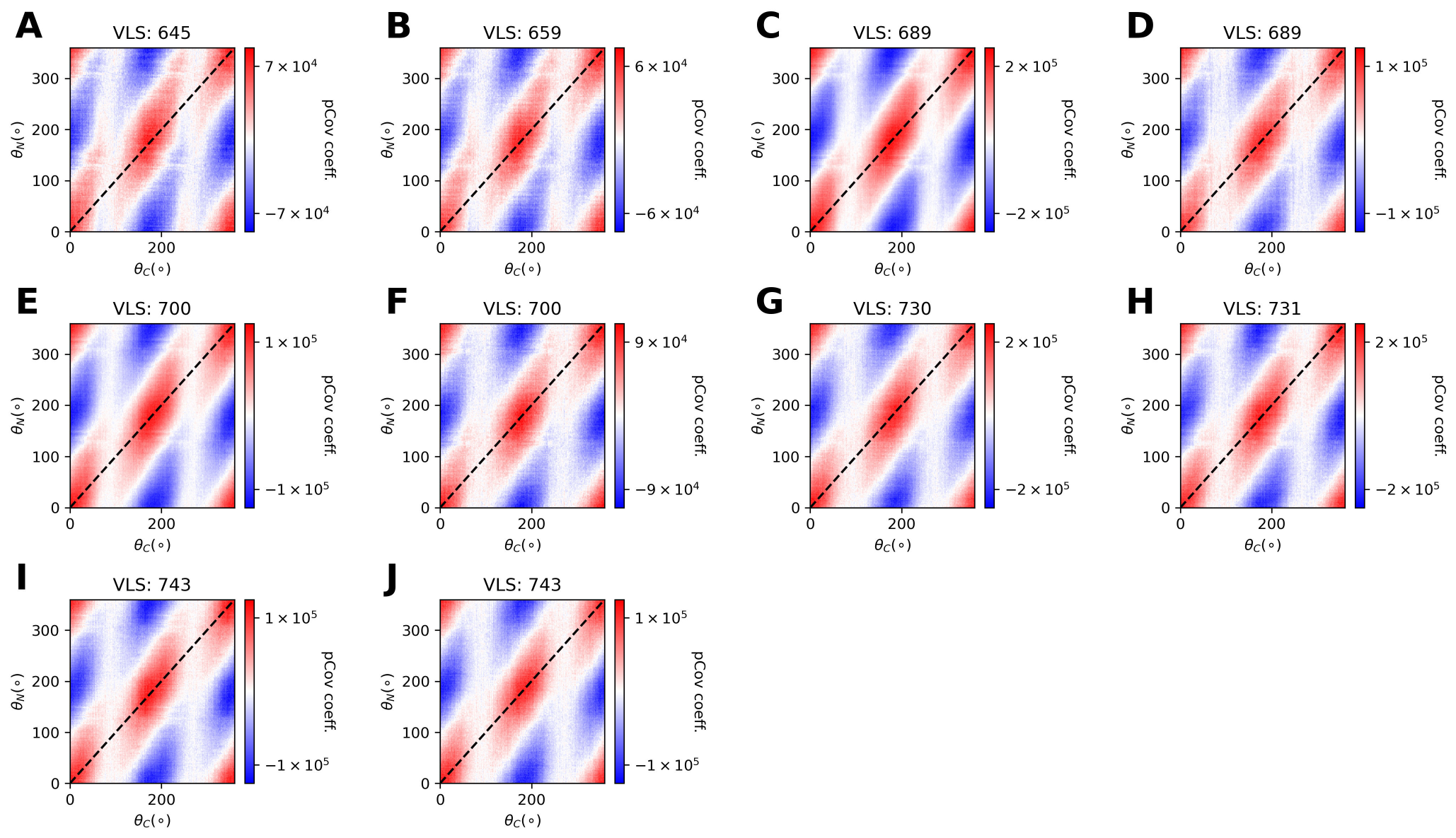}
    \caption{Partial covariance maps calculated on pyridine data.
    The false-color map in each panel shows the partial covariance map calculated on streaked c-VMI data with VLS centroids in a certain bin. 
    The title of each panel shows the mean value of VLS centroids of data in this bin. 
    Data in panels \textbf{A, B, C, E, H, I} were taken with the MCP voltages listed in Table~\ref{sm:table:cVMI}. 
    Data in panels \textbf{D, F, G, J} were taken with 1.1 times the MCP voltages listed in Table~\ref{sm:table:cVMI}.}
    \label{fig:s_pCov_PY}
\end{figure}

\begin{figure}[h!]
    \centering
    \includegraphics[width=1\textwidth]{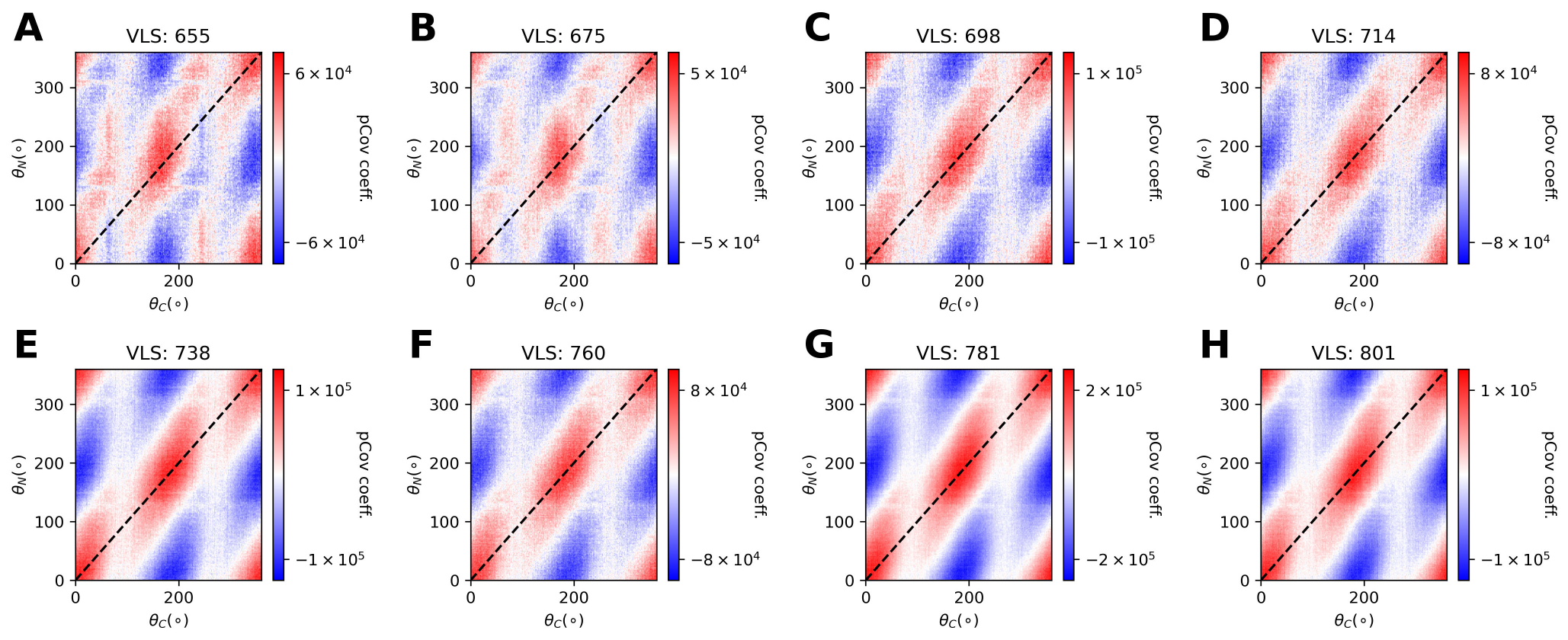}
    \caption{Partial covariance maps calculated on pyrimidine data. 
    The false-color map in each panel shows the partial covariance map calculated on streaked c-VMI data with VLS centroids in a certain bin. 
    The title of each panel shows the mean value of VLS centroids of data in this bin.}
    \label{fig:s_pCov_MY}
\end{figure}

\begin{figure}[h!]
    \centering
    \includegraphics[width=1\textwidth]{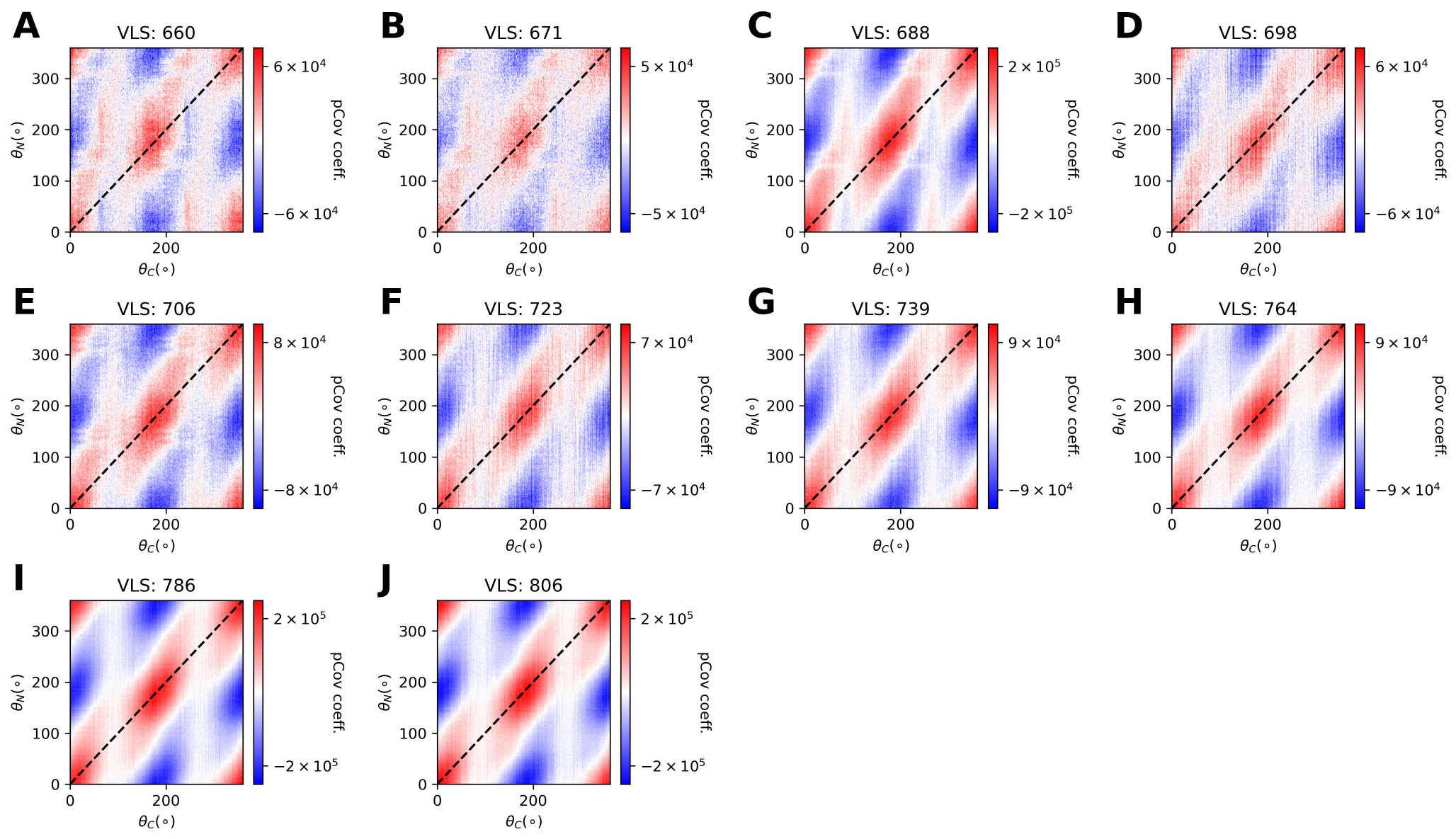}
    \caption{Partial covariance maps calculated on $s$-triazine data.
    The false-color map in each panel shows the partial covariance map calculated on streaked c-VMI data with VLS centroids in a certain bin. 
    The title of each panel shows the mean value of VLS centroids of data in this bin.}
    \label{fig:s_pCov_TR}
\end{figure}
 
For each dataset, we calculate the spectral center-of-mass of each single shot, and bin these data into 2 groups based on the spectral centers-of-mass. 
For data in each group, we calculate a 2-D partial covariance map based on Eq.~\ref{eq_s:pCov}. 
Figures~\ref{fig:s_pCov_PY}, \ref{fig:s_pCov_MY} and \ref{fig:s_pCov_TR} show partial covariance maps for all delay points from pyridine, pyrimidine and \textit{s}-triazine data. 
When the X-ray central photon energy is close to the nitrogen K-edge, the asymmetry parameter $\beta_2$ (see Sec.~\ref{sec_s:beta_2_from_cov_analysis}) of the N-1s photoemission feature is close to 0. 
The angular distribution of the N-1s photoemission feature is approximately isotropic, and all partial covariance maps therefore do not have a $\cos^2\theta_N$ shape along the $\theta_N$-axis. 
On the other hand, the C-1s photoemission features in all datasets have above $100~$eV kinetic energies and their asymmetry parameters $\beta_2$ are always close to 2. 
The angular distribution of the C-1s photoemission feature therefore has a dipolar shape, and the shapes of all partial covariance maps are modulated by a $\cos^2\theta_C$ feature along the $\theta_C$-axis.

We extract the photoionization delays by focusing on the best-correlated area on partial covariance maps. 
Every such partial covariance map has a high-value area (i.e. the area with the best join variability) around the diagonal line $\theta_N = \theta_C$. 
When the XFEL central photon energy gradually decreases to the nitrogen K-edge (i.e. smaller value in VLS center of mass), the most correlated area in partial covariance maps gradually deviate from the diagonal line and moves to $\theta_N = \theta_C+\Delta \vartheta$. 
This diagonal shift feature in the partial covariance map has an intuitive interpretation. 
Given a photionization delay $\Delta \tau = T_L \times \Delta\vartheta/2\pi$, the N-1s photoemission feature is streaked to $\theta+\Delta\vartheta$ when the C-1s photoemission feature is streaked to $\theta$, no matter what the detector angle $\theta$ is.
Therefore, the streaking-induced signal change at $\theta+\Delta\vartheta$ in the N-1s photoemission feature has the best joint variability with the streaking-induced signal changes at $\theta$ in the C-1s photoemission feature. 

\begin{figure}[h!]
    \centering
    \includegraphics[width=\textwidth]{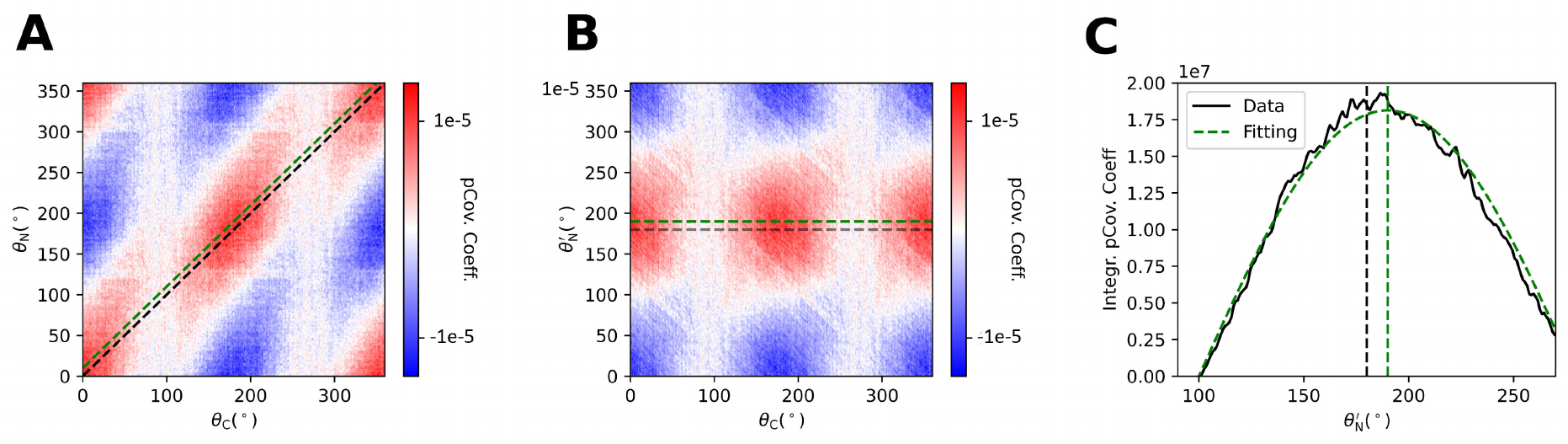}
    \caption{Demonstration of fitting photoionization delay on partial covariance map.  
    Panel~{(A)} shows the partial covariance map calculated on pyrimidine data at XFEL central photon energy $410~$eV, which is the same as Fig.~\ref{fig:s_pCov_MY}C. 
    Panel~{(B)} shows the rolled map of the partial covariance map in panel (A), which is the same as Fig.~2C in the main text. 
    The diagonal line is rolled to horizontal. The most correlated region, demonstrating the photoionization delay, clearly deviates from the horizontal line at $180^\circ$. 
    Panel~{(C)} shows the 1-D trace by integrating the rolled map in panel (B), and its sine fitting. 
    The maximum of the 1-D trace deviates from $180^\circ$ by approximately $10^\circ$, which corresponds to approximately $120~$as in photoionization delay.}
    \label{fig:s_Delay_Fit_Rolling_and_Fitting}
\end{figure}

We extract the photoionization delay from the diagonal shift feature in partial covariance maps. 
As shown in Fig.~\ref{fig:s_Delay_Fit_Rolling_and_Fitting}, panels (A) and (B), we first rotate the $\operatorname{pCov}[\tilde{X}(\theta_C), \tilde{Y}(\theta_N); I(\omega)]$ into another map $\operatorname{pCov}[\tilde{X}(\theta_C), \tilde{Y}(\theta^{\prime}_N); I(\omega)]$ by using the transformation $\theta^{\prime}_N = \theta_N - \theta_C+ 180^\circ$. 
By comparing Fig.~\ref{fig:s_Delay_Fit_Rolling_and_Fitting}B to Fig.~\ref{fig:s_Delay_Fit_Rolling_and_Fitting}A, the diagonal line in Fig.~\ref{fig:s_Delay_Fit_Rolling_and_Fitting}A becomes the horizontal line in Fig.~\ref{fig:s_Delay_Fit_Rolling_and_Fitting}B. 
As a result, the photoionization-delay-induced diagonal shift in Fig.~\ref{fig:s_Delay_Fit_Rolling_and_Fitting}A results in a vertical shift in Fig.~\ref{fig:s_Delay_Fit_Rolling_and_Fitting}B. 
We then integrate along the $\theta_C$-axis of the rolled partial covariance map to get an 1-D trace $f(\theta^{\prime}_N)$. 
The integration is taken on $\theta_C \in [-75^\circ, 75^\circ] \cup [105^\circ, 255^\circ]$ to avoid the nodes along the $\theta_C$-axis. 
Such an integration is valid since the N-1s photoemission feature is almost isotropic. 
In the last step, we perform a sine fitting on the 1-D trace $f(\theta^{\prime}_N)$.
We extract an angular shift $\Delta \vartheta$ between the peak of the sine fitting curve and $\theta^{\prime}_N = 180^\circ$. 
The value of $T_L \times \Delta \vartheta/2\pi$ is used as the photoionization delay in the partial covariance analysis. 

We use a bootstrapping approach to estimate the error in the covariance analysis of photoionization delays. 
For each delay point, we perform 100 times of resampling with replacement on data in this group to generate 100 unique datasets. 
For each generated dataset, we perform the aforementioned partial covariance analysis procedure and extract a photoionization delay. 
The errorbar of each delay point shown in Fig.~3A in the main text represents 3 times the standard deviation of 100 delays generated from the bootstrapping method.

\clearpage

\subsubsection{Center-of-mass analysis}\label{sec_s:com_delay_analysis}

As a complementary method, we developed the center-of-mass (CM) analysis, which also extracts photoionization delays from our experimental measurements. 
Here, we demonstrate the method based on pyrimidine at an XFEL central photon energy of approximately 417 eV, which corresponds to three separate runs, 
where each run contains about 34000 shots.

\begin{figure}[h!]
    \centering
    \includegraphics[width=\textwidth]{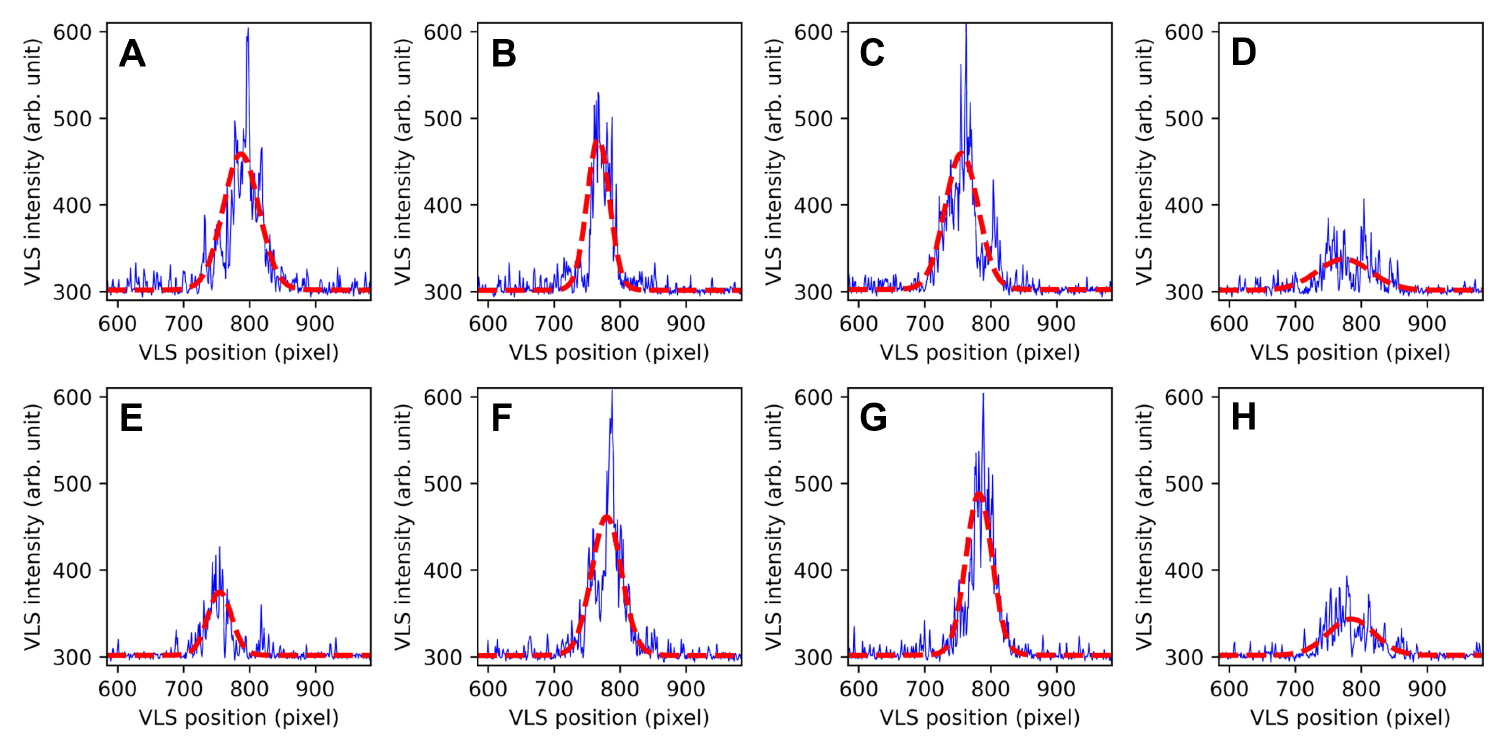}
    \caption{VLS spectra of 8 randomly chosen shots in the same run. The blue solid curve is the original signal, while the red dashed curve is the fit, the center of which is fixed to the CM of the corresponding spectrum and the shape is determined by the Voigt fitting.}
    \label{fig:VLS_CM}
\end{figure}

The central energy and pulse intensity can be characterized by fitting the VLS spectra with a Voigt profile, as demonstrated in Fig. \ref{fig:VLS_CM}, where the center of the peak is fixed at the CM of the spectrum. Within the same run, the shot-to-shot central wavelength and intensity change is not negligible. Therefore, intensity selection, central-wavelength grouping, and shot-to-shot electron kinetic energy adjustment are performed.
    
\begin{figure}[h!]
    \centering
    \includegraphics[width=0.65\textwidth]{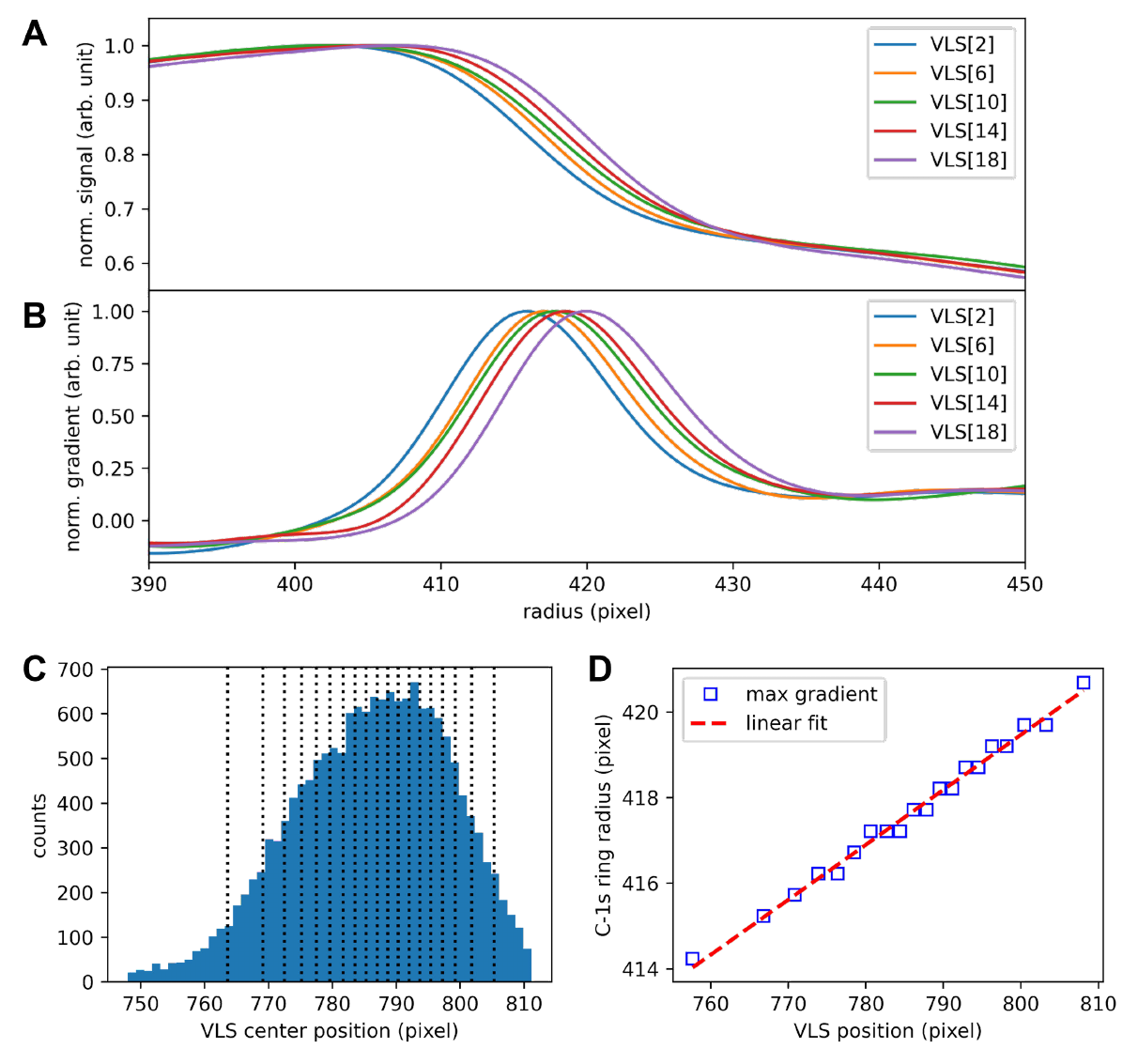}
    \caption{VLS-radius calibration by the X-ray-IR temporally unoverlapped shots. (A) The normalized signals near the C-1s edge for various VLS bins. (B) The normalized gradients (reversed) of the corresponding signals in (A). (C) The histogram of the VLS center positions for the run, where the dotted vertical lines indicate the division of the VLS bins. (D) The VLS-radius calibration. The blue squares correspond to the centers of the VLS bins and the corresponding maximal-gradient radii, which has a minimal increment of 0.5 pixel, while the red dashed line corresponds to the linear fitting using all the VLS bins except the first and the last bins.}
    \label{fig:VLS_radius}
\end{figure}

The kinetic energy of both C-1s and N-1s electrons increases as the photon energy increases, which leads to its projection on the c-VMI with a greater radius. Here we use the C-1s electron for the VLS-radius calibration, where a clear falling edge is visible, and its position can be determined by the greatest gradient, as shown in Fig. \ref{fig:VLS_radius}. Within the same run, the relation between the VLS CM and the C-1s radius can be approximated by a linear function.

\begin{figure}[h!]
    \centering
    \includegraphics[width=0.8\textwidth]{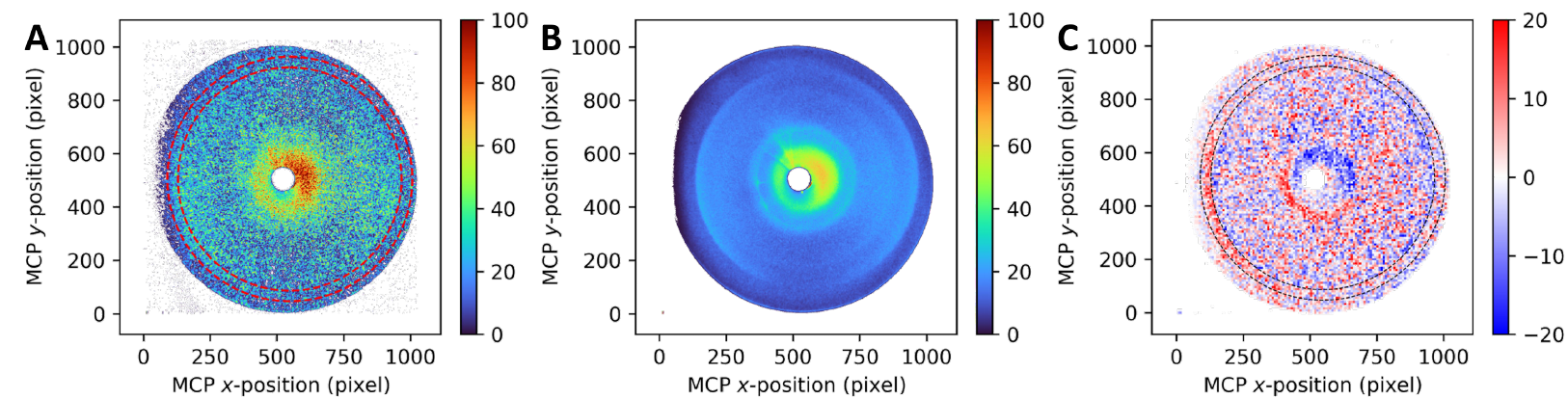}
    \caption{Principle of CM fitting. (A) A typical single-shot image at 416.4 eV, where the CM is taken within the area between the two dashed circles. (B) Average image of the X-ray-IR unoverlapped shots. (C) Differential image of (A) subtracted by (B). The image is binned and averaged by $8\times8$ pixels to reduce fluctuation. The dashed circles correspond to the ones in (A).}
    \label{fig:ring_cut}
\end{figure}

When the X-ray and the IR are spatially and temporally overlapped, the streaking effect causes the shift of the electron distribution on the c-VMI, which increases the signal beyond the radius determined by the shots in which X-ray and IR pulses were not temporally overlapping. Therefore, the CM of the ring region outside the C-1s radius, which is shot-to-shot calibrated according to the single-shot VLS spectrum, is sensitive to the streaking direction. Figure \ref{fig:ring_cut} demonstrates the principle with a relatively strongly streaked shot, where its differential plot subtracted by the averaged unstreaked image reveals the increase of the signal in the ring region. The ratio between the ring-region signal and the full-image signal can be used as a parameter to characterize the strength of streaking.

\begin{figure}[h!]
    \centering
    \includegraphics[width=0.9\textwidth]{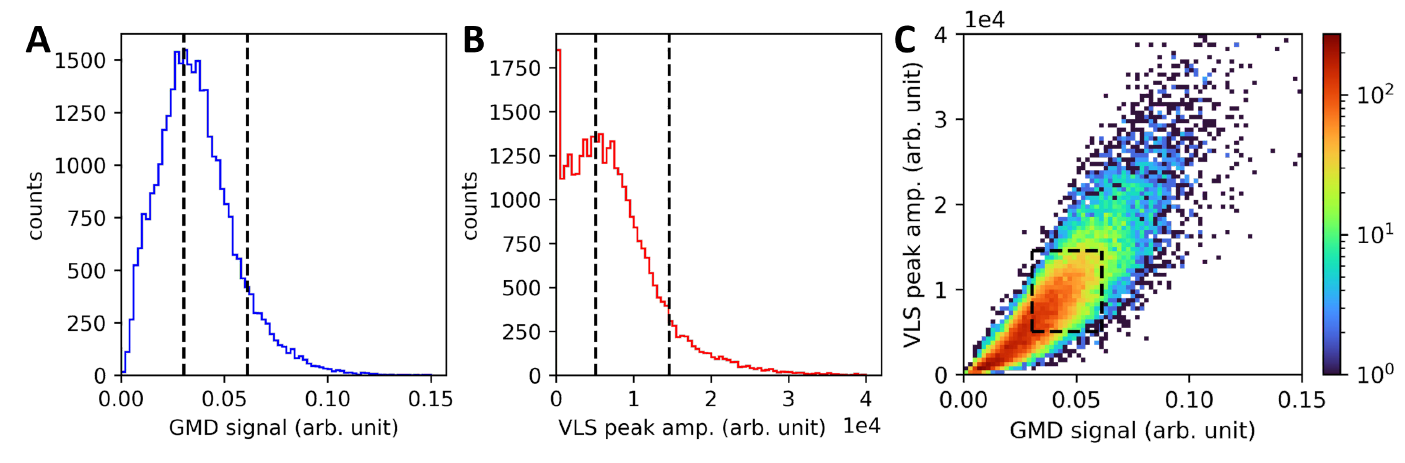}
    \caption{Shot selection based on GMD and VLS amplitude. (A) Distribution of the GMD signal. The vertical dashed lines correspond to 40\% and 90\% of the distribution, between which the shots are selected. (B) Distribution of the integrated VLS peak fitted as shown in Fig. \ref{fig:VLS_CM}. The vertical dashed lines correspond to 40\% and 90\% of the distribution, between which the shots are selected. (C) The two-dimensional histogram of the GMD signal and the VLS peak amplitude, where the selection criterion is indicated by the black dashed box.}
    \label{fig:GMD_VLS}
\end{figure}

\begin{figure}[h!]
    \centering
    \includegraphics[width=0.70\textwidth]{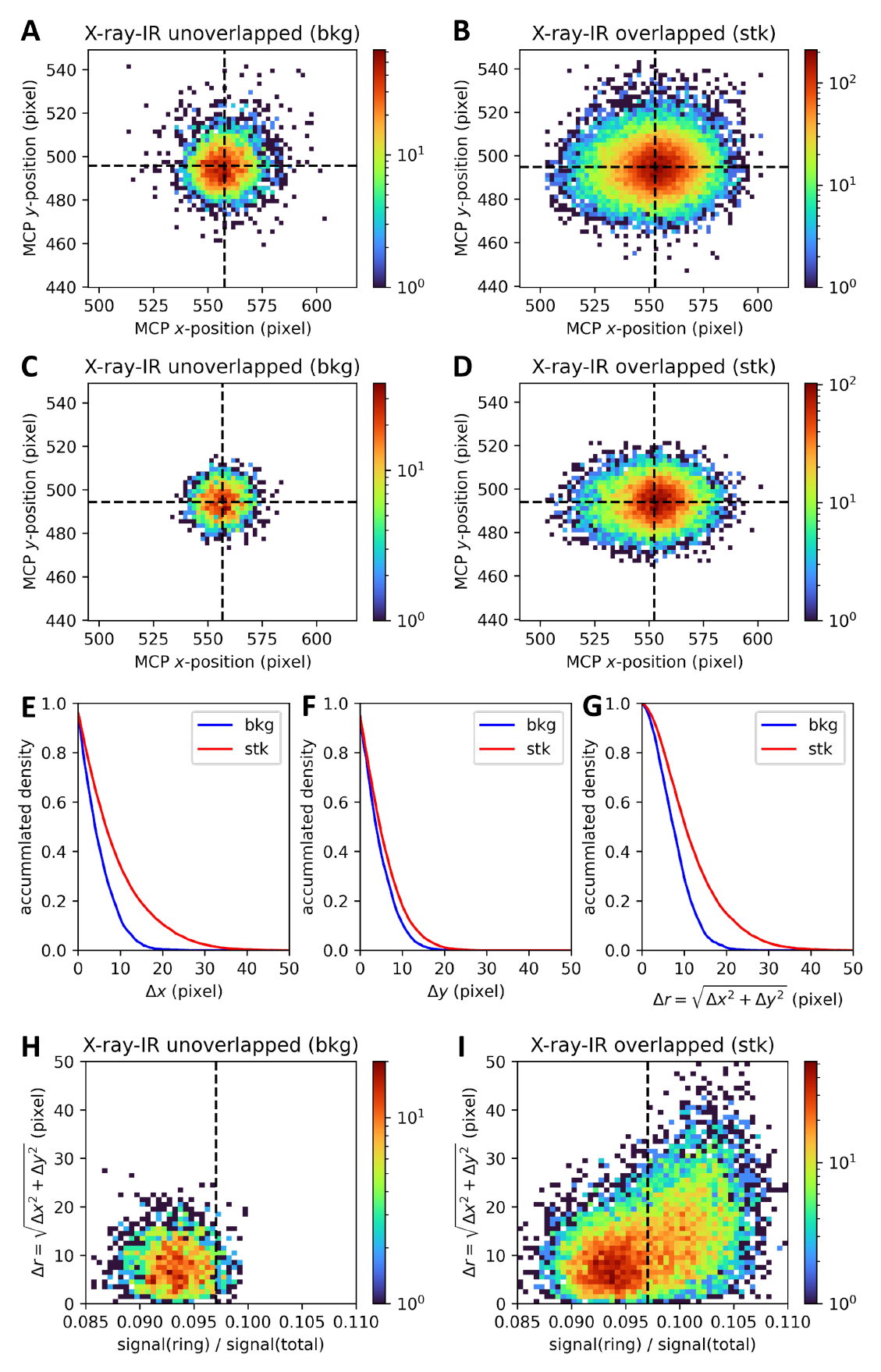}
    \caption{CM distributions of the X-ray-IR temporally unoverlapped (bkg) shots and the X-ray-IR temporally overlapped (stk) shots. (A,B) CM distribution of the whole run. (C,D) CM distribution of the run after GMD and VLS-amplitude selections. The black dashed lines in (A-D) indicate the average CM in $x$- or $y$-dimension. (E,F,G) Accumulated density of the CM shift (the vertical value corresponds to the proportion of shots with CM shifts smaller than indicated by the horizontal value) regarding $\Delta x$, $\Delta y$, and $\Delta r = \sqrt{{\Delta x}^2 + {\Delta y}^2}$. The GMD and VLS-amplitude selections are applied. (H,I) Ratio between the ring-region defined in Fig. \ref{fig:ring_cut} and the full image. The vertical dashed lines correspond to the 95\% level of the bkg shots, which is used as the threshold of streaking-strength selection.}
    \label{fig:xy_shift}
\end{figure}

The single-shot intensity can be characterized by both gas monitor detector (GMD) and VLS peak amplitude. The former method measures the ionization counts from a gas cell, and the latter method measures the signal intensity on the camera. The distributions are compared in Fig. \ref{fig:GMD_VLS}, where a clear correlation can be found. For each run, shots with GMD and VLS amplitude between 40\% and 90\% are selected, in order to exclude the images with too weak signal, where the noise level is relatively high, as well as images with too strong signal, which may suffer from the detector saturation. 
The CM distributions before and after the GMD and VLS amplitude selections are compared in Fig. \ref{fig:xy_shift}. Since the CM fitting is less stable with weaker signal, the distributions without selection shows some random scattering feature. The X-ray-IR unoverlapped (bkg) CM after selection has a circular distribution, which corresponds to the random error of the CM fitting, while the X-ray-IR overlapped (stk) CM after selection shows larger shifts in both $x$- and $y$-directions with an elliptical distribution, which corresponds to the dipole distribution of the C-1s electron.
A distinct separation between the weakly and strongly streaked shots is possible with the streaking-strength parameter defined as the ring-region signal divided by the full-image signal. As shown in (H) and (I) in Fig. \ref{fig:xy_shift}, the stk distribution contains one part that is similar to the bkg distribution which corresponds to the weakly streaked shots and another part that is above the threshold (95\% of bkg) and corresponds to a significant proportion of electrons moving into the ring region. This threshold is used as another criterion for shot selection.

\begin{figure}[h!]
    \centering
    \includegraphics[width=0.85\textwidth]{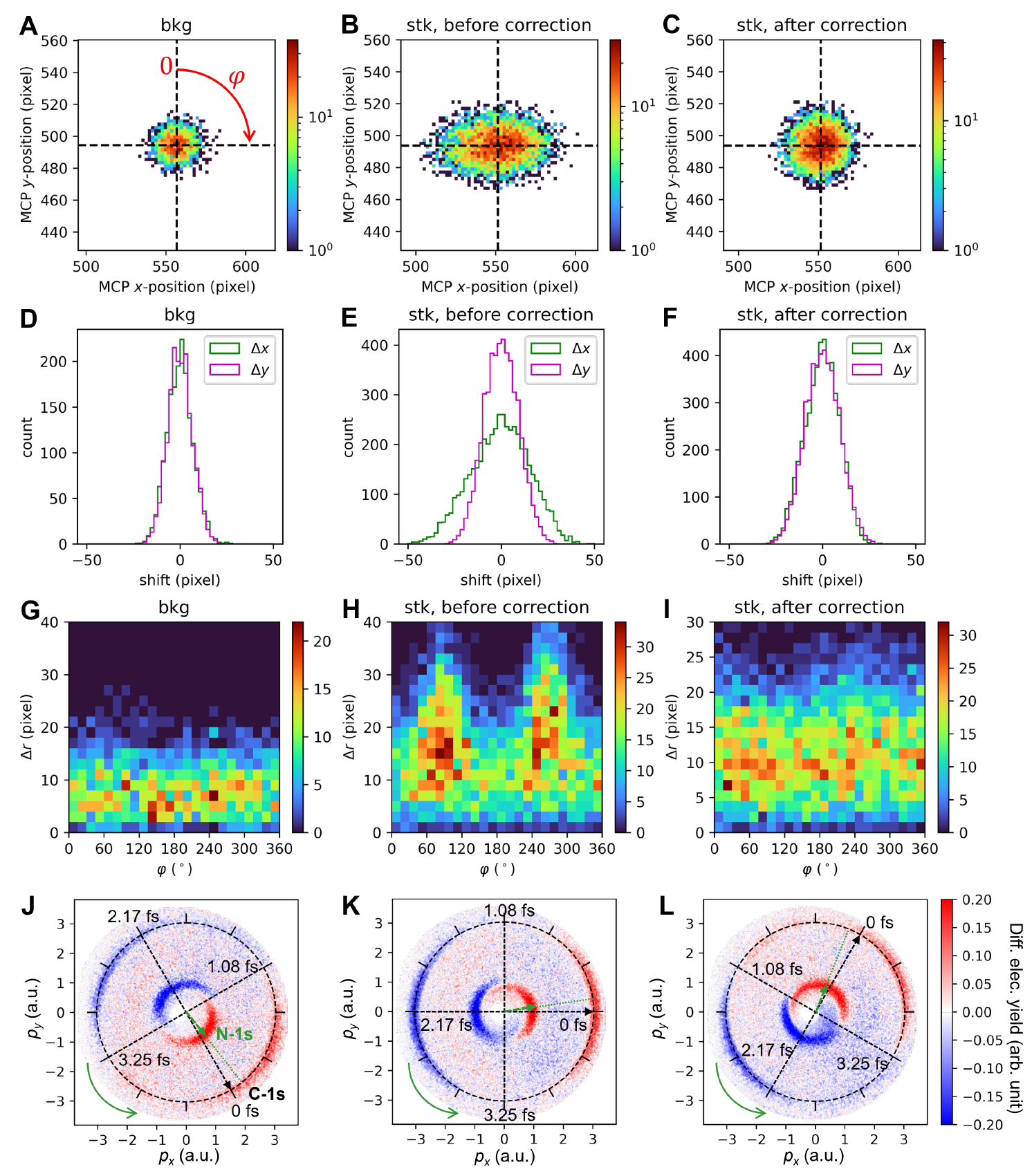}
    \caption{Dipole correction and streaking-direction determination. The GMD and VLS-amplitude selections and the streaking-strength selection are applied. (A) CM distribution of bkg. (B) CM distribution of stk before dipole correction. (C) CM distribution of stk after dipole distribution. (D,E,F) The $x$- and $y$-projections of the corresponding CM distributions. (G,H,I) CM distribution in the polar coordinates corresponding to (A-C), where the azimuth angle is defined as indicated in (A). 
    \textcolor{black}{
    (J,K,L) Difference images between streaking within a narrow range of angles and the average over all streaking angles, revealing the angular offset of the N-1s and C-1s electrons, which directly encodes their time delay, where the streaking directions have been corrected.}}
    \label{fig:xy_correction}
\end{figure}

Applying the selection rules mentioned above, one gets an elliptical distribution whose central part is also elliptical, as shown in Fig. \ref{fig:xy_correction}B, instead of the circular feature in Fig. \ref{fig:xy_shift}D that corresponds to the contribution of the weakly streaked shots. In order to retrieve the streaking direction, a factor of ${\rm std}(\Delta y)/{\rm std}(\Delta x)$ is multiplied by $\Delta x$, so that the elliptical distribution is stretched into the circular distribution, and the streaking direction can be inferred, as shown in Fig. \ref{fig:xy_correction}. The retrieved streaking direction is almost uniformly distributed after the correction, which agrees with the physical insight that the X-ray-IR time delay jitters randomly across the $2\pi$-phase. 
\textcolor{black}{
We note that by applying this correction, the retrieved streaking direction is not necessarily the intensity maximum of the C-1s ring, as compared in Fig. \ref{fig:xy_shift}J-L. This is indeed the case if one closely examines the scheme plot in Fig.~1B of the main text.}

In order to obtain the relative time delay between the C-1s and N-1s electrons, the phase analysis is performed with the shots grouped according to the determined streaking direction, as illustrated in Fig. \ref{fig:cart_polar_diff}. In Fig. \ref{fig:cart_polar_diff}A-C the subtle shift of the electron within a group can be visualized by the differential plot subtracting the average of all streaking directions, as compared in Fig. \ref{fig:cart_polar_diff}D-F, where one can find the C-1s electron shifted towards the streaking direction, causing a positive edge in the differential plot and a negative edge in the opposite direction. The N-1s electron overlaps with the above-threshold ionization (ATI) electron from the valence shell in the inner part of the image. However, since the ATI electron distribution is insensitive to the relative time delay between X-ray and IR, the differential plots exclusively manifests the shift of the N-1s photoelectron. 
We note that fixing the streaking direction and investigating the signal counterclockwise on the electron edge is equivalent to fixing a position on the electron edge and investigating the signal variation while rotating the streaking direction clockwise, while the latter has an advantage that the phase is unaffected by the inhomogeneity of the detector. For example, by fixing two boxes at the N-1s and C-1s edges, as illustrated in Fig. \ref{fig:cart_polar_diff}H, one obtains the oscillating signals as functions of the streaking angle $\vartheta$, as plotted in Fig. \ref{fig:cart_polar_diff}J and Fig. \ref{fig:cart_polar_diff}K for the summed data and individual data from each run, respectively, where the relative time delay between N-1s and C-1s can be extracted by fitting the phases of the cosine-like oscillations. The same procedure is repeatedly performed at various detector angles, and the average delay and the uncertainty can be determined. 

\begin{figure}[h!]
    \centering
    \includegraphics[width=\textwidth]{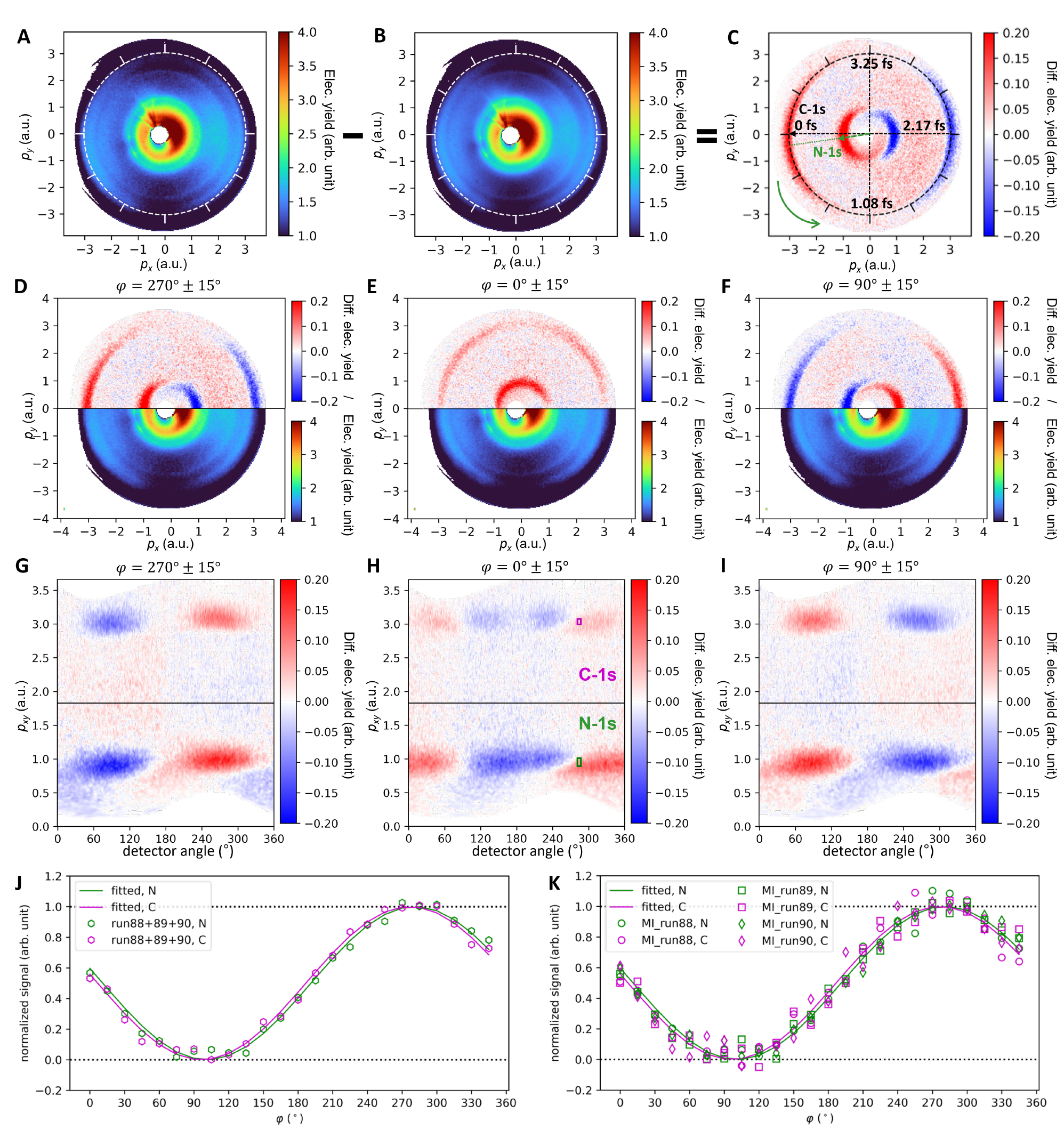}
    \caption{(Caption on the following page.)}
    \label{fig:cart_polar_diff}
\end{figure}
\begin{figure}[t]
    \contcaption{Illustration of the center-of-mass analysis method for the extraction of the photoionization delays. (A) Averaged image of stk shots with determined streaking angle between $270^\circ \pm 15^\circ$. (B) Averaged image of stk shots with all streaking angles. (C) Differential image between (A) and (B). (D,E,F) The averaged streaked images at different streaking angles $\vartheta$ sorted by post-selection, where the lower panels correspond to the original images, whilst the upper panels are subtracted by the averaged image over all streaking angles. The images are normalized to arbitrary unit, while the normalization is consistent between the original and the differential images. (G,H,I) Differential images converted into the polar coordinates, with the upper and lower panels corresponding to the C-edge-corrected and N-edge-corrected images, respectively. (J) Signal variation as a function of the streaking angle $\vartheta$ at a fixed area on the detector indicated by the green box for N-1s and the magenta box for C-1s in (H). The signals are normalized and subtracted by the mean value. The curves correspond to the cosine fitting. (K) Same as (J) but the markers correspond to three individual runs with independent data processing, while the three runs are summed in (A-J).}
\end{figure}


\clearpage
\subsubsection{Comparison of the two methods}

\begin{figure}[h!]
    \centering
    \includegraphics[width=\textwidth]{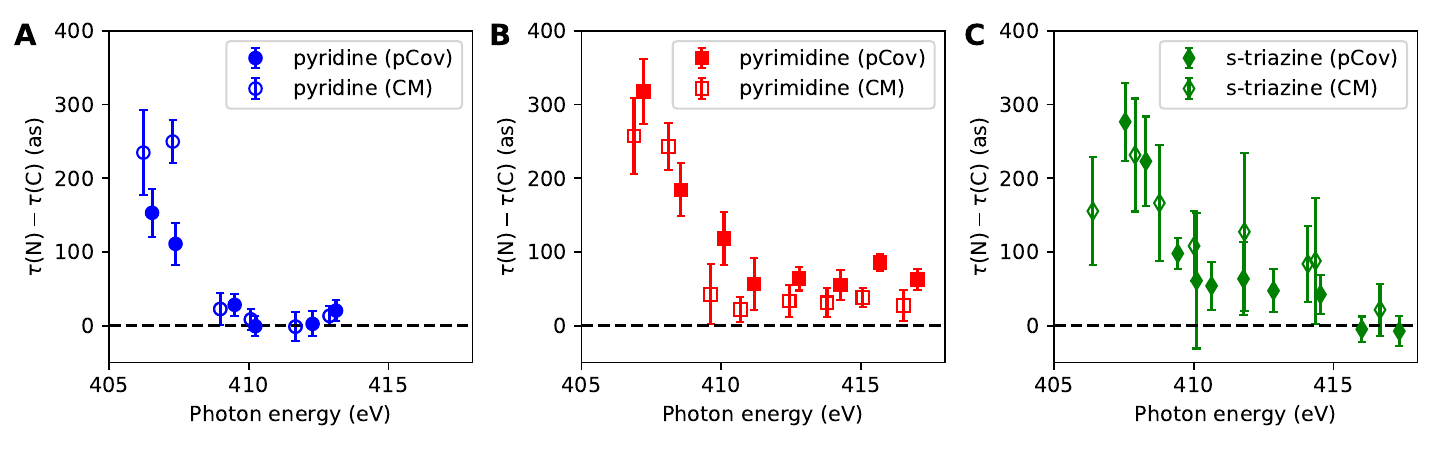}
    \caption{Comparison of the extracted time delays with the partial covariance analysis (pCov) and the center-of-mass method (CM). (A) Pyridine (B) Pyrimidine (C) $s$-triazine.}
    \label{fig:CM_vs_corr}
\end{figure}

We are now in the position to compare these two independent methods of extracting photoionization delays -- the partial covariance and the center-of-mass method. Figure~\ref{fig:CM_vs_corr} directly compares the photoionization delays by showing the results of the partial-covariance method with filled symbols and those of the center-of-mass method with empty symbols. The results from the two methods agree very well, practically always within the respective error bars. Since the error bars of the partial-covariance method are smaller, particularly in the case of $s$-traizine, we have chosen to show those data in the main text.

\clearpage

\subsection{Determination of Photoelectron Asymmetry Parameters}\label{sec_s:beta_2_from_cov_analysis}

We perform a covariance analysis to extract the photoelectron asymmetry parameter $\beta_2$ as a function of X-ray photon energy for pyridine, pyrimidine, and \textit{s}-triazine. 
The data analysis of $\beta_2$ is performed on unstreaked c-VMI images. 

\begin{figure}[h!]
    \centering
    \includegraphics[width=1\textwidth]{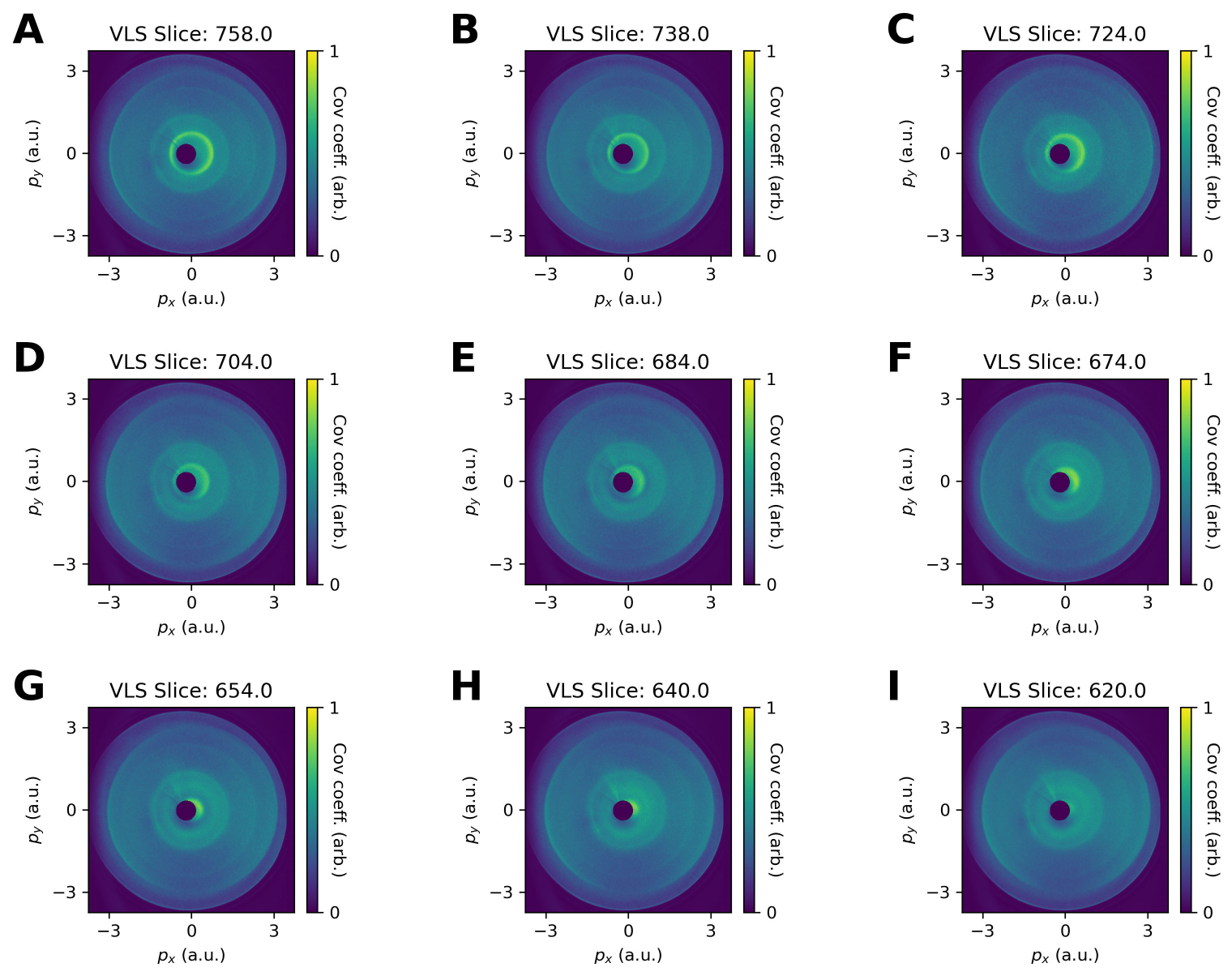}
    \caption{Covariance maps calculated between c-VMI pixels and VLS slices. 
    Panels \textbf{(A)-(I)} represent covariance maps calculated on pyrimidine data. 
    The title of each panel labels the center of the VLS slice (i.e. uncalibrated X-ray spectrum) used in the calculation of the covariance map in this panel.}
    \label{fig:s_Cov_VMI_VLS_pyrimidine}
\end{figure}

In the first step, we calculate the simple covariance $\operatorname{Cov}\left[ M(p_x, p_y), I(\omega)\right]$ between the c-VMI pixel value $M(p_x, p_y)$ and the VLS pixel $I(\omega)$. 
Given a X-ray photon energy $\hbar\omega$, the 2-D covariance map $\operatorname{Cov}\left[ M(p_x, p_y), I(\omega)\right]$ as a function of $(p_x, p_y)$ contains the X-ray-photon-energy-dependent features of C-1s and N-1s photoelectrons. 
The signal from ATI is removed in these 2-D covariance maps, because ATI electrons were generated by the streaking laser and not correlated with X-ray photon energy. 
The white noise in c-VMI images is removed in the simple covariance maps. 
In the practical data analysis procedure, we calculate the simple covariance between the c-VMI pixel value $M(p_x, p_y)$ and VLS slices of 10 VLS pixels each to improve the signal-to-noise ratio. 
The covariance maps calculated on pyrimidine data are shown in Fig. \ref{fig:s_Cov_VMI_VLS_pyrimidine}. 
From panels A to I in Fig.~\ref{fig:s_Cov_VMI_VLS_pyrimidine}, radii of both photoemission features gradually decrease (i.e. lower kinetic energy) when using the VLS slice with smaller value (i.e. lower X-ray photon energy) to calculate 2-D covariance maps.

These 2-D covariance maps still contain some unwanted features that are not parts of C-1s and N-1s photoemission features and are detrimental to our analysis of $\beta_2$. 
For example, we can observe a uniform background in each panel of Fig.~\ref{fig:s_Cov_VMI_VLS_pyrimidine}. 
This uniform background comes from carbon and nitrogen K-edge Auger-Meitner electron yield. 
The simple covariance map $\operatorname{Cov}\left[ M(p_x, p_y), I(\omega)\right]$ still contains signals from Auger-Meitner electron yield because Auger-Meitner electron yield depends on X-ray pulse energies. 
For the same reason, these simple covariance maps $\operatorname{Cov}\left[ M(p_x, p_y), I(\omega)\right]$ also contains X-ray artifacts, which appear as sharp rings with the same shapes in every panel of Fig.~\ref{fig:s_Cov_VMI_VLS_pyrimidine}. 

To better extract $\beta_2$ from these simple covariance maps, we utilize the fact that both the Auger-Meitner electron yield and the X-ray artifacts do not sensitively depend on the X-ray photon energy. 
To this end, we calculate the differential map between 2 covariance maps calculated with adjacent VLS slices after appropriate normalization. 
By calculating such a differential map, we can remove contributions from the Auger-Meitner electron yield and the X-ray artifacts. 
The resultant differential map only contains the differential traces of C-1s and N-1s photoemission features. 
By applying an inverse Abel transform to the differential map, we can retrieve two 1-D traces, namely $\Delta I_0(p_r)$ and $\Delta I_2(p_r)$, from the differential map. 
Since $\Delta I_0(p_r)$ and $\Delta I_2(p_r)$ are retrieved from a differential map, they contain both positive and negative features. 
To avoid the singularity in calculating $\int \Delta I_2(p_r)/\int \Delta I_0(p_r)$, we fit the value of $\beta_2$ by minimizing the following objective,
\begin{equation}
    f(\beta_2) = \int_{\mathrm{ROI}} \mathrm{d}p_r~\left(\Delta I_2(p_r) - \beta_2 \Delta I_0(p_r)\right)^2,\label{eq:s_beta2_fitting}
\end{equation}
which is equivalent to a linear regression between $\Delta I_0(p_r)$ and $\Delta I_2(p_r)$ with the bias term set to $0$.

\begin{figure}[h!]
    \centering
    \includegraphics[width=0.8\textwidth]{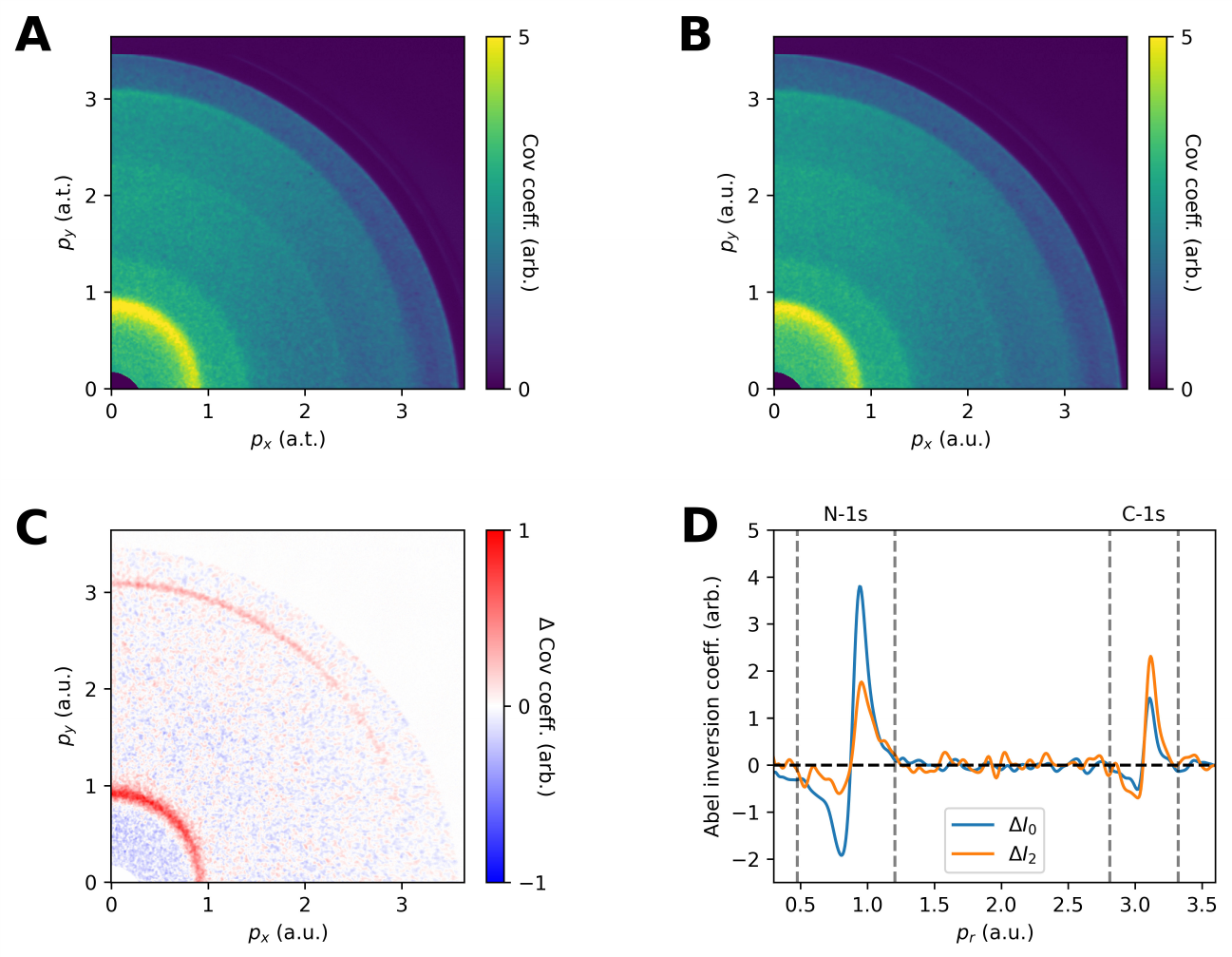}
    \caption{Use of the relative difference in covariance maps to determine $\beta_2$. 
    Panel~{(A)} shows the covariance map calculated with the VLS slice at 812. 
    Panel~{(B)} shows the covariance map calculated with the VLS slice at 802. 
    Panel~{(C)} shows the differential map between panels (A) and (B). 
    Panel~{(D)} shows the inverse Abel transform of the quadrant shown in panel (C).
    }
    \label{fig:s_Demonstrating_Beta2_Fitting}
\end{figure}

When the X-ray central photon energy is around $450~$eV, all four quadrants of c-VMI images are complete and can be used in the photoelectron asymmetry parameter analysis. 
We calculate the values of $\beta_2$ on all four quadrants separately. 
The average of $\beta_2$ calculated on four quadrants is used as the mean value, and the standard deviation of $\beta_2$ calculated on four quadrants is used as the error. 
When the X-ray central photon energy is close to the nitrogen K-edge, only the top right quadrant (orientation shown in Fig.~\ref{fig:s_Cov_VMI_VLS_pyrimidine}) is used in the photoelectron asymmetry parameter analysis. 
The errorbars of N-1s $\beta_2$ from $407~$eV to $418~$eV are taken from the numerical uncertainty of the fitting in Eq.~\ref{eq:s_beta2_fitting}.



\clearpage

\section{Theoretical Methodology}

\subsection{Core-level photoionization calculations}
\begin{figure}[h]
    \centering
    \includegraphics[width=0.9\textwidth]{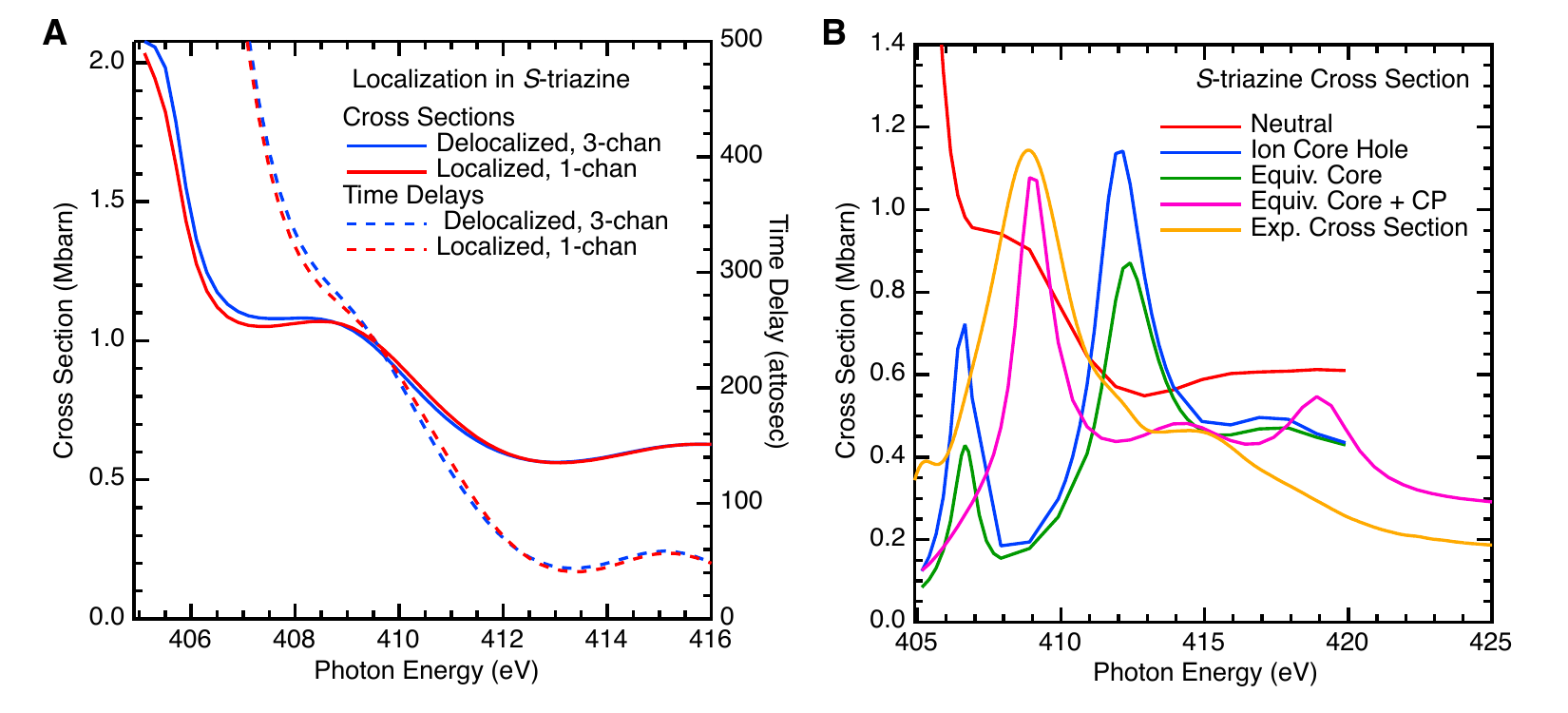}
    \caption{Approximations used in the computation of the core-level photoionization calculations. (A) Difference between a three-channel calculation of the N$(1{\rm s})^{-1}$  photionization using  a three-channel calculation for ionizing the three nearly degenerate N$(1{\rm s})^{-1}$ hole states of $s$-triazine using delocalized Hartree-Fock orbitals of the neutral molecule (blue lines) and a one-channel photoionization calculation using a core orbital localized on one N-atom. (B) Total photoionization cross section of $s$-triazine computed with different choices of the orbitals used: localized neutral orbitals (red), orbitals from an open-shell Hartree-Fock calculation on the ion with a vacancy on one of the N-atoms (blue), orbitals obtained from a equivalent-core calculation (green),  orbitals obtained from a equivalent-core calculation with an added correlation-polarization potential (pink), and experimental absorption data\cite{vall08a} (tan).}
       \label{fig:TheorySteps} 
\end{figure}

The N $(1{\rm s})^{-1}$  photoionization of the three azabenzenes considered here were studied using two methods, the basis-set complex Kohn method \cite{Orel1990a} and \texttt{ePolyScat} \cite{Gianturco1994a,Natalense1999a}.  The \texttt{ePolyScat} calculations used a static-exchange potential that was modified by the addition of an approximate local potential that represented the correlation and polarization interactions felt by the emitted photoelectron due to its interaction with the residual ion of the molecule \cite{Perdew1981}. The one-electron basis set used to obtain the occupied molecular orbitals was a correlation consistent valence triple zeta \cite{Dunning1989a} and the geometry used minimized the Hartree-Fock energy. For the systems with equivalent N-atoms, a one-channel localized hole calculations from \texttt{ePolyScat} were found to be equivalent to calculations where the ionization from the different the N $(1{\rm s})^{-1}$ holes were coupled together in a multichannel calculation using the complex-Kohn method \cite{Marante2020a} as seen in Fig.~\ref{fig:TheorySteps}A for the ionization of $s$-triazine.  The orientation-averaged one-photon time delays are also seen to be nearly the same for these two calculations in Fig.~\ref{fig:TheorySteps}A. In all calculations the initial state and in the ionized state were constructed using the same set of orbitals.  The positions of the resonances in the photoionization cross sections were sensitive to the orbitals used in the calculation.  In Fig.~\ref{fig:TheorySteps}B we compare the cross section for $s$-triazine  using the orbitals from the neutral initial state, orbitals from a calculation where the open shell ion state with an N $(1{\rm s})^{-1}$ vacancy is localized on one of the N-atoms, and orbitals from an equivalent-core calculation \cite{Hoshino2018a}. Additionally, we have plotted the cross section where we have added a correlation-polarization (CP) potential to the equivalent-core calculation and scaled the CP potential by a factor of 2.0. This shifted the position of the largest resonance to an energy that agrees with the peak in the experimental absorption cross section \cite{vall08a}.  In Fig.~\ref{fig:s2} we can see that the same CP potential shifts the peak seen in the N $(1{\rm s})^{-1}$ ionization in all three molecules considered here to agree with experimental absorption cross sections \cite{vall08a}.

\subsection{Characterization of the resonances above the N K-edge of the azabenzenes.}
\begin{figure}[h]
    \centering
    \includegraphics[width=\textwidth]{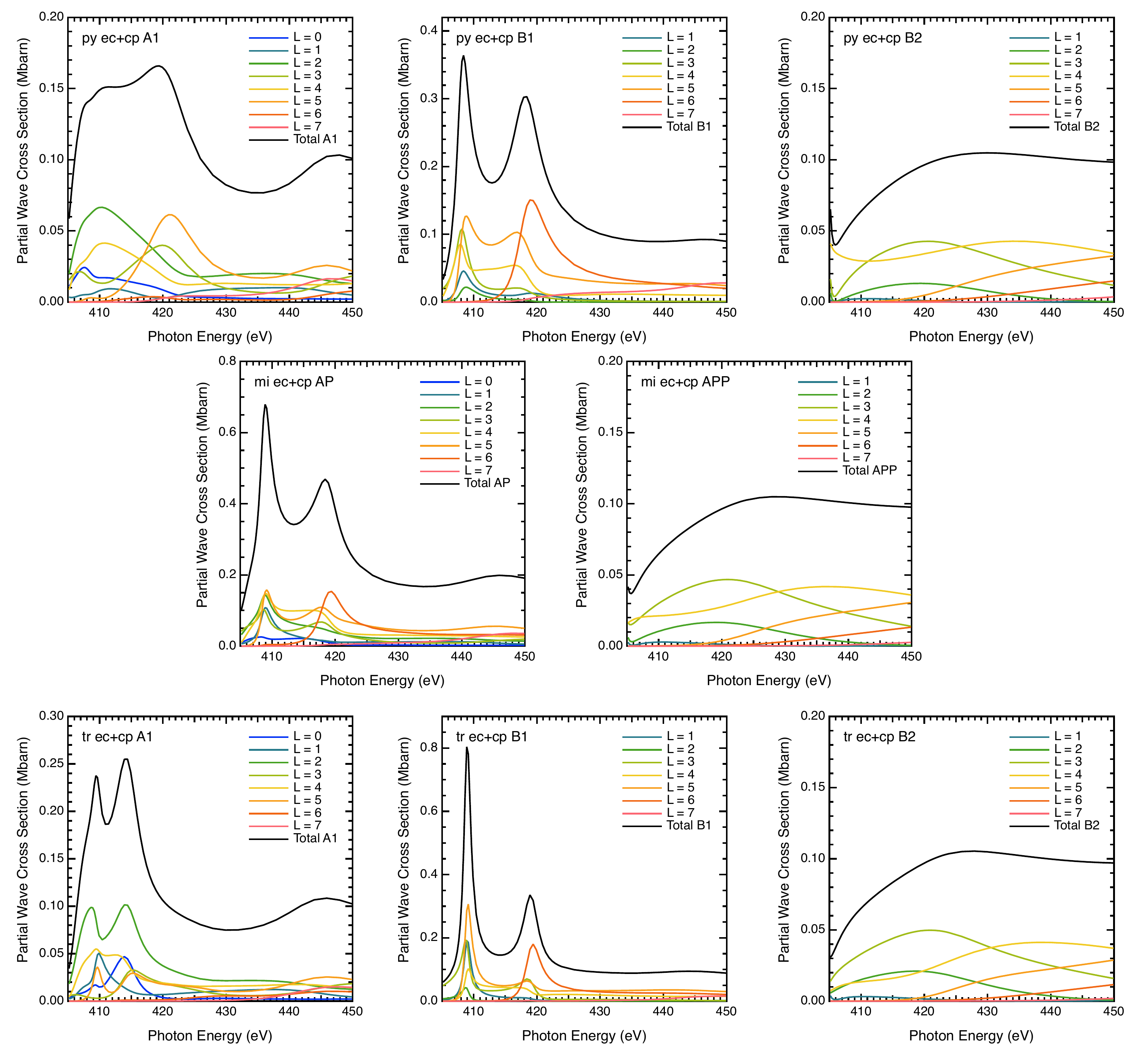}
    \caption{Decomposition of the total photoionization cross section above the nitrogen K-edge, calculated with the CP method, into the partial-wave cross sections of each dipole-accessible irreducible representation (irrep) in the point groups of the single-site ionized azabenzene molecules. (top) pyridine with C$_{\rm 2v}$ irreps A$_1$, B$_1$ and B$_2$, (middle) pyrimidine with C$_{\rm s}$ irreps A$^{\prime}$ and A$^{\prime\prime}$, (bottom) $s$-triazine with C$_{\rm 2v}$ irreps A$_1$, B$_1$ and B$_2$.}
       \label{fig:s9} 
\end{figure}

In the photoionization calculation, we found evidence of several shape resonances in each molecule.  The partial cross sections for each irreducable representation and each of the three molecules considered here are presented in Fig.~\ref{fig:s9}.  In the orientation of the molecules used here, for pyridine and $s$-triazine in the equivalent-core approximation, the molecules have C$_{\rm 2v}$ symmetry so that the orbitals with A$_1$ and B$_1$ symmetry are $\sigma$ type orbitals in the plane of the molecule and the B$_2$ orbitals are $\pi$ orbitals with a node in the plane of the molecules. For pyrimidine in the equivalent core approximation the system has C$_s$ symmetry so the A$'$ orbitals are $\sigma$ orbitals and the A$''$ orbitals are $\pi$ orbitals.  Immediately we see that there are no $\pi$ resonances  in the continuum.  In pyridine, there are two noticeable resonances below 430~eV with the lower one, around 410 eV, having a peak cross section in the $L=2$ partial wave, where the upper resonance near 420 eV has a peak cross section in the $L=5$ partial wave.  In B$_1$ symmetry, the lower resonance near 408 eV has significant contributions from $L=5$ whereas the higher resonance has a significant contribution from $L=6$.  In the cross sections for pyrimidine, there are again resonances around 409 eV and 420 eV, but due to the lower symmetry in this system it is difficult to distinguish the different resonances in the partial-wave analysis. In $s$-triazine, the A$_1$ symmetry cross section again has two clear resonances, however in this case the upper resonance has a lower energy, at about 415 eV, compared to the upper resonance in the B$_1$ symmetry resonances. In the B$_1$ partial cross sections, the lower resonance has the largest contribution from the $L=5$ partial waves and the upper resonance has the largest contribution in the $L=6$ partial wave.

To analyze these shape resonances in more detail, we performed calculations to obtain the position and wave functions of the resonant states directly. In these calculations we used a local approximation to electron-molecule interaction potential where the exchange potential was replaced by the free-electron-gas exchange (FEGE) potential, using a fixed scattering energy of 15 eV in the computation of the exchange potential \cite{Lucchese1996a,Natalense1999a}. Using the static potential, plus the FEGE exchange potential, plus the CP potential, truncated at a fixed $r$ of $8.5$~\AA, which is the distance from the center of the molecule to where the magnitude of the orbitals are less than $10^{-6}$ in atomic units. We then solved the scattering equations at complex energies with negative imaginary parts on the unphysical sheet, i.e.  $\textrm{Im}\,(k) < 0$, where k is the complex-valued momentum of the photoelectron, to locate the poles of the $S$-matrix. The wave functions at the poles will then have only outgoing flux, unlike wave functions at real energy which have both incoming and outgoing flux. These states which are Siegert states with a cutoff potential, also known as Kapur-Peierls eigenstates \cite{Lane1966,Meyer1982,Kukulin1989}, are then used to visualize resonances including the location of the resonance on the molecule and their angular momentum composition which can indicate the mechanism for trapping the metastable resonance state. These wavefunctions have the asymptotic form $\exp{(ikR)} = \exp{\left(ik_rR\right)}\exp{\left(k_iR\right)}$, where $k_r=\textrm{Re}(k)$, $k_i=-\textrm{Im}\,(k)>0$. This explains their divergence at large $R$ seen in Fig. 5 of the main text and in Fig. \ref{fig:s5} below.

\begin{figure}[h]
    \centering
    \includegraphics[width=1.0\textwidth]{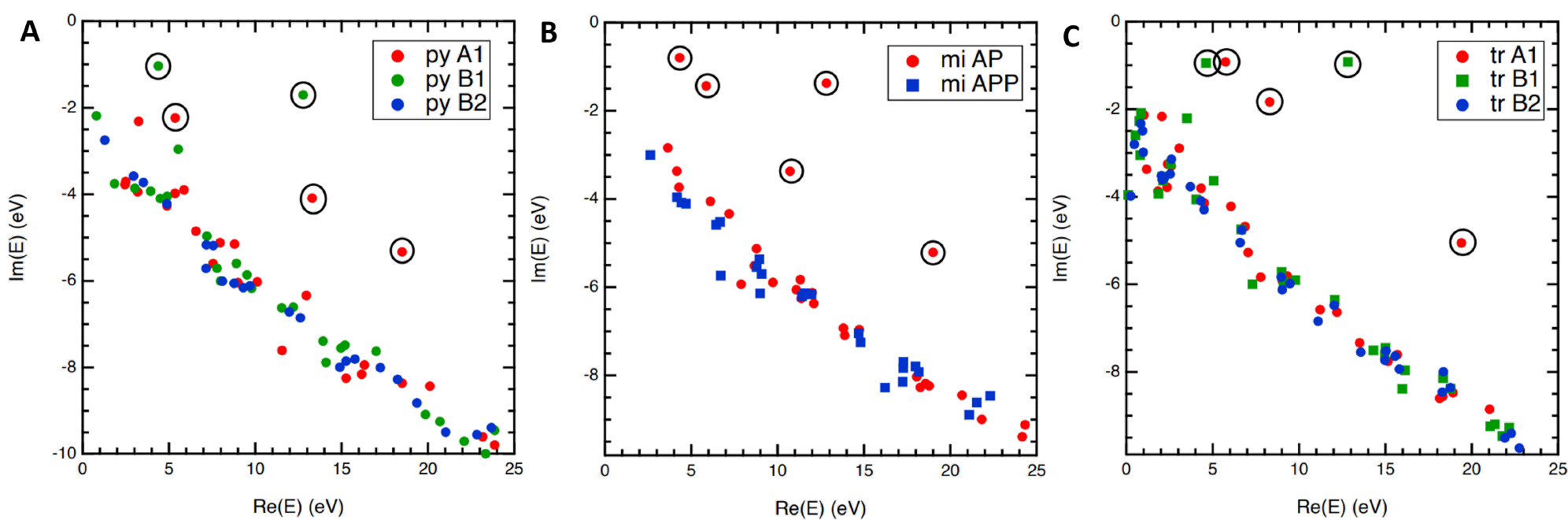}
    \caption{$S$-matrix pole locations of (A) pyridine, (B) pyrimidine and (C) $s$-triazine in the vicinity of the nitrogen K-edge. The resonance positions are identified with circles. The legends indicate the molecule and the dipole-accessible irreps in the point group of each single-site-ionized molecule (C$_{\rm 2v}$ for pyridine and $s$-triazine with irreps A$_1$, B$_1$ and B$_2$, C$_{\rm s}$ for pyrimidine with irreps A$^{\prime}$ and A$^{\prime\prime}$).}
    \label{fig:s3}
\end{figure}

The location of the poles found using the Kapur-Peierls method for the three molecules considered are shown in Fig.~\ref{fig:s3}. Each plot shows many poles along a line, which correspond to non-resonant scattering \cite{Meyer1982} and some poles located closer to the real energy axis away from this line which correspond to resonances.  In Fig.~\ref{fig:s3} we have circled five poles for each molecule which are well separated from the non-resonant scattering poles.  Although the treatment of exchange is different in the pole-search results shown in Fig.~\ref{fig:s3} and calculated cross sections in Fig.~\ref{fig:s9}, there is a very good correspondence between the resonance features in cross sections given in Fig.~\ref{fig:s9} and the location of the resonance poles identified in Fig.~\ref{fig:s3}.

\begin{figure}
    \centering
    \includegraphics[width=0.70\textwidth]{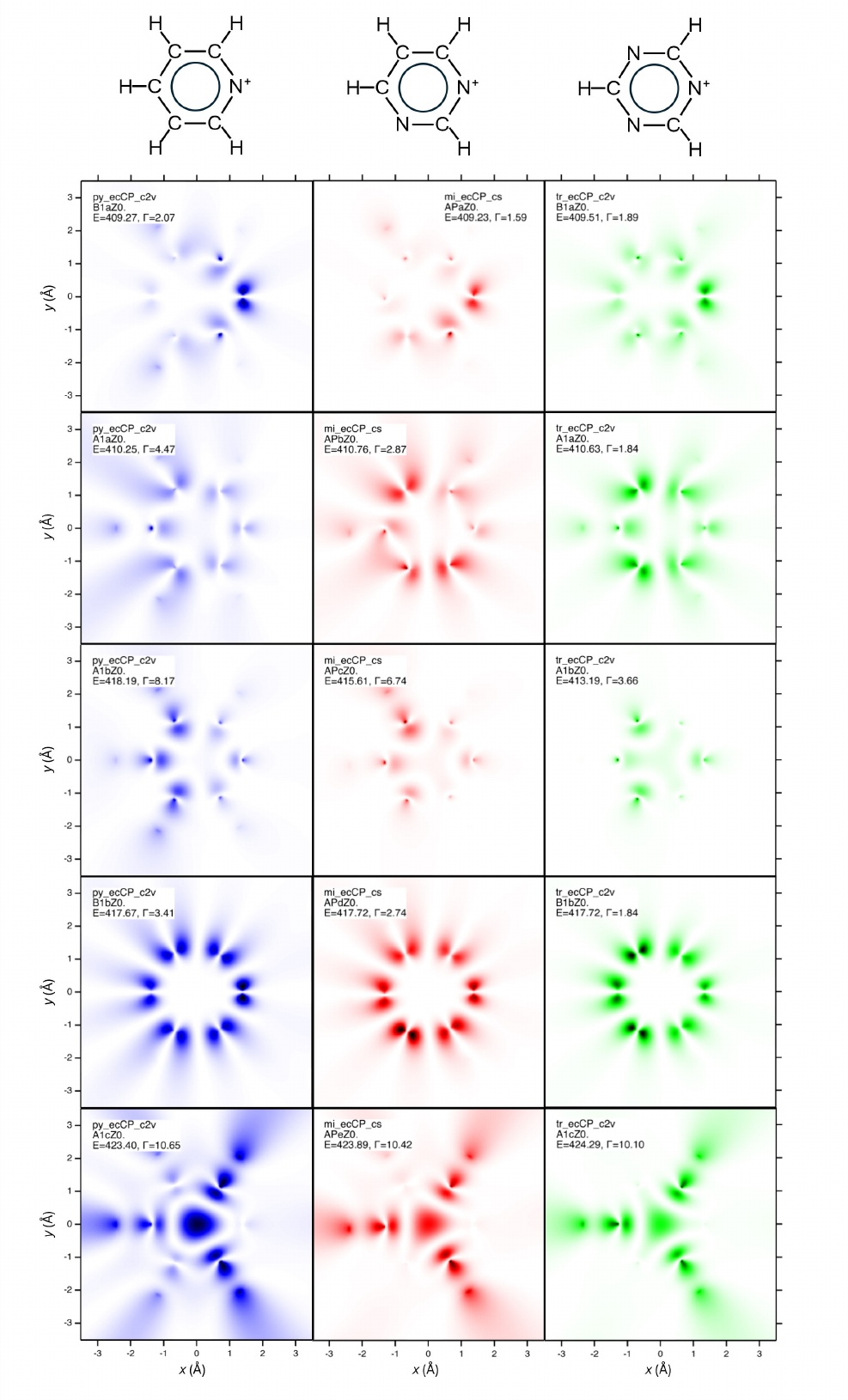}
    \caption{Resonant wave functions (Kapur-Peierls states) of all five resonances of each azabenzene molecule identified in Fig. \ref{fig:s3}. The left, middle, and right columns correspond to pyridine (blue), pyrimidine (red), and $s$-triazine (green), respectively, with their molecular scaffolds displayed on the top of the corresponding columns.}
    \label{fig:s4}
\end{figure}

For each pole shown in Fig.~\ref{fig:s3}, in Fig.~\ref{fig:s4} we plot the absolute square of the resonance wave function in the plane of the molecule.  For each molecule, the orientation of the molecule has the N-atom which is ionized on the right-hand side of the plots.  In each case, the resonance wave function only contains outgoing waves, so that if one were to extend these plots to larger distances from the molecule, the wave functions would grow exponentially. The plots then indicate the part of the wave functions where the electron is trapped by angular momentum barriers around the region of the molecule.  The leftmost column contains the wave function for the strong resonances which occur near threshold.  It is apparent that the strength of these resonances is due to the fact that the resonant state has a large amplitude at the N-atom which was ionized.  In contrast, resonances in the right-hand column have very little amplitude near where the ionized N-atom is located leading to the absence of this resonance in the computed cross sections shown in Fig.~\ref{fig:s9}.  The next-to-the-last column on the right-hand side gives the resonant wave functions for the well isolated resonances occurring at a photon energy of 417 eV which corresponds to a photoelectron kinetic energy of 12 eV.  These resonances clearly have a nodal structure corresponding to $L=6$ in agreement to the partial wave contributions to the cross sections shown in Fig.~\ref{fig:s9}.


As an alternative to the qualitative analysis of the resonance wave functions shown in Fig.~\ref{fig:s4}, one can also plot the absolute square amplitudes of the partial-wave radial functions, as given in Fig.~\ref{fig:s5}.  In this figure, for a given value of $L$, the square amplitudes of all the radial wave functions are given. The most important radial wave functions are then highlighted using color.  The resonances in the left column clearly show resonance wave functions which are primarily trapped in an $L=5$ angular momentum barrier from which the electron tunnels leading to a large amplitude in the $L=5$ photoionization amplitude.  In a similar fashion, the resonances shown in the next-to-last column on the right show $L=6$ resonances which then tunnel through the corresponding angular momentum barrier leading to strong $L=6$ photoionization amplitudes as seen in Fig.~\ref{fig:s9}.


\begin{figure}[h]
    \centering
    \includegraphics[width=0.7\textwidth]{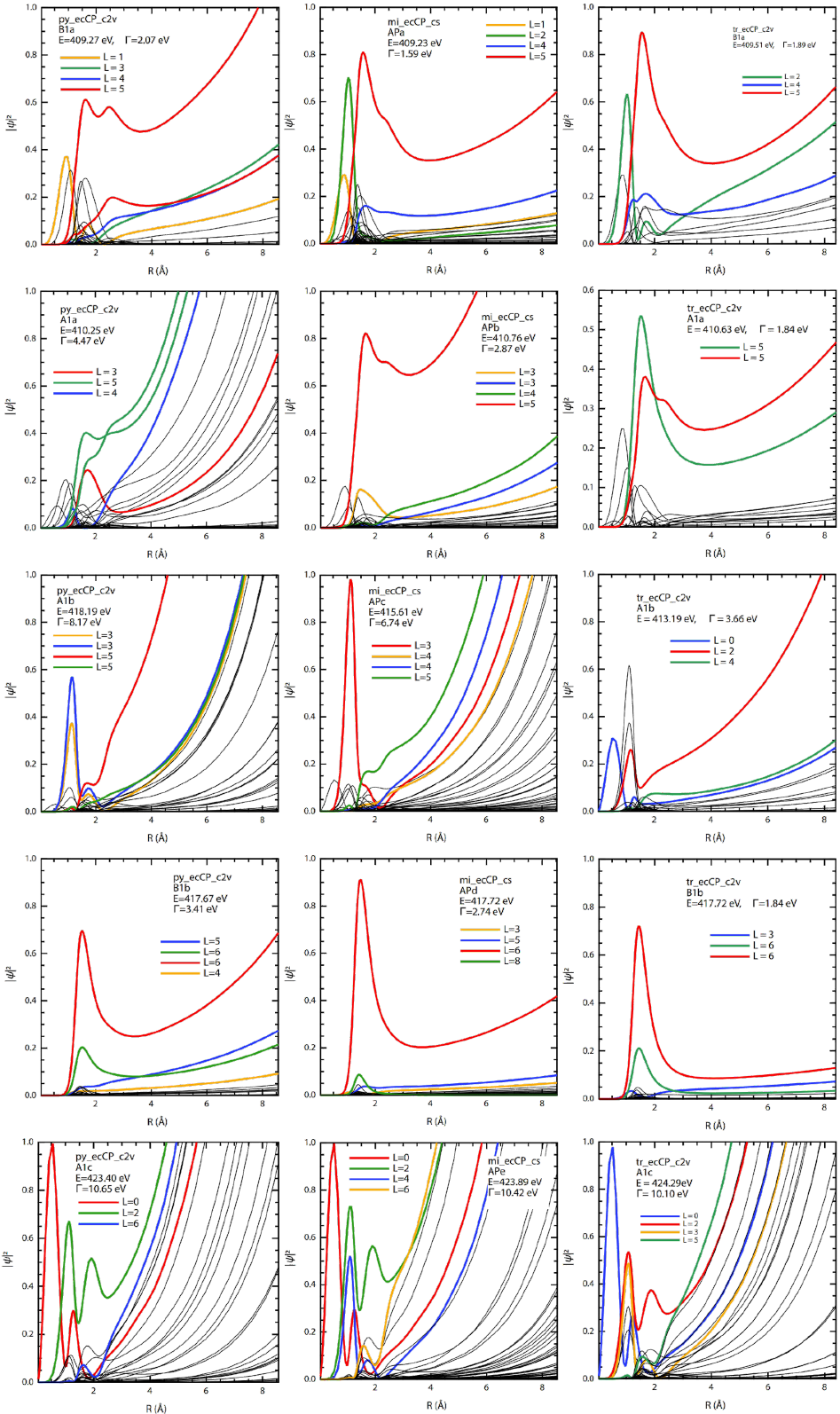}
    \caption{Overview of the radial parts of the partial-wave resonant wave functions (Kapur-Peierls states) of all 5 resonances identified in each of the 3 molecules in Fig. \ref{fig:s3}. The text in each panel identifies the molecule (left: pyridine, middle: pyrimidine, right: $s$-triazine), the irreducible representation, the energy location of the resonance ($E$) and the width of the resonance ($\Gamma$), which can be referenced to Figs.~\ref{fig:s3} and \ref{fig:s4}.}
    \label{fig:s5}
\end{figure}

\clearpage

\section{Additional comparisons between experiment and theory}

\subsection{Comparison of experimental and theoretical cross sections and asymmetry parameters}
\begin{figure}[h]
    \centering
    \includegraphics[width=0.6\textwidth]{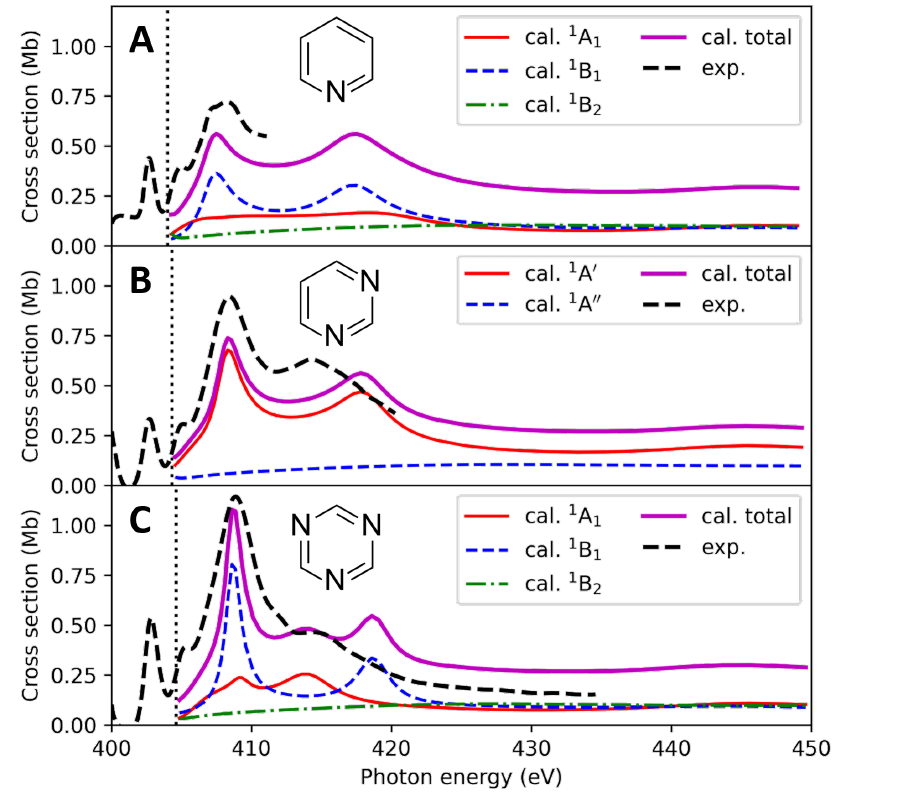}
    \caption{Comparison of experimental (Ref.~\cite{vall08a}) and theoretical (this work) photoionization cross sections at the nitrogen K-edge, where the ionization potential is indicated by the vertical dotted line. The full magenta line is the total photoionization cross section, whereas the other colored lines are the symmetry-resolved partial cross sections. The calculations shown in this figure have been shifted to lower energies by 0.9~eV, 0.6~eV and 0.3~eV in the cases of pyridine, pyrimidine and $s$-triazine, respectively, as mentioned in the text.}
    \label{fig:s2}
\end{figure}

This section contains additional comparisons between our experimental measurements and calculations. First, we discuss the comparison of the experimental X-ray absorption cross sections in the vicinity of the nitrogen K-edge from the literature with our calculations. Such a comparison in the case of pyrimidine is shown in Fig.~3 of the main text. Figure~\ref{fig:s2} shows such comparisons for all three molecules. The experimental cross sections have been taken from Ref. \cite{vall08a}. 
In the case of pyridine (Fig.~\ref{fig:s2}A), the calculations reveal two clear local maxima above the ionization threshold, whereby the location of the lower one quantitatively agrees with the experiment. 
In the case of pyrimidine (Fig.~\ref{fig:s2}B), both experiment and theory show two clear above-threshold local maxima, the lower of which is in quantitative agreement. The calculated position of the upper maximum is overestimated by $\sim$3.5~eV, as discussed and indicated in the main text (Fig.~3). 
The experimental cross section of $s$-triazine (Fig.~\ref{fig:s2}C) reveals two local maxima, whereas the calculations display three. Based on the shift of $\sim$3.5~eV encountered in the case of pyrimidine and the similarity of the two molecules, we propose that the higher-lying observed maximum is assigned to the highest-lying of the three calculated maxima. 

The comparison shown in Fig.~\ref{fig:s2} was also used to determine accurate values of the N-1s ionization energies, which were used for consistency in the remainder of this work. Specifically, the calculated photoionization cross sections were shifted to best match the measured X-ray absorption spectra and these shifts were used to determine the N-1s ionization energies. In this way, we obtained N-1s binding energies of 404.0~eV, 404.3~eV and 404.6~eV for pyridine, pyrimidine and $s$-triazine, respectively. These values are in reasonable agreement with the literature values \cite{vall08a}. The overall shifts of the CP (ECO) calculations, that were carried out without initially taking into account the N-1s chemical shifts, thus amounted to 0.9~eV (4.6~eV), 0.6~eV (4.3~eV) and 0.3~eV (4.0~eV). These shifts were applied in all comparisons of experiment and theory shown in this work.

\begin{figure}[h]
    \centering
    \includegraphics[width=\textwidth]{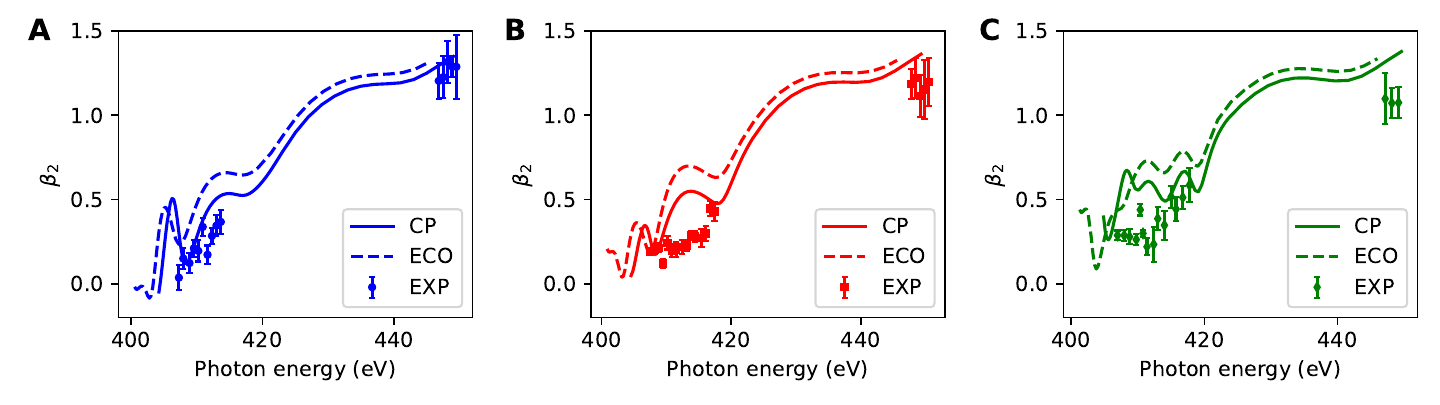}
    \caption{Asymmetry parameters ($\beta_2$) of nitrogen 1s electrons of pyridine (A), pyrimidine (B), and $s$-triazine (C), with the correlation-polarization potential (CP) or the equivalent-core orbitals (ECO) calculations compared to the experimental (EXP) values from this work extracted by Abel inversion. The CP calculations have been shifted by the same amount as in Fig.~\ref{fig:s2} and the ECO calculations have been shifted by an additional 3.7~eV, as mentioned in the text.}
    \label{fig:beta}
\end{figure}

In addition to determining time delays, the present work has also accessed the N-1s photoelectron asymmetry parameters of the azabenzene molecules, which have not been reported in the literature so far. Figure~\ref{fig:beta} compares the asymmetry parameters ($\beta_2$) measured in this work with the CP and ECO calculations. The $\beta_2$-values are found to increase from close to zero near threshold to values ranging from 1-1.25 around 450~eV. In all three molecules, we find reasonable agreement between the measured and calculated asymmetry parameters and a slightly better agreement for the CP calculations, which further justifies our choice made in the main text to focus the comparisons and discussions on these calculations.

\clearpage

\subsection{Comparison of Coulomb-laser coupling delays with TDSE and experiments}

\begin{figure}[h]
    \centering
    \includegraphics[width=\textwidth]{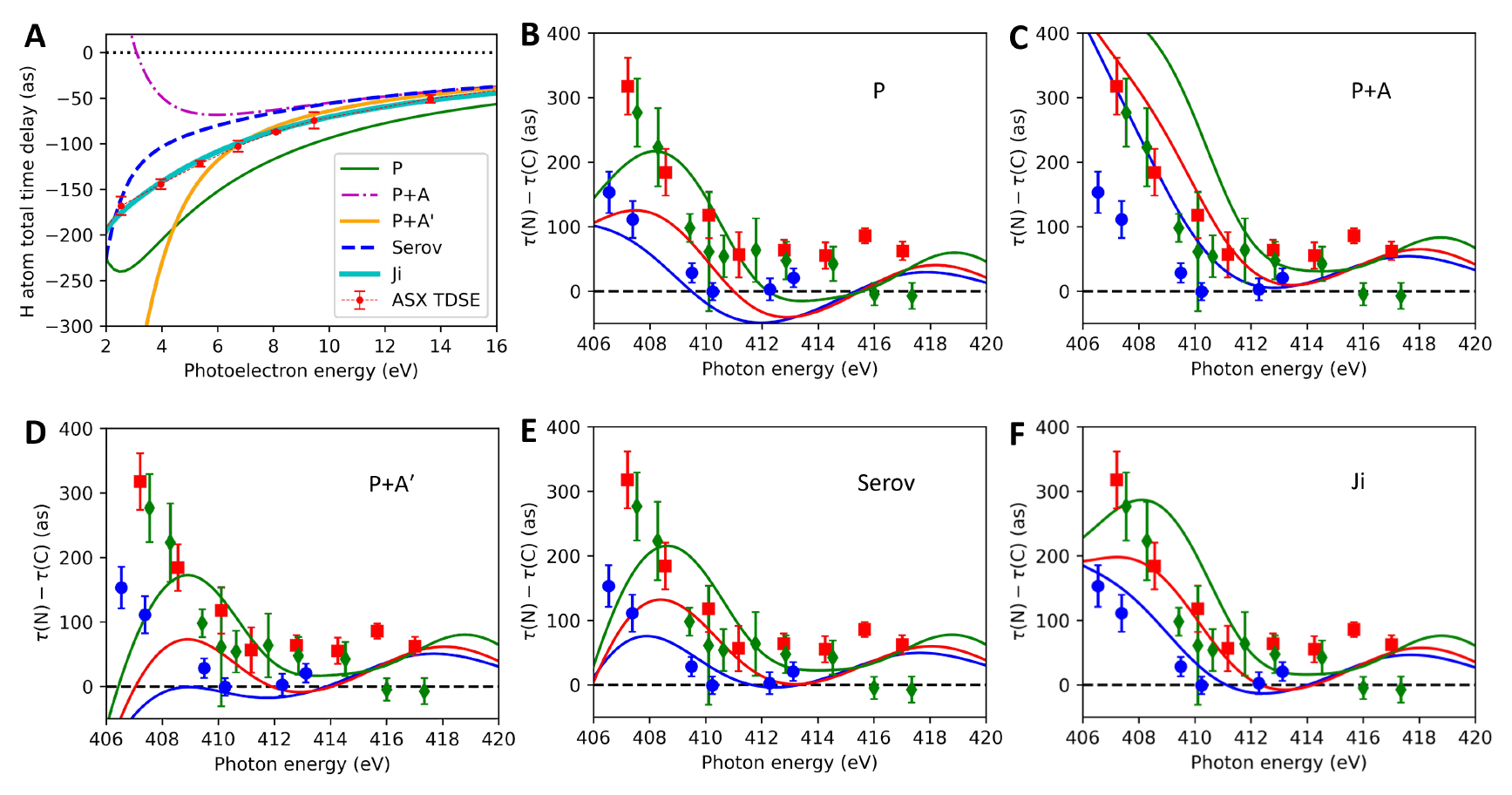}
    \caption{Comparison of different versions of the CLC delays with TDSE and experimental results. The results from TDSE calculations of attosecond angular streaking of atomic hydrogen (``ASX TDSE'') are compared in (A) to the sum of the atomic Wigner delay and the different variants of the CLC delays. ``Serov'' refers to the expression of the CLC delays given in Eq. (14.41) of Ref. \cite{serov_interpretation_2015}. In (B-F) the full lines are the total calculated delays, i.e. a sum of the one-photon-ionization delays calculated with the CP method and the CLC delays calculated according to different CLC formulae, specifically (B) P, (C) P+A, (D) P+A$^{\prime}$ in Ref. \cite{Dahlstrom2013}, (E) derived by Serov {\it et al} \cite{serov_interpretation_2015}, \textcolor{black}{and (F) derived by Ji {\it et al} \cite{ji2024analytical}.}}
    \label{fig:s6}
\end{figure}

Next, we discuss the performance of various analytical expressions of the CLC delays, which have been shown to be equivalent to the continuum-continuum delays of RABBIT \cite{pazourek15a}. The most widely used and simplest version consists of only the phase of the hydrogenic continuum-continuum-transition matrix element and is labeled ``P'' in Refs. \cite{dahlstrom12a,Dahlstrom2013}. A better approximation is obtained by adding correction factors for the long-range amplitudes of the continuum wave functions. This gives rise to the improved ``P+A'' (for phase+amplitude) version of the CLC delays. A third version of the CLC delays is obtained by moving the starting point of the integration of the radial continuum wave functions away from the origin, which is known as ``P+A$^{\prime}$'' \cite{Dahlstrom2013}. Besides, Serov et al. introduced yet another analytical expression of the CLC delays, which differs from the other ones, notably in depending on the Wigner delays (see Eq.(14.41) in Ref.~\cite{serov_interpretation_2015}). 
\textcolor{black}{
Recently, Ji et al. derived a new analytical formula for the radial integral of the continuum-continuum transition from hydrogen-like atoms, which shows perfect agreement with TDSE at low kinetic energies \cite{ji2024analytical}. Although the results presented in Ref. \cite{ji2024analytical} are based on the RABBIT scheme, the radial integral is generally applicable. For the streaking experiment, considering that the kinetic energy change of the photoelectron is a continuous quantity instead of plus or minus one photon energy as in the RABBIT scheme, the definition of the CLC delay is slightly modified as: 
\begin{equation}
    \tau_{\rm CLC}(E;|\Delta E|) = \frac{1}{2}
    \left( \frac{\partial}{\partial |\Delta E|} \phi_{\rm CC}^{\rm emi}(E;|\Delta E|) - \frac{\partial}{\partial |\Delta E|} \phi_{\rm CC}^{\rm abs}(E;|\Delta E|) \right) ~ ,
    \label{eq:CLC_delay_def}
\end{equation}
where $\phi_{\rm CC}^{\rm emi}(E;|\Delta E|)$ is the phase of the continuum-continuum transition of the emission pathway from $E + |\Delta E|$ to $E$, while $\phi_{\rm CC}^{\rm abs}(E;|\Delta E|)$ is the phase of the absorption pathway from $E - |\Delta E|$ to $E$. We note that the derivatives in Eq. (\ref{eq:CLC_delay_def}) have weak dependence on $|\Delta E|$, which may correspond to the weak intensity dependence of the streaking phase, as pointed out by Ref. \cite{kheifets_ionization_2022}. Here we have adjusted the energy change $|\Delta E| = 0.28$ eV in order to best match the TDSE simulation under similar conditions, as shown in Fig.~\ref{fig:s6}A, 
which compares the different versions of the CLC delays from Ref. \cite{dahlstrom12a}, added to the analytically known photoionization delays of the hydrogen atom, with TDSE calculations of attosecond angular streaking of atomic hydrogen ionized by an attosecond short-wavelength pulse \cite{kheifets_ionization_2022}. 
This comparison shows that the CLC delay defined in Eq. (\ref{eq:CLC_delay_def}) and calculated according to Ref. \cite{ji2024analytical} is indeed consistent with the other approaches at higher kinetic energies, while at lower kinetic energies it quantitatively reproduces the TDSE values.} 

Panels B-F of Fig.~\ref{fig:s6} show the comparison of the experimental data from this work with the azabenzene delays from the CP calculations to which we added the different variants of the CLC delays. As expected on the basis of the results shown in panel A, we find that our method yields the best agreement between theory and experiment. Therefore, we have chosen this approach to describing the CLC delays for the results shown in the main text.

\clearpage

\subsection{The role of post-collision interactions}

Finally, we discuss the role of the Auger-Meitner (AM) decay and the post-collision interaction (PCI) between the photoelectron and the AM electron. Following N-1s ionization, KVV AM decay creates AM electrons with kinetic energies of $\sim 380$~eV, which are not detected in our experiments because their momenta are larger that the maximal momentum that is projected on the cVMI detector. The 1s vacancy of nitrogen has an AM lifetime of 5.1~fs \cite{nicolas_lifetime_2012}. The N-1s photoelectron thus initially leaves behind a singly charged molecular cation. Since the AM electron is much faster than the photoelectron, the former will overtake the latter, which then evolves in the long-range potential of a doubly-charged cation. Since a rigorous quantum-mechanical description of the effect of AM decay and PCI on photoionization delays is not yet available and is beyond the scope of this work, we here resort to simple estimations.

\begin{figure}[h]
    \centering
    \includegraphics[width=\textwidth]{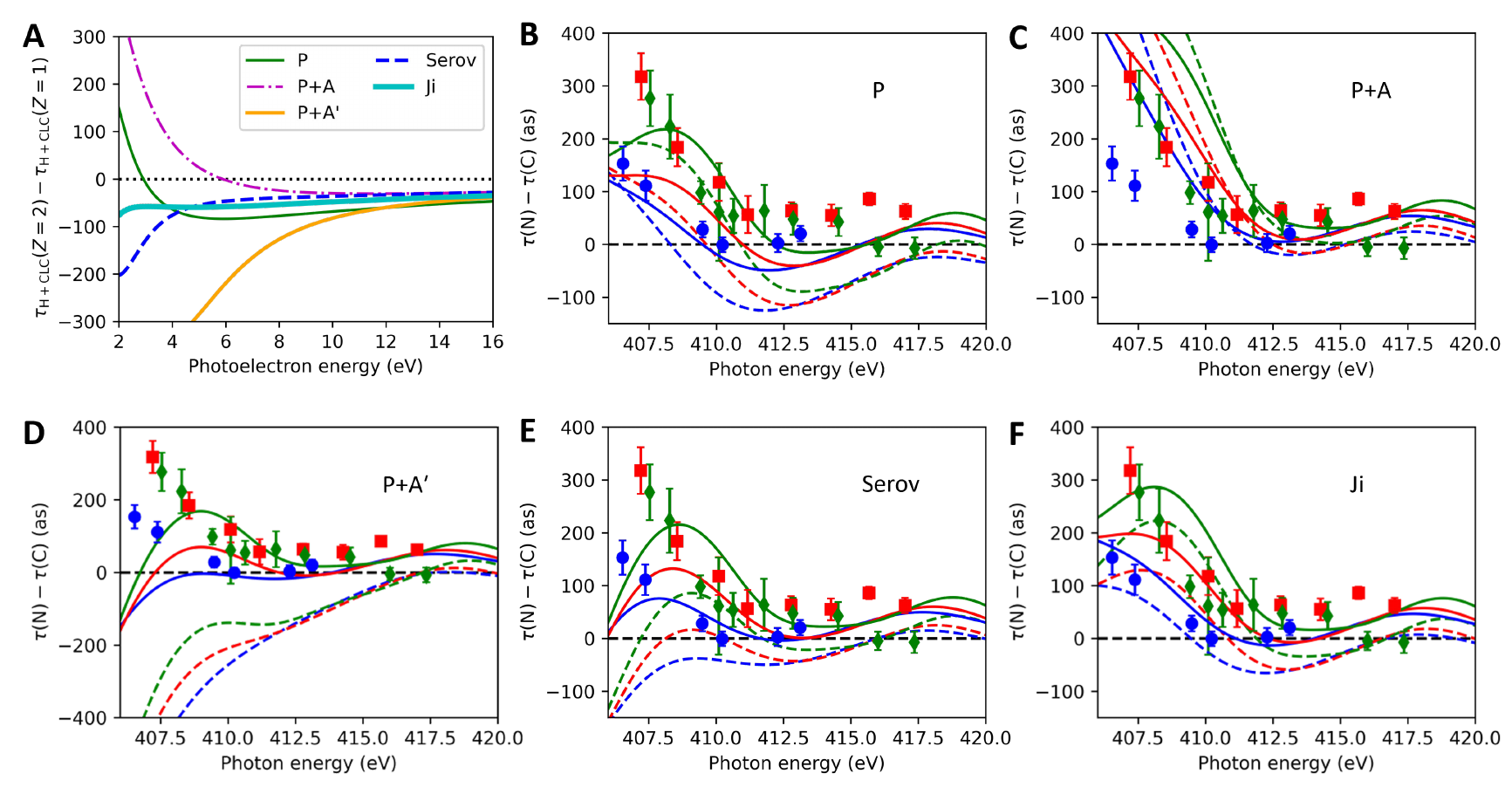}
    \caption{Illustration of the effect of PCI on photoionization delays. A) Difference of the total photoionization delays (Wigner + CLC) for a hydrogenic system with $Z=2$ vs. $Z=1$ ($L=0$ in both cases), i.e. $\tau_{{\rm H}, Z=2}^{\rm Wigner}$+$\tau_{Z=2}^{\rm CLC}$-$\tau_{{\rm H}, Z=1}^{\rm Wigner}$-$\tau_{Z=1}^{\rm CLC}$. B-F) The symbols with error bars are the experimental data and the full lines are the calculations, both reproduced from Fig.~3A of the main text. The dashed lines are obtained from the solid lines as follows: $\tau_{\rm dashed}=\tau_{\rm solid}+\tau_{{\rm H}, Z=2}^{\rm Wigner}+\tau_{Z=2}^{\rm CLC}-\tau_{{\rm H}, Z=1}^{\rm Wigner}-\tau_{Z=1}^{\rm CLC}$.
    }
    \label{fig:CLCPCI}
\end{figure}

First, it is worth recalling that photoionization (Wigner) delays can be decomposed into the sum of a delay originating from the Coulomb potential and a delay originating from the short-range potential \cite{pazourek15a}. Since both the Coulomb delays and the CLC delays are known analytically, we can express the change in these delays from $Z=1$ to $Z=2$ analytically. These results are shown in Fig. \ref{fig:CLCPCI}A, which shows the difference between the total photoionization delay (Wigner+CLC) for $Z=2$ and $Z=1$ for the different variants of the CLC correction discussed in the previous section. We find that these corrections all agree well above kinetic energies of $\sim$10~eV, but diverge at lower kinetic energies.

Next, exploiting the additivity of the short-range, Coulomb and CLC delays, we can ''correct'' the calculated time-delay results shown in the main text for the occurrence of the PCI effect, i.e. for the switch from $Z=1$ to $Z=2$ when the AM electron overtakes the photoelectron. For this purpose, we take the calculated Wigner delays (as shown, e.g. in Fig.~4D,E,F of the main text), add the difference of Coulomb delays for $Z=2$ and $Z=1$ ($L=0$ in both cases for 1s electrons) and add the CLC delay for $Z=2$ \cite{russek_post-collision_1986}. The resulting delays are shown as dashed lines in Fig. \ref{fig:CLCPCI}B-F, together with the results that ignore the PCI effect (from Fig.~\ref{fig:s6}) as full lines. The dashed and full lines can thus be viewed as the two limiting cases of an infinite and a vanishing AM lifetime, respectively. For most variants of the CLC delays, the discussed correction is minor, especially for the CLC \textcolor{black}{formula defined in Eq. (\ref{eq:CLC_delay_def}) (see section 4.2) that best agrees with the TDSE results (Fig.~\ref{fig:s6}F)}. We therefore conclude that the possible effect of PCI on the measured time delays is minor and can be safely neglected in the interpretation of the present results.

\clearpage
\clearpage
\bibliography{attobib,theory,references}